\documentclass [11pt, proquest] {uwthesis}[2016/11/22]
 
\usepackage{amssymb, amsthm}
\usepackage{amsmath}    
\usepackage{mathrsfs}   
\usepackage{graphicx}   
\usepackage{color}      
\usepackage{footnote}   

\usepackage{slashed}    
\usepackage{subfigure}  
\usepackage{hyperref}   

\usepackage{rotating}   
\usepackage{multirow}   

\begin{document}
 
\prelimpages
 
\Title{Possible Beyond the Standard Model Physics Motivated by Muonic Puzzles}
\Author{Yu-Sheng Liu}
\Year{2017}
\Program{Physics}

\Chair{Gerald A. Miller}{Prof.}{Physics}
\Signature{Sanjay K. Reddy}
\Signature{Jason Detwiler}

\titlepage

\newpage

\copyrightpage

\setcounter{page}{-1}

\abstract{
Recent measurements of the proton radius using the Lamb shift in muonic hydrogen are troublingly discrepant with values extracted from hydrogen spectroscopy and electron-proton scattering experiments. This discrepancy, which differs by more than five standard deviations, may be a signal of new physics caused by a violation of lepton universality. Another candidate for a new physics signal is the muon anomalous magnetic moment. The measurement at BNL differs from the standard model prediction by at least three standard deviations. Motivated by these two puzzles, first we use polarized lepton-nucleon elastic scattering to search for a new scalar boson, and furthermore we suggest new measurements of the nucleon form factors. Next, we display a method to analyze the beam dump experiments without using approximation on phase space, and we use it to constrain all possible new spin-0 and spin-1 particles, and a variety of other measurements to study the possibility of the new physics. Finally, assuming a new scalar boson can solve the two puzzles simultaneously, we present a general model-independent analysis and constrain the existence of the new physics.
}

\tableofcontents
\listoffigures




\textpages
 
\chapter{Introduction}

Interest in a new light bosons arises from recent studies of the proton radius using the Lamb shift in the $2{\rm S}_{1/2}-2{\rm P}_{3/2}$ transition in muonic hydrogen \cite{Pohl:2010zza,Antognini:1900ns}. The value of the proton radius was reported to be 0.84087(39) fm, whereas the CODATA (2014) value \cite{Mohr:2015ccw} 0.8751(61) fm differs by five standard deviations. The major difference between these two reported data is that the former extracts the proton radius from muonic hydrogen and the latter from electronic hydrogen and electron-proton scattering experiments. Although the different proton radius may arise due to subtle lepton-nucleon non-perturbative effects within the standard model \cite{Hill:2011wy,Pohl:2013yb} or the extraction mehthod from electron scattering \cite{Hill:2010yb}, it could also be a signal of new physics caused by a violation of lepton universality. A new boson which couples preferably to the muon and proton \cite{TuckerSmith:2010ra} could be the explanation.

Another candidate for a new physics signal is the muon anomalous magnetic moment, defined as $a_\mu =\frac{(g-2)_\mu}{2}$. The measurement at BNL \cite{Blum:2013xva} differs from the standard model prediction by at least three standard deviations\footnote{The experimental and the standard model uncertainty are $63\times 10^{-11}$ and $\simeq 49\times 10^{-11}$, respectively.}
\begin{align}
\Delta a_\mu=a_\mu^{\rm exp}-a_\mu^{\rm th}=&(287\pm 80)\times 10^{-11}\;\cite{Davier:2010nc}\\
=&(261\pm 78)\times 10^{-11}\;\cite{Hagiwara:2011af}
\end{align}
where the different values depend on the choice of lowest order hadronic contribution. This discrepancy is also possibly explained by the same new physics as in the proton radius puzzle.

Motivated by these two muonic puzzles, we look for new physics as a possible solution. In the first example (chapter \ref{ch:LN scattering}), lepton-nucleon elastic scattering experiments, using the one-photon and one-scalar-boson exchange mechanisms considering all possible polarizations, is used to study searches for a new scalar boson. We show that the scalar boson produces relatively large effects in certain kinematic region when using sufficient control of lepton and nucleon spin polarization. Furthermore we suggest new measurements of the nucleon form factors. We generalize current techniques to measure the ratio $G_E/G_M$ and present a new method to separately measure $G^2_M$ and $G^2_E$ using polarized incoming and outgoing muons.

In chapter \ref{ch:beam dump}, we investigate beam dump experiments used to search for for new physics. Beam dump experiments have been used to search for new particles with null results interpreted in terms of limits on masses $m_\phi$ and coupling constants $\epsilon$. However these limits have been obtained by using approximations [including the Weizs\"{a}cker-Williams (WW) approximation] or Monte-Carlo simulations. We display methods to obtain the cross section (for all spin-0 and spin-1 particles) and the resulting particle production numbers without using approximations or Monte-Carlo simulations. We show that the approximations cannot be used to obtain accurate values of cross sections. The corresponding exclusion plots differ by substantial amounts when seen on a linear scale. To understand the impact of a discovery, we generate pseudodata (assuming given values of $m_\phi$ and $\epsilon$) in the currently allowed regions of parameter space. The use of approximations to analyze the pseudodata for the future experiments is shown to lead to considerable errors in determining the parameters. Furthermore, we find that a new region of parameter space can be explored without using one of the common approximations, $m_\phi\gg m_e$. Our method can be used as a consistency check for Monte-Carlo simulations.

In chapter \ref{ch:boson exclusion}, we assume that the existence of a new particle, $\phi$, resolves the source of these two puzzles. To study this hypothetical $\phi$, we present a systematic method to analyze the parameter space of the new physics. For spin-0 and spin-1 particles, the only possible candidate is scalar boson because the pseudoscalar and axial-vector bosons have contribution to $(g-2)_\mu$ with the wrong sign, and the vector boson contribution to the proton radius puzzle is ruled out by the hyperfine splitting of muonic hydrogen. Using a variety of measurements, we constrain the mass of the new scalar boson and their couplings to the electron, muon, neutron, and proton. Making no assumptions about the underlying model, these constraints and the requirement that it solve both problems limit the mass of the bosons to between about 100 keV and 100 MeV. We identify unexplored regions in the coupling constant-mass plane to be covered by experiments. Potential future experiments and their implications for theories with mass-weighted lepton couplings are discussed.

A conclusion is presented in chapter \ref{ch:conclusion}.

\chapter{Polarized Lepton-Nucleon Elastic Scattering}\label{ch:LN scattering}

\section{Introduction}

Lepton-nucleon elastic scattering is important both in theory and experiment. The use of polarization techniques has yielded much new information. In this chapter, we study lepton-nucleon elastic scattering using the one-photon and one-scalar-boson exchange mechanism by considering all possible polarizations. The differential cross sections are calculated in a general reference frame. There are two applications of our results. The first one is searching for a new light scalar boson. The second one, using only the one-photon exchange contribution, is to provide new ways to measure the nucleon form factor: generalizing current techniques to measure the ratio $G_E/G_M$ and presenting a new method to separately measure $G_M^2$ and $G_E^2$ using polarized incoming and outgoing muons.

It is known that a light scalar boson with mass around 1 MeV is a candidate to explain both proton radius and muon anomalous magnetic moment puzzles simultaneously \cite{Pohl:2013yb,TuckerSmith:2010ra,Izaguirre:2014cza}. The non-relativistic potential between lepton and nucleon caused by exchange of a scalar boson is written as
\begin{align}
V_\phi(r)=-\frac{g_lg_N}{4\pi}\frac{e^{-m_\phi r}}{r}=\epsilon_l\epsilon_N V_{EM}(r)e^{-m_\phi r},
\end{align}
where $m_\phi$ is the mass of the scalar boson; $g_l$ and $g_N$ are scalar bosons coupling to lepton and nucleon, respectively; $\epsilon=g/e$ is the relative coupling strength. We adopt the constraints in \cite{TuckerSmith:2010ra} that for $m_\phi=1$ MeV
\begin{align}\label{eq:constraints on g}
\epsilon_e\lesssim 2.3\times10^{-4},\;\epsilon_p=\epsilon_\mu\lesssim 1.3\times10^{-3},\;\epsilon_n\lesssim 6.7\times10^{-5}.
\end{align}
The potential of the scalar boson is suppressed by $\epsilon_l\epsilon_N$ and an exponentially decay factor comparing with the Coulomb potential. The fact that the potential of the scalar boson is intrinsically much smaller than the Coulomb potential makes it hard to find this new boson.

There are proposals to search for the light scalar boson, such as a direct detection method \cite{Izaguirre:2014cza}. In this chapter, we study the cross section of elastic lepton-nucleon scattering caused by one-photon and one-scalar-boson exchange. Since the muon is much heavier than the electron, the lepton mass can not be neglected, as is done for electron-proton scattering. The two-photon exchange contribution is expected (from perturbation expansion) and measured to be at a few percent level compared with one-photon exchange (OPE) \cite{Blunden:2005ew,Afanasev:2005mp,Arrington:2007ux,Carlson:2007sp}. Therefore, the effect of one-scalar-boson exchange must be greater than few percent of OPE to be observed. We consider unpolarized and polarized elastic scattering cross sections. We find the following two cases that for certain kinematic regions where the effects of the scalar boson are dominant and potentially observable: electron-neutron cross section with incoming and outgoing electrons polarized, and muon-neutron cross section with incoming and outgoing neutrons polarized. 

There are several facilities that can measure the lepton-nucleon elastic scattering, for electron, such as JLab, and for muon, MUSE at PSI and the J-PARC muon facility. Based on our results and the current experimental setup and capability \cite{SoLID:2013fta,MUSE}, although the polar angle resolution is sufficient, the kinematic region where scalar bosons may have significant effects is beyond the current polar angle measured range; the electron-neutron scattering is more promising than muon-neutron scattering due to the high intensity of the electron flux. Further estimates and discussions are in section \ref{sec:searching for phi_leptons polarized} and \ref{sec:searching for phi_nucleons polarized}.

The polarized cross section with one-photon exchange is used to measure the Sachs form factors, $G_M$ and $G_E$ \cite{Rock:2001zi,Arrington:2011kb,Punjabi:2014tna}. The standard technique of measurement is using elastic electron-nucleon scattering using the following experiments: unpolarized Rosenbluth separation \cite{Rosenbluth:1950yq}, polarized lepton beam and nucleon target, and polarized lepton beam and recoil nucleon. The latter two experiments, using the ratio technique (also known as polarization transfer method), were first developed in \cite{Akhiezer:1974em} and later discussed in more detail in \cite{Arnold:1980zj}. We explore other possible ways to determine form factors by including lepton masses and other lepton polarization configurations different from conventional ones. In section \ref{sec:form factor ij}, we generalize the current method to measure the ratio of form factors, $G_E/G_M$, by including lepton mass, non-longitudinal lepton polarization, and more polarization configurations (polarized one lepton and one nucleon, either they are incoming or outgoing). In section \ref{sec:form factor 13}, we present a new method to measure $G_E^2$ or $G_M^2$ directly for certain kinematic conditions in elastic muon-nucleon scattering cross section with polarized incoming and outgoing muons.

The outline of this chapter is as follows. In section \ref{sec:setup}, we set up the formalism needed to compute cross sections. In section \ref{sec:LN cross section}, the cross sections are calculated in a general reference frame with all possible polarizations, and the massless lepton limit is slso discussed. In section \ref{sec:searching for a scalar boson}, we show that there are two cases in which the scalar boson is potentially observable: electron-neutron cross section with incoming and outgoing electrons polarized, and muon-neutron cross section with incoming and outgoing neutrons polarized. In section \ref{sec:measuring form factors}, comparing with the current method, we give a more general result for the ratio of form factors, $G_E/G_M$, by including lepton mass, non-longitudinal lepton polarization, and more polarization configurations. We also discuss new measurements to selectively obtain the contribution from $G_E^2$ or $G_M^2$ in the cross section with polarized incoming and outgoing muons. A discussion is presented in section \ref{sec:conclusions}.

\section{Setup\label{sec:setup}}
The elastic lepton-nucleon scattering process is denoted as
\begin{align}
l(p_1,s_1)+N(p_2,s_2)\rightarrow l(p_3,s_3)+N(p_4,s_4),
\end{align}
where $l$ and $N$ stand for lepton and nucleon; $p$ and $s$ are momentum and spin polarization; the number 1, 2, 3, and 4 label the incoming lepton, incoming nucleon, outgoing lepton, and outgoing nucleon. This notation is used throughout the entire paper. In the lowest order, we consider the interaction by exchanging a scalar boson or a photon between the lepton and the nucleon. There are three contributions to the cross section: one-photon exchange, one-scalar-boson exchange, and the interference terms.

\subsection{Kinematics}
In the nucleon rest frame (lab frame), we choose the the coordinate such that $\mathbf{p}_1$ is along $z$ axis and $\mathbf{p}_3$ is in $x$-$z$ plane to exploit the symmetry. With these choices, all the kinematics quantities which we need can be expressed in terms of only two variables, $|\mathbf{p}_1|$ and the scattering angle $\theta$. The scattering angle is an angle between outgoing and incoming lepton, and defined as $\cos\theta=\hat{\mathbf{p}}_1\cdotp\hat{\mathbf{p}}_3$. In the lab frame, the external momenta and space-like momentum transfer $q$ can be expressed as follows
\begin{align}
\mathbf{p}_1&=|\mathbf{p}_1|\hat{\mathbf{z}},\; E_1=\sqrt{|\mathbf{p}_1|^2+m_l^2},\;\mathbf{p}_2=\mathbf{0},\; E_2=m_N,\\
\mathbf{p}_3&=\frac{(m_l^2+E_1m_N)\cos\theta+(E_1+m_N)\sqrt{m_N^2-m_l^2\sin^2\theta}}{(E_1+m_N)^2-\cos^2\theta|\mathbf{p}_1|^2}|\mathbf{p}_1|\hat{\mathbf{p}_3},\\
\hat{\mathbf{p}_3}&=\sin\theta\hat{\mathbf{x}}+\cos\theta\hat{\mathbf{z}},\\
E_3&=\frac{(E_1+m_N)(m_l^2+E_1m_N)+\cos\theta|\mathbf{p}_1|^2\sqrt{m_N^2-m_l^2\sin^2\theta}}{(E_1+m_N)^2-\cos^2\theta|\mathbf{p}_1|^2}\\
\mathbf{p}_4&=\mathbf{p}_1-\mathbf{p}_3,\; E_4=E_1+m_N-E_3,\\
q&=p_1-p_3=p_4-p_2,\\
q^2&=\frac{2|\mathbf{p}_1|^2m_N(m_N+E_1\sin^2\theta-\cos\theta\sqrt{m_N^2-m_l^2\sin^2\theta})}{(E_1+m_N)^2-\cos^2\theta|\mathbf{p}_1|^2}.
\end{align}
We use the mostly-plus metric, so that $q^2>0$ if $q$ is space-like. If we take spin polarization into account, each polarized particle needs two angular variables to specify the spin direction, see section \ref{sec:polarization} below.

\subsection{Dynamics}
\begin{figure}
\centering
\includegraphics[scale=1.2]{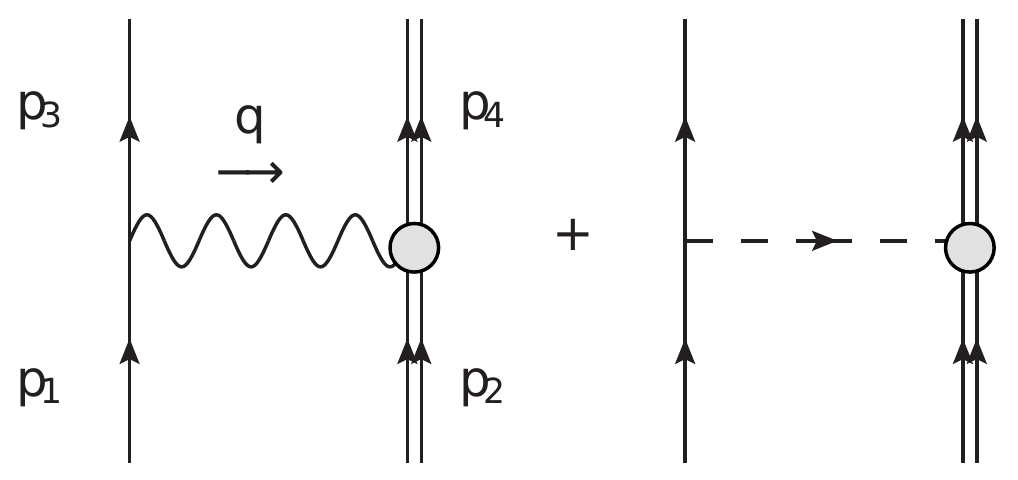}
\caption{\label{fig:amplitude} The tree level amplitudes of lepton-nucleon elastic scattering: the single (double) line on the left (right) denotes lepton (nucleon); the wavy (dashed) line denotes photon (scalar boson).}
\end{figure}

The scalar boson interacts with leptons and nucleons through Yukawa couplings
\begin{align}
\mathcal{L}_\phi\supset-\frac{1}{2}(\partial\phi)^2-\frac{1}{2}m_\phi^2\phi^2+e\epsilon_l\phi\bar{\psi}_l\psi_l+e\epsilon_N\phi\bar{\psi}_N\psi_N,
\end{align}
where $\epsilon=g/e$; $g_l$ and $g_N$ are Yukawa couplings between lepton-scalar and nucleon-scalar. In general, the couplings $g_l$ and $g_N$ can be either positive and negative. There are two lowest order diagrams: one-photon and one-scalar-boson exchange, see figure \ref{fig:amplitude}.

The amplitude squared is given by
\begin{align}\label{eq:amplitudes}
|\mathcal{T}|^2=e_l^2 e_N^2\left(I_{\gamma\gamma}+\lambda I_{\gamma\phi}+\lambda^2I_{\phi\phi}\right),
\end{align}
where $e_l$ and $e_N$ are $U(1)_{EM}$ couplings of leptons and nucleons, respectively; $\lambda$ and $I_{xy}$ using the mostly-plus metric are given by
\begin{align}
\lambda(q^2)=&\epsilon_l\epsilon_N\frac{q^2}{q^2+m_\phi^2}\label{eq:lambda}\\
I_{\gamma\gamma}=&\frac{1}{q^4}{\rm Tr}(u_3\bar{u}_3\gamma_\mu u_1\bar{u}_1\gamma_\nu){\rm Tr}(u_4\bar{u}_4 V^\mu u_2\bar{u}_2 V^\nu)\\
I_{\gamma\phi}=&\frac{1}{q^4}\Big[{\rm Tr}(u_3\bar{u}_3\gamma_\mu u_1\bar{u}_1){\rm Tr}(u_4\bar{u}_4 V^\mu u_2\bar{u}_2)\nonumber\\
&\quad+{\rm Tr}(u_3\bar{u}_3 u_1\bar{u}_1\gamma_\mu){\rm Tr}(u_4\bar{u}_4 u_2\bar{u}_2 V^\mu)\Big]\\
I_{\phi\phi}=&\frac{1}{q^4}{\rm Tr}(u_3\bar{u}_3 u_1\bar{u}_1){\rm Tr}(u_4\bar{u}_4 u_2\bar{u}_2)\\
V^\mu(q)=&F_1(q^2)\gamma^\mu-\frac{i}{2m_N}F_2(q^2)\sigma^{\mu\nu}q_\nu
\end{align}
where $m_\phi$ is mass of the scalar boson; $\sigma^{\mu\nu}=\frac{i}{2}[\gamma^\mu,\gamma^\nu]$; $F_1(q^2)$ and $F_2(q^2)$ are the form factors of nucleon $U(1)_{EM}$ coupling. In general, there is a form factor, $F_3(q^2)$, of scalar-nucleon coupling, however, it always appears with $g_N$, therefore we can include it into the constraint of $g_N$.

\subsection{Polarization \label{sec:polarization}}
The product of spinors $u\bar{u}$ in the mostly-plus metric is given by
\begin{align}\label{eq:uu}
u\bar{u}=\frac{1}{2}(1-\gamma_5\slashed{s})(-\slashed{p}+m)
\end{align}
where $s$ is the spin polarization which satisfies $s\cdotp p=0$ and $s^2=1$. We can solve for the spin polarization using the two constraints
\begin{align}
s=z\frac{E}{\sqrt{m^2+|\mathbf{p}|^2\sin^2\alpha}}\left(\frac{|\mathbf{p}|}{E}\cos\alpha,\hat{\mathbf{s}}\right),
\end{align}
where $\alpha$ and $\beta$ are the polar and azimuthal angle of $\mathbf{s}$ with respect to $\mathbf{p}$; $\cos\alpha=\hat{\mathbf{p}}\cdotp\hat{\mathbf{s}}$; $z=\pm 1$ and becomes helicity if the particle is longitudinally polarized.

The polarization of leptons and nucleons in the lab frame in the coordinate system we chose can be expressed as
\begin{align}
\hat{\mathbf{s}}_i=&\sin\alpha_i\cos\beta_i\hat{\mathbf{x}}+\sin\alpha_i\sin\beta_i\hat{\mathbf{y}}+\cos\alpha_i\hat{\mathbf{z}}\quad\text{for $i=$1 or 2}\\
\hat{\mathbf{s}}_3=&\sin\alpha_3\cos\beta_3(\hat{\mathbf{y}}\times\hat{\mathbf{p}}_3)+\sin\alpha_3\sin\beta_3\hat{\mathbf{y}}+\cos\alpha_3\hat{\mathbf{p}}_3\nonumber\\
=&(\sin\alpha_3\cos\beta_3\cos\theta+\cos\alpha_3\sin\theta)\hat{\mathbf{x}}+\sin\alpha_3\sin\beta_3\hat{\mathbf{y}}\nonumber\\
&+(-\sin\alpha_3\cos\beta_3\sin\theta+\cos\alpha_3\cos\theta)\hat{\mathbf{\mathbf{z}}}\\
\hat{\mathbf{s}}_4=&(\sin\alpha_4\cos\beta_4\cos\theta'-\cos\alpha_4\sin\theta')\hat{\mathbf{x}}+\sin\alpha_4\sin\beta_4\hat{\mathbf{y}}\nonumber\\
&+(\sin\alpha_4\cos\beta_4\sin\theta'+\cos\alpha_4\cos\theta')\hat{\mathbf{\mathbf{z}}}
\end{align}
where $\theta'$ is angle between $\mathbf{p}_4$ and $\mathbf{p}_1$; $\cos\theta'=\hat{\mathbf{p}}_4\cdotp\hat{\mathbf{p}}_1=\frac{|\mathbf{p}_1|-|\mathbf{p}_3|\cos\theta}{\sqrt{E_4^2-m_N^2}}$.

There are two important special cases. For transverse polarization $\alpha_T=\frac{\pi}{2}$, $s_T=z(0,\hat{\mathbf{s}})$. For longitudinal polarization $\alpha_L=0$,
\begin{align}\label{eq:sL}
s_L=z\frac{E}{m}\left(\frac{|\mathbf{p}|}{E},\hat{\mathbf{p}}\right).
\end{align}
This is useful for a light lepton because if its mass is small compared with the beam energy, the light lepton is naturally longitudinally polarized. 

\subsection{Massless Particle \label{sec:polarization of massless particle}}
The massless lepton limit is important for electron scattering or the high energy limit. In general, naively taking $m_l\rightarrow 0$ in physical quantities does not give us the correct result (see section \ref{sec:massless lepton} below), because the longitudinal polarization (\ref{eq:sL}) blows up. The correct way to take the limit is to set the polarization to be longitudinal, $s_\mu\rightarrow p_\mu/m_l$, then take the massless limit. Then the product of spinors (\ref{eq:uu}) becomes
\begin{align}
u\bar{u}=-\frac{1}{2}(1+z\gamma_5)\slashed{p}{\rm\; (massless)},
\end{align}
where $z=\pm 1$ is the helicity of the lepton.

\section{Cross Section\label{sec:LN cross section}}
Combining with the amplitude squared (\ref{eq:amplitudes}), the differential cross section in a general reference frame is
\begin{align}\label{eq:cross section}
\frac{d\sigma}{d\Omega}=A(I_{\gamma\gamma}+\lambda I_{\gamma\phi}+\lambda^2 I_{\phi\phi})=\left(\frac{d\sigma}{d\Omega}\right)_{\gamma\gamma}+\left(\frac{d\sigma}{d\Omega}\right)_{\gamma\phi}+\left(\frac{d\sigma}{d\Omega}\right)_{\phi\phi},
\end{align}
where
\begin{align}
A=\frac{\alpha^2}{4}\frac{|\mathbf{p}_3|}{|E_2\mathbf{p}_1-E_1\mathbf{p}_2|}\left[E_1+E_2-E_3\frac{(\mathbf{p}_1+\mathbf{p}_2)\cdotp\hat{\mathbf{p}}_3}{|\mathbf{p}_3|}\right]^{-1}.
\end{align}
In the lab frame, the kinematic factor $A$ becomes
\begin{align}
A_{\rm lab}=\frac{\alpha^2}{4m_N}\frac{|\mathbf{p}_3|}{|\mathbf{p}_1|}\left[E_1+m_N-E_3\frac{|\mathbf{p}_1|}{|\mathbf{p}_3|}\cos\theta\right]^{-1}.
\end{align}

In (\ref{eq:cross section}), it is worth mentioning that the pure one-photon or one-scalar-boson exchange contribution, $\left(\frac{d\sigma}{d\Omega}\right)_{\gamma\gamma}$ and $\left(\frac{d\sigma}{d\Omega}\right)_{\phi\phi}$, is always greater or equal to zero, whereas the interference term $\left(\frac{d\sigma}{d\Omega}\right)_{\gamma\phi}$ can be positive or negative depending on the signs of Yukawa couplings $g$.

\subsection{Unpolarized}
The cross section (\ref{eq:cross section}) is proportional to $I_{\gamma\gamma}$, $I_{\gamma\phi}$, and $I_{\phi\phi}$, so it is sufficient to show these three $I$'s instead of full cross section. We will use superscript to indicate which particle is polarized, e.g., $I^{1,4}$ means incoming lepton and outgoing nucleon are polarized, $I^{1,2,3,4}$ stands for fully polarized, etc. The unpolarized $I$'s are defined as
\begin{align}
I^{\rm u}=\left(\frac{1}{2}\sum_{z_1}\right)\left(\frac{1}{2}\sum_{z_2}\right)\sum_{z_3}\sum_{z_4}I^{1,2,3,4}
\end{align}
where the superscript u refers to unpolarized. In a general reference frame,
\begin{align}
&I^{\rm u}_{\gamma\gamma}=\frac{G_E^2+\tau G_M^2}{1+\tau}\left(R^2-\frac{1+\tau}{\tau}\right)+G_M^2\left(2-\frac{m_l^2}{\tau m_N^2}\right)\\
&I^{\rm u}_{\gamma\phi}=-2G_E R\frac{m_l}{\tau m_N}\label{eq:I_gp unpol}\\
&I^{\rm u}_{\phi\phi}=\frac{1+\tau}{\tau}\left(1+\frac{m_l^2}{\tau m_N^2}\right)
\end{align}
where $G_M$ and $G_E$ are Sachs form factors
\begin{align}
G_M(q^2)=F_1(q^2)+F_2(q^2),\; G_E(q^2)=F_1(q^2)-\tau F_2(q^2);
\end{align}
$\tau=\frac{q^2}{4m_N^2}$ is defined in the usual way\footnote{Note that there is no minus sign in mostly-plus metric, and $\tau>0$ if $q$ is space-like.}; $R$ depends on Mandelstam variables\footnote{$s=-(p_1+p_2)^2$, $t=-(p_1-p_3)^2=-q^2$, and $u=-(p_1-p_4)^2$.}
\begin{align}
R=\frac{u-s}{t}=\frac{(p_1+p_3)\cdotp p_2}{(p_1-p_3)\cdotp p_2}=-\frac{4 p_1\cdotp p_2}{q^2}-1.
\end{align}
As an example, in the lab frame, $R_{\rm lab}=\frac{E_1+E_3}{E_1-E_3}$, with massless lepton limit ($m_l\rightarrow0$), the Rosenbluth cross section is 
\begin{align}
\left(\frac{d\sigma}{d\Omega}\right)_{\rm Rosenbluth}=\left(AI^{\rm u}_{\gamma\gamma}\right)_{{\rm lab},m_l\rightarrow0}.
\end{align}

The interference term (\ref{eq:I_gp unpol}) is proportional to lepton mass $m_l$. The interference term for muon is about two order of magnitude bigger than for electron, see figure \ref{fig:unpolarized cross section}.

\subsection{Partially Polarized\label{sec:partially polarized}}
There is no contribution to cross section if only one particle is polarized in exchanging one-photon and one-scalar-boson. The reason is the following. The parity flips momentum and time reversal flips momentum and angular momentum. One can only flip spin polarization by combining parity and time reversal, and it is equivalent to change the overall sign of the polarization. Therefore, if the theory conserves $PT$, the spin asymmetry part can depend only on product of even number of spin polarizations to remain invariant under $PT$. For example, if we polarize three particles, there is no term depending on the product of all three polarizations, but there are terms depending on product of two polarizations, see section \ref{sec:fully polarized}. On the other hand, one can consider an interaction which breaks time reversal to have an additional dependence on odd number of spin polarizations, such as exchanging a Z boson. In conclusion, in exchanging a photon and a scalar boson, one needs to polarize at least two particles to have a spin dependent part.

\subsubsection{One Lepton and One Nucleon Polarized}
First we study the case that one lepton and one nucleon are polarized (each lepton and nucleon can be incoming or outgoing). In order to combine all four cases into one expression $I^{i,j}$, we require that the first superscript, $i$, to be lepton (1 or 3); the second superscript, $j$, to be nucleon (2 or 4).
\begin{align}
I^{i,j}=a^{i,j}\sum_{z_{k}}\sum_{z_{l}}I^{i,j,k,l}=a^{i,j}\Big[I^{\rm u}+s_\mu^i s_\nu^j(J^{i,j})^{\mu\nu}\Big]
\end{align}
where $I^{\rm u}$ is the unpolarized part; $i$, $j$ are particle labels and not summed; $a^{i,j}$ takes care of the factors $\frac{1}{2}$ due to the difference between average or sum over initial or final states
\begin{align}
a^{i,j}=\begin{cases}
1 & \text{if }(i,j)=(1,2)\\
\frac{1}{2} & \text{if }(i,j)=(1,4)\text{ or }(3,2)\\
\frac{1}{4} & \text{if }(i,j)=(3,4)\\
\end{cases} 
\end{align}
or more compactly $a^{i,j}=\left(\frac{1}{2}\right)^{\frac{i+j-3}{2}}$. The results are 
\begin{align}
(J_{\gamma\gamma}^{i,j})^{\mu\nu}=&2\frac{m_l}{m_N}G_M\Bigg[b^j F_2\frac{(p_2+p_4)^\mu q^\nu}{q^2}-\frac{G_E}{\tau}\Bigg(g^{\mu\nu}-\frac{q^\mu q^\nu}{q^2}\Bigg)\Bigg]\label{eq:I_gg^ij 1}\\
(J_{\gamma\phi}^{i,j})^{\mu\nu}=&2G_M\Bigg[R\Bigg(g^{\mu\nu}-\frac{q^\mu q^\nu}{q^2}\Bigg)+\frac{(p_2+p_4)^\mu(p_1+p_3)^\nu}{q^2}\Bigg]\\
(J_{\phi\phi}^{i,j})^{\mu\nu}=&0
\end{align}
where $b^j$ depends only on nucleon label
\begin{align}
b^j=\begin{cases}
1 & \text{if } j=2\\
-1 & \text{if } j=4\\
\end{cases} 
\end{align}
or more compactly $b^j=(-1)^{\frac{j}{2}+1}$. The expression of $(J_{\gamma\gamma}^{i,j})^{\mu\nu}$ (\ref{eq:I_gg^ij 1}) is the most compact form we found. However, the following expression for $(J_{\gamma\gamma}^{i,j})^{\mu\nu}$
\begin{align}\label{eq:I_gg^ij 2nd}
(J_{\gamma\gamma}^{i,j})^{\mu\nu}=2\frac{m_l}{m_N}\frac{G_M}{1+\tau}\Bigg[&b^j G_M\frac{(p_2+p_4)^\mu q^\nu}{q^2}-G_E\Bigg(2b^j \frac{p_j^\mu q^\nu}{q^2}+\frac{1+\tau}{\tau}g^{\mu\nu}-\frac{q^\mu q^\nu}{\tau q^2}\Bigg)\Bigg]
\end{align}
is more useful for measuring the Sachs form factors, $G_M$ and $G_E$, by polarized method, which is discussed in section \ref{sec:measuring form factors}.

It is worth noticing that it seems that in the massless lepton limit $(J_{\gamma\gamma}^{i,j})^{\mu\nu}$ vanishes, however it is not true because there is another $m_l$ in the denominator when the lepton spin polarization becomes longitudinal (\ref{eq:sL}). Therefore, this expression is well-behaved in the massless limit. We discuss the massless limit in detail in section \ref{sec:massless lepton}.

\subsubsection{Incoming and Outgoing Leptons Polarized}
For polarized incoming and outgoing leptons, the observable quantity is
\begin{align}
I^{1,3}=\left(\frac{1}{2}\sum_{z_2}\right)\sum_{z_4}I^{1,2,3,4}=\frac{1}{2}(1+s^1\cdotp s^3)I^{\rm u}+s_\mu^1 s_\nu^3(J^{1,3})^{\mu\nu}.
\end{align}
The results are
\begin{align}
(J_{\gamma\gamma}^{1,3})^{\mu\nu}=&-G_M^2 g^{\mu\nu}+2\frac{G_E^2+\tau G_M^2}{1+\tau}\Bigg(\frac{q^\mu q^\nu}{2\tau q^2}-\frac{p_2^{\{\mu} p_4^{\nu\}}+R q_{\phantom{1}}^{[\mu} p_2^{\nu]}}{q^2}\Bigg)\\
(J_{\gamma\phi}^{1,3})^{\mu\nu}=&2\frac{m_l}{\tau m_N}G_E \frac{q_{\phantom{1}}^{[\mu} p_2^{\nu]}}{q^2}\\
(J_{\phi\phi}^{1,3})^{\mu\nu}=&-\frac{1+\tau}{\tau} \frac{q^\mu q^\nu}{q^2}
\end{align}
where the curly and square brackets are symmetrization and anti-symmetrization notation
\begin{align}
p^{\{\mu}q^{\nu\}}=p^\mu q^\nu+p^\nu q^\mu,\; p^{[\mu}q^{\nu]}=p^\mu q^\nu-p^\nu q^\mu.
\end{align}
The expression in terms of $(J_{\gamma\gamma}^{1,3})^{\mu\nu}$ is the most compact form we found. The following expression of $I_{\gamma\gamma}^{1,3}$
\begin{align}\label{eq:I_gg^13 2nd}
I_{\gamma\gamma}^{1,3}=\frac{1}{2}I^{\rm u}_{\gamma\gamma}+s_\mu^1 s_\nu^3\Bigg\{&-\frac{1}{2}G_M^2\frac{m_l^2}{\tau m_N^2} g^{\mu\nu}+2\frac{G_E^2+\tau G_M^2}{1+\tau}\Bigg[\nonumber\\
&\frac{g^{\mu\nu}}{4}\Bigg(R^2-\frac{1+\tau}{\tau}\Bigg)+\frac{q^\mu q^\nu}{2\tau q^2}-\frac{p_2^{\{\mu} p_4^{\nu\}}+R q_{\phantom{1}}^{[\mu} p_2^{\nu]}}{q^2}\Bigg]\Bigg\}
\end{align}
is useful to show that $G_M^2$ and $G_E^2$ are separated, as unpolarized case, in spin asymmetry part. By choosing kinematics conditions, we can separately measure $G_E^2$ or $G_M^2$. Further discussions are in section \ref{sec:form factor 13}.

\subsubsection{Incoming and Outgoing Nucleons Polarized}
For polarized incoming and outgoing nucleons,
\begin{align}
I^{2,4}=\left(\frac{1}{2}\sum_{z_1}\right)\sum_{z_3}I^{1,2,3,4}=\frac{1}{2}(1+s^2\cdotp s^4)I^{\rm u}+s_\mu^2 s_\nu^4(J^{2,4})^{\mu\nu}.
\end{align}
The results are
\begin{align}
(J_{\gamma\gamma}^{2,4})^{\mu\nu}=&-G_M^2\Bigg(g^{\mu\nu}-\frac{m_l^2}{\tau m_N^2}\frac{q^\mu q^\nu}{q^2}+2\frac{p_1^\mu p_1^\nu+p_3^\mu p_3^\nu}{q^2}\Bigg)\nonumber\\
&-F_2^2\tau\left(R^2-\frac{1+\tau}{\tau}\right)\frac{q^\mu q^\nu}{q^2}+2G_M F_1\frac{q^\mu q^\nu+Rq_{\phantom{1}}^{[\mu} p_1^{\nu]}}{q^2}\\
(J_{\gamma\phi}^{2,4})^{\mu\nu}=&-2\frac{m_l}{m_N}\Bigg(F_2 R\frac{q^\mu q^\nu}{q^2}+\frac{G_M}{\tau}\frac{q_{\phantom{1}}^{[\mu} p_1^{\nu]}}{q^2}\Bigg)\\
(J_{\phi\phi}^{2,4})^{\mu\nu}=&-\left(1+\frac{m_l^2}{\tau m_N^2}\right)\frac{q^\mu q^\nu}{q^2}.
\end{align}
The asymmetry part of $I_{\gamma\gamma}^{2,4}$ contains the $G_M G_E$ cross term, therefore it is harder to use it to measure the form factors separately.

\subsection{Fully Polarized\label{sec:fully polarized}}
From the argument at the beginning of section \ref{sec:partially polarized}, the fully polarized cross section depends on the product of even number of spin polarizations. One can separate the unpolarized and partially polarized contributions, then leave the part which depends on all four polarizations,
\begin{align}
I_{\gamma\gamma}^{1,2,3,4}=&-\frac{1}{4}\left[5+s^1\cdotp s^3+s^2\cdotp s^4+(s^1\cdotp s^3)(s^2\cdotp s^4)\right]I^{\rm u}_{\gamma\gamma}\nonumber\\
&+\frac{1}{2}\left[(1+s^2\cdotp s^4)I_{\gamma\gamma}^{1,3}+(1+s^1\cdotp s^3)I_{\gamma\gamma}^{2,4}\right]+\frac{1}{4}I_{\gamma\gamma}^{1,2}+\frac{1}{2}(I_{\gamma\gamma}^{1,4}+I_{\gamma\gamma}^{3,2})+I_{\gamma\gamma}^{3,4}\nonumber\\
&+2G_M^2s^1_\mu s^2_\nu s^3_\rho s^4_\sigma\Bigg\{-\frac{F_2^2}{G_M^2}\frac{q^\nu q^\sigma}{q^2}\Bigg(\frac{q^\mu q^\rho}{2q^2}-\tau\frac{p_2^{\{\mu} p_4^{\rho\}}+Rq_{\phantom{1}}^{[\mu} p_2^{\rho]}}{q^2}\Bigg)\nonumber\\
&-\frac{F_1}{G_M}\Bigg[\frac{(p_2+p_4)^{\{\mu} g^{\rho\}[\nu}q^{\sigma]}+Rq^{[\mu}g^{\rho][\nu}q^{\sigma]}}{4q^2}+\frac{q^\mu q^\nu q^\rho q^\sigma+q_{\phantom{1}}^{[\mu} p_2^{\rho]}q_{\phantom{1}}^{[\nu} p_1^{\sigma]}}{q^4}\Bigg]\nonumber\\
&+\frac{g^{\mu\nu}g^{\rho\sigma}+g^{\mu\sigma}g^{\nu\rho}}{4}+\frac{q^{[\mu}g^{\sigma][\nu}q^{\rho]}+g^{\nu[\mu}q^{\rho]}p_1^\sigma+g^{\sigma[\mu}q^{\rho]}p_3^\nu}{2q^2}\Bigg\}\label{eq:I_gg fully}\\
I_{\gamma\phi}^{1,2,3,4}=&-I^{\rm u}_{\gamma\phi}+\frac{I_{\gamma\phi}^{1,3}I_{\gamma\phi}^{2,4}}{I^{\rm u}_{\gamma\phi}}+\frac{1}{4}I_{\gamma\phi}^{1,2}+\frac{1}{2}(I_{\gamma\phi}^{1,4}+I_{\gamma\phi}^{3,2})+I_{\gamma\phi}^{3,4}\nonumber\\
&-s^1_\mu s^2_\nu s^3_\rho s^4_\sigma G_M\frac{m_l}{\tau m_N}\Bigg(\frac{q^{[\rho} g^{\mu][\nu}q^{\sigma]}}{2q^2}+\frac{2}{R}\frac{q_{\phantom{1}}^{[\mu} p_2^{\rho]}q_{\phantom{1}}^{[\nu}p_1^{\sigma]}}{q^4}\Bigg)\\
I_{\phi\phi}^{1,2,3,4}=&\frac{I_{\phi\phi}^{1,3}I_{\phi\phi}^{2,4}}{I^{\rm u}_{\phi\phi}}.
\end{align}

\subsection{Massless Lepton Limit}\label{sec:massless lepton}
As we discussed in section \ref{sec:polarization of massless particle}, naively taking the limit $m_l\rightarrow 0$ does not yield correct results. Only some results are still safe when we take the limit $m_l\rightarrow 0$, such as $I^{\rm u}$, $I^{i,j}_{\gamma\gamma}$, $I^{2,4}$, etc. We discuss the general massless lepton limit in this section.

For a lepton of negligible mass, from either direct calculation or conservation of angular momentum, the interference term $I_{\gamma\phi}$ vanishes for all cases. Therefore it is sufficient to display $I_{\gamma\gamma}$ and $I_{\phi\phi}$.

\begin{enumerate}
\item
The unpolarized $I^{\rm u}$ is safe in the limit $m_l\rightarrow0$,
\begin{align}
&I^{\rm u}_{\gamma\gamma}=\frac{G_E^2+\tau G_M^2}{1+\tau}\left(R^2-\frac{1+\tau}{\tau}\right)+2G_M^2\\
&I^{\rm u}_{\phi\phi}=\frac{1+\tau}{\tau}.
\end{align}

\item For one lepton and one nucleon polarized,
\begin{align}
I^{i,j}=a^{i,j}\sum_{z_{k}}\sum_{z_{l}}I^{i,j,k,l}=a^{i,j}\Big[I^{\rm u}-z^i s_\mu^j (J^j)^\mu\Big]
\end{align}
the result can also be obtained by setting the lepton polarization to be longitudinal $s^i_\mu\rightarrow p^i_\mu/m_l$ then taking $m_l\rightarrow0$ limit,
\begin{align}
(J_{\gamma\gamma}^j)^\mu=&G_M\Bigg[b^j F_2 R\frac{q^\mu}{m_N}-\frac{G_E}{\tau}\frac{(p_1+p_3)^\mu}{m_N}\Bigg]\\
(J_{\phi\phi}^j)^\mu=&0.
\end{align}

\item For incoming and outgoing leptons polarized, one can not directly take the limit $m_l\rightarrow0$, but the results are
\begin{align}
I_{\gamma\gamma}^{1,3}=&\frac{1}{2}(1+z^1z^3)I_{\gamma\gamma}^{\rm u}\\
I_{\phi\phi}^{1,3}=&\frac{1}{2}(1-z^1z^3)I_{\phi\phi}^{\rm u}.
\end{align}

\item For incoming and outgoing nucleons polarized, it is safe to take the limit $m_l\rightarrow0$,
\begin{align}
I^{2,4}=\left(\frac{1}{2}\sum_{z_1}\right)\sum_{z_3}I^{1,2,3,4}=\frac{1}{2}(1+s^2\cdotp s^4)I^{\rm u}+s_\mu^2 s_\nu^4(J^{2,4})^{\mu\nu}.
\end{align}
The results are
\begin{align}
(J_{\gamma\gamma}^{2,4})^{\mu\nu}=&-G_M^2\Bigg(g^{\mu\nu}+2\frac{p_1^\mu p_1^\nu+p_3^\mu p_3^\nu}{q^2}\Bigg)\nonumber\\
&-F_2^2\tau\left(R^2-\frac{1+\tau}{\tau}\right)\frac{q^\mu q^\nu}{q^2}+2G_M F_1\frac{q^\mu q^\nu+Rq_{\phantom{1}}^{[\mu} p_1^{\nu]}}{q^2}\\
(J_{\phi\phi}^{2,4})^{\mu\nu}=&-\frac{q^\mu q^\nu}{q^2}.
\end{align}

\item Finally, for fully polarized case, the result is much simpler than (\ref{eq:I_gg fully})
\begin{align}
I_{\gamma\gamma}^{1,2,3,4}=&-I^{\rm u}_{\gamma\gamma}+\frac{1}{2}(1+z^1 z^3)I_{\gamma\gamma}^{2,4}+\frac{1}{4}I_{\gamma\gamma}^{1,2}+\frac{1}{2}(I_{\gamma\gamma}^{1,4}+I_{\gamma\gamma}^{3,2})+I_{\gamma\gamma}^{3,4}\\
I_{\phi\phi}^{1,2,3,4}=&\frac{I_{\phi\phi}^{1,3}I_{\phi\phi}^{2,4}}{I^{\rm u}_{\phi\phi}}.
\end{align}
\end{enumerate}

\section{Searching for a Scalar Boson\label{sec:searching for a scalar boson}}
If we want to see a new physics signal in elastic lepton-nucleon scattering using one photon and one-scalar-boson exchange, it needs to be greater than two-photon exchange (TPE) contribution since TPE is not fully understood because of the internal nucleon propagator. The TPE contribution, which is the next leading order contribution of the standard model, is generically suppressed by one more fine structure constant $\alpha$ than one-photon exchange. Therefore, the contribution involving scalar boson, $\left(\frac{d\sigma}{d\Omega}\right)_{\gamma\phi}$ and $\left(\frac{d\sigma}{d\Omega}\right)_{\phi\phi}$, must be greater than at least few percent of $\left(\frac{d\sigma}{d\Omega}\right)_{\gamma\gamma}$ to be detected.

We adopt the constraints in \cite{TuckerSmith:2010ra} that for $m_\phi=1$ MeV
\begin{align}
\epsilon_e\lesssim 2.3\times10^{-4},\;\epsilon_p=\epsilon_\mu\lesssim 1.3\times10^{-3},\;\epsilon_n\lesssim 6.7\times10^{-5}.
\end{align}
Therefore, recall (\ref{eq:lambda}), $|\lambda|\lesssim10^{-6}$. In general, $\left(\frac{d\sigma}{d\Omega}\right)_{\gamma\phi}$ and $\left(\frac{d\sigma}{d\Omega}\right)_{\phi\phi}$ are suppressed by $\lambda$ and $\lambda^2$ with respect to the leading standard model contribution $\left(\frac{d\sigma}{d\Omega}\right)_{\gamma\gamma}$, respectively. Although it seems to be hard to observe scalar boson in elastic scattering, we still find some kinematic regions where the scalar boson may be found. This is because for such kinematic region, $\left(\frac{d\sigma}{d\Omega}\right)_{\gamma\gamma}$ goes or approaches to zero, whereas $\left(\frac{d\sigma}{d\Omega}\right)_{\gamma\phi}$ and $\left(\frac{d\sigma}{d\Omega}\right)_{\phi\phi}$ go or approach to zero slower and eventually dominate. Those regions are usually narrow in parameter space and hard to measure due to polar angle measuring range and flux intensity.

\subsection{Unpolarized, One Lepton and One Nucleon Polarized}
\begin{figure}
\centering
\includegraphics[scale=1.7]{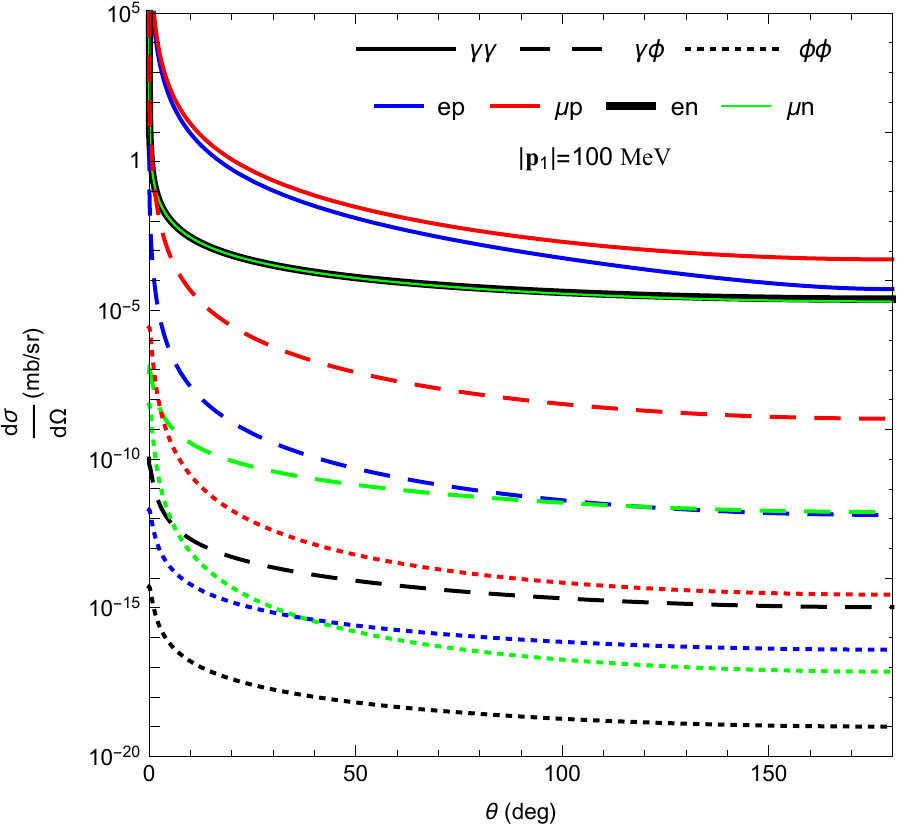}
\caption{\label{fig:unpolarized cross section} Unpolarized differential cross section in the lab frame for $|\mathbf{p}_1|$=100 MeV: the solid, dashed, and dotted lines correspond to $\left(\frac{d\sigma}{d\Omega}\right)_{\gamma\gamma}$, $\left(\frac{d\sigma}{d\Omega}\right)_{\gamma\phi}$, and $\left(\frac{d\sigma}{d\Omega}\right)_{\phi\phi}$, respectively. It shows that $\left(\frac{d\sigma}{d\Omega}\right)_{\gamma\phi}$ and $\left(\frac{d\sigma}{d\Omega}\right)_{\phi\phi}$ are in general suppressed by $|\lambda|\lesssim10^{-6}$ and $|\lambda^2|\lesssim10^{-12}$, respectively. $\left(\frac{d\sigma}{d\Omega}\right)_{\gamma\phi}$ with muon is greater about two order of magnitudes than $\left(\frac{d\sigma}{d\Omega}\right)_{\gamma\phi}$ with electron, and it can be understood by $\left(\frac{d\sigma}{d\Omega}\right)_{\gamma\phi}$ containing $\frac{m_l}{m_N}$ factor in Eq. (\ref{eq:I_gp unpol}). Note that there can be an overall minus sign in front of $\left(\frac{d\sigma}{d\Omega}\right)_{\gamma\phi}$ depending on the sign of $\lambda$. $\left(\frac{d\sigma}{d\Omega}\right)_{\gamma\gamma}$ and $\left(\frac{d\sigma}{d\Omega}\right)_{\phi\phi}$ are always greater or equal to zero.}
\end{figure}

The unpolarized differential cross section for $|\mathbf{p}_1|$=100 MeV is shown in figure \ref{fig:unpolarized cross section}. One can see that generally $\left(\frac{d\sigma}{d\Omega}\right)_{\gamma\phi}$ and $\left(\frac{d\sigma}{d\Omega}\right)_{\phi\phi}$ are suppressed by $|\lambda|\lesssim10^{-6}$ and $|\lambda^2|\lesssim10^{-12}$, respectively. Also, there is about two orders of magnitude enhancement in $\left(\frac{d\sigma}{d\Omega}\right)_{\gamma\phi}$ if the lepton is muon instead of electron. This can be understood by $\left(\frac{d\sigma}{d\Omega}\right)_{\gamma\phi}$ containing $\frac{m_l}{m_N}$ factor in (\ref{eq:I_gp unpol}). We do not observe any kinematic region, which depends only on $\theta$ for the unpolarized case, for the scalar boson to be observable.

For one lepton and one nucleon polarized, by examining the parameter space $(\theta,\alpha_i,\beta_i,\alpha_j,\beta_j)$, we do not find any kinematic region where the effect of the scalar boson is significant either.

\subsection{Incoming and Outgoing Leptons Polarized\label{sec:searching for phi_leptons polarized}}
\begin{figure}
\centering
\includegraphics[scale=1.5]{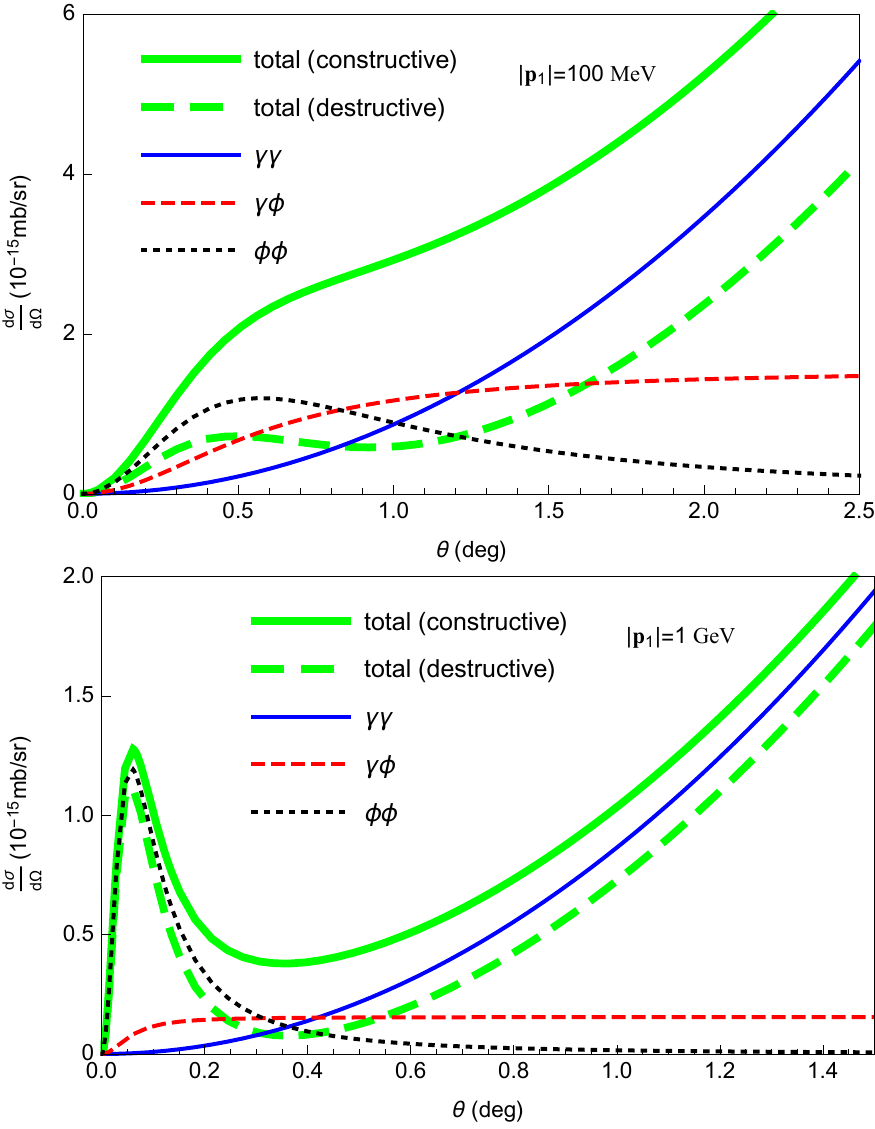}
\caption{\label{fig:cross_section_13} Elastic electron-neutron forward scattering with electron helicity flip ($\alpha_1=0$, $\alpha_3=180^\circ$, and $\theta\rightarrow0$) in the lab frame: the upper and lower figures correspond to incoming momentum $|\mathbf{p}_1|=100$ MeV and 1 GeV; the blue, dashed red, dotted black, and thick green lines correspond to $\left(\frac{d\sigma}{d\Omega}\right)_{\gamma\gamma}$, $\left(\frac{d\sigma}{d\Omega}\right)_{\gamma\phi}$, $\left(\frac{d\sigma}{d\Omega}\right)_{\phi\phi}$, and total differential cross section, respectively. There can be an overall minus sign in front of $\left(\frac{d\sigma}{d\Omega}\right)_{\gamma\phi}$ depending on the sign of $\lambda$; based on the same reason, there are two green lines depending on the sign of $\lambda$.}
\end{figure}

The differential cross section of polarized incoming and outgoing leptons in the lab frame may be good for finding scalar boson in electron-neutron scattering. For helicity flip forward scattering ($\alpha_1=0$, $\alpha_3=180^\circ$, and $\theta\rightarrow0$), although the differential cross section goes to zero, $\left(\frac{d\sigma}{d\Omega}\right)_{\gamma\phi}$ and $\left(\frac{d\sigma}{d\Omega}\right)_{\phi\phi}$ go to zero much slower than $\left(\frac{d\sigma}{d\Omega}\right)_{\gamma\gamma}$ and dominate at small angle. See figure \ref{fig:cross_section_13}, we can look for a bump to observe the scalar boson. One can find that $|\mathbf{p}_1|$ around 100 MeV may be a good choice. If $|\mathbf{p}_1|$ is smaller, $\left(\frac{d\sigma}{d\Omega}\right)_{\gamma\gamma}$ dominates; if $|\mathbf{p}_1|$ is bigger, the region where $\left(\frac{d\sigma}{d\Omega}\right)_{\gamma\phi}$ and $\left(\frac{d\sigma}{d\Omega}\right)_{\phi\phi}$ dominate is closer to $\theta=0$ and the region where scalar contribution dominates becomes narrower.

One facility to measure electron nucleon elastic scattering is the SoLID (Solenoidal Large Intensity Device) in Hall A at JLab \cite{SoLID:2013fta}. The polar angle resolution is around 1 mr, so the precision is sufficient to find the bump. To estimate the production rate, the unpolarized luminosity is $1.3\times10^{39}$ $\rm cm^{-2}s^{-1}$ for $\rm LD_2$ target; the differential cross section is around $10^{-15}$ mb/sr for the scattering angle from $0^\circ$ to $2^\circ$ (figure \ref{fig:cross_section_13} for $|\mathbf{p}_1|=100$ MeV). Combining everything above, it will be about 2 days for an event to occur.

Although the number seems promising, the experiment faces several obstacles: the polar angle is outside the minimum equipment coverage range which is about $8^\circ$; polarizing the incoming electron reduces the luminosity; measuring the polarization of the the outgoing electron is very difficult. Moreover, because the target is liquid deuterium, to separate electron-neutron scattering signals from much bigger electron-proton scattering signals is difficult. The differential cross section for electron-proton scattering blows up in the forward direction, and it is about $10^{-4}$ ($10^{-2}$) mb/sr around $1^\circ$ ($0.1^\circ$).

\subsection{Incoming and Outgoing Nucleons Polarized\label{sec:searching for phi_nucleons polarized}}
\begin{figure}
\centering
\includegraphics[scale=1.5]{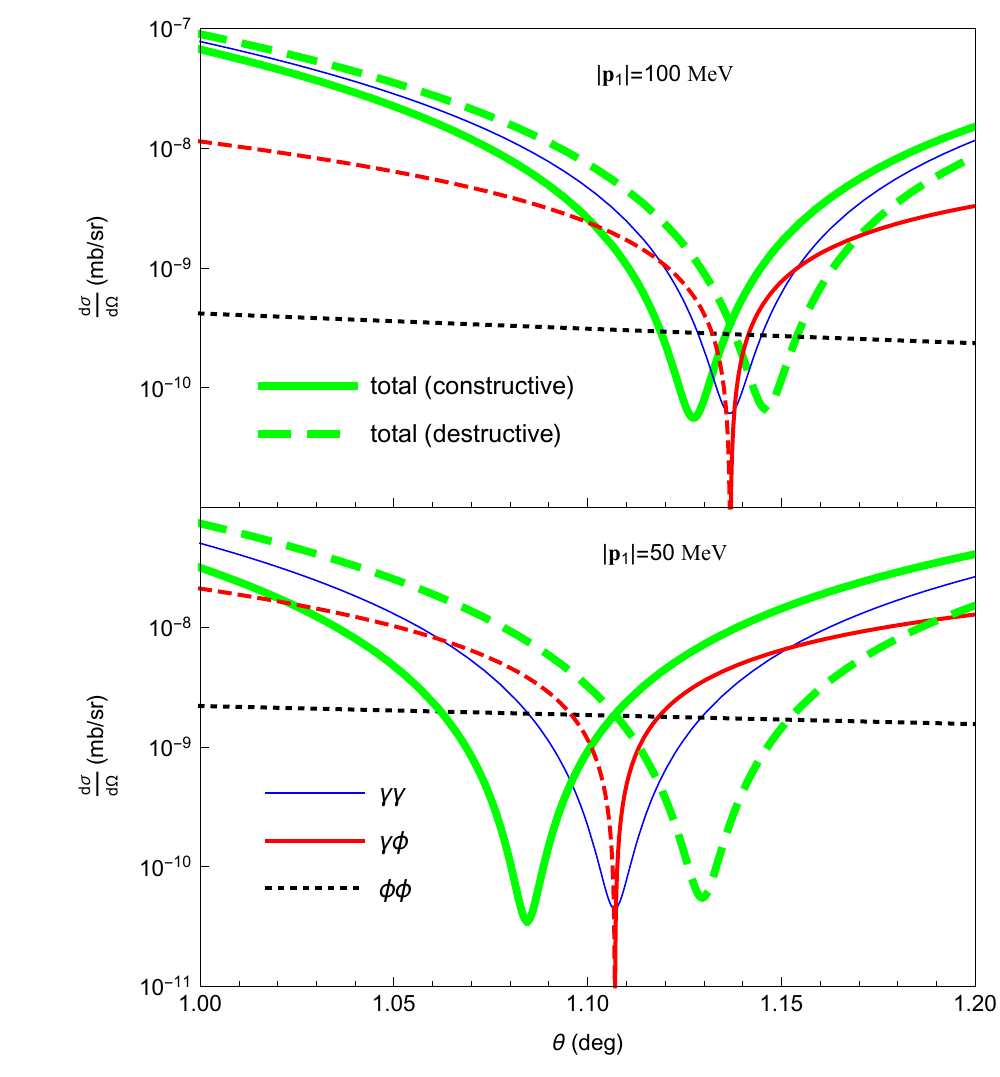}
\caption{\label{fig:cross_section_24} Elastic muon-neutron scattering for incoming and outgoing neutrons polarized $(\alpha_2,\beta_2,\alpha_4,\beta_4)=(90.5^\circ,0^\circ,180^\circ,0^\circ)$ in the lab frame: the upper and lower figures correspond to incoming momentum $|\mathbf{p}_1|=100$ MeV and 50 MeV; the thin blue, (solid and dashed) red, dotted black, and thick green lines correspond to $\left(\frac{d\sigma}{d\Omega}\right)_{\gamma\gamma}$, $\left(\frac{d\sigma}{d\Omega}\right)_{\gamma\phi}$, $\left(\frac{d\sigma}{d\Omega}\right)_{\phi\phi}$, and total differential cross section, respectively. The solid and dashed red lines means they are differed by a minus sign, and there can be an overall minus sign in front of $\left(\frac{d\sigma}{d\Omega}\right)_{\gamma\phi}$ depending on the sign of $\lambda$; based on the same reason, there are two green lines depending on the sign of $\lambda$.}
\end{figure}

The differential cross section of polarized incoming and outgoing nucleons in the lab frame may be good for finding scalar boson in muon-neutron scattering in neutron rest frame. For the spin polarization in the scattering plane ($\beta_2=0$ and $\beta_4=0$), one can choose $\alpha_2=90.5^\circ$ and $\alpha_4=180^\circ$, then there is a region near $0^\circ$ (depending on $|\mathbf{p}_1|$) where the contribution of $\left(\frac{d\sigma}{d\Omega}\right)_{\gamma\phi}$, and $\left(\frac{d\sigma}{d\Omega}\right)_{\phi\phi}$ dominates. See figure \ref{fig:cross_section_24}, we can look for the shift of the local minimum predicted by standard model. One can find that $|\mathbf{p}_1|$ around 50 MeV may be a good choice. If $|\mathbf{p}_1|$ is smaller, the region which $\left(\frac{d\sigma}{d\Omega}\right)_{\gamma\phi}$ and $\left(\frac{d\sigma}{d\Omega}\right)_{\phi\phi}$ dominate is closer to $\theta=0$; if $|\mathbf{p}_1|$ is bigger, $\left(\frac{d\sigma}{d\Omega}\right)_{\gamma\gamma}$ dominates.

For MUSE at PSI \cite{MUSE}, the polar angle resolution is around 1 mr, which is a sufficient precision to observe the shift of the local minimum if we combine muon-neutron and anti-muon-neutron scattering. To estimate the production rate, the muon unpolarized flux about $2\times10^5$ Hz at momentum 115 MeV; the target $\rm LD_2$ has thickness 4 cm and density 162.4 $\rm kg/m^3$; the differential cross section is around $10^{-9}$ mb/sr for the scattering angle from $1.0^\circ$ to $1.2^\circ$ (figure \ref{fig:cross_section_24} for $|\mathbf{p}_1|=50$ MeV). Combining all numbers above, it will take about 1000 years for an event to occur.

The main obstacles for this experiment is that the intensity of muon flux is much smaller than electron flux. It also suffers from other difficulties: the polar angle is outside the equipment coverage $20^\circ$ to $100^\circ$; because the target is liquid deuterium, to separate muon-neutron scattering signals from much bigger muon-proton scattering signals is difficult (the differential cross section is about $10^6$ mb/sr around $1.1^\circ$).

\section{Measuring $G_E$ and $G_M$\label{sec:measuring form factors}}
In this section, we generalize and discuss the currently used methods \cite{Rock:2001zi,Arrington:2011kb,Punjabi:2014tna}, and examine alternative ways to measure form factors when incoming and outgoing leptons are polarized.

\subsection{Polarize One Lepton and One Nucleon\label{sec:form factor ij}}
From (\ref{eq:I_gg^ij 2nd}), one can choose kinematics conditions such that the spin asymmetry part,
\begin{align}
P^{i,j}=I_{\gamma\gamma}^{i,j}-a^{i,j}I^{\rm u}_{\gamma\gamma}=a^{i,j}s_\mu^i s_\nu^j(J^{i,j})^{\mu\nu},
\end{align}
can selectively receive $G_M^2$ or $G_MG_E$ contributions. In the lab frame, if the spin polarization vector $\mathbf{s}^j$ of nucleon is perpendicular or parallel to momentum transfer vector $\mathbf{q}$, their corresponding contribution to asymmetry part of $I_{\gamma\gamma}^{i,j}$ are $P^{i,j}_T$ and $P^{i,j}_L$,
\begin{align}
P^{i,j}_T=&-2a^{i,j}G_MG_E\frac{m_l}{\tau m_N}s^i\cdotp s^j_T\\
P^{i,j}_L=&2a^{i,j}b^jG_M^2\frac{m_l}{m_N}\frac{s_i\cdotp(p_2+p_4)s^j_L\cdotp q}{|\mathbf{q}|^2},
\end{align}
which leads to the most general from of so-called ratio technique or polarization transfer method to measure the form factors,
\begin{align}\label{eq:ratio tech}
\frac{G_E}{G_M}=-b^j\Bigg(\frac{P_T}{P_L}\Bigg)^{i,j}\frac{\tau}{|\mathbf{q}|^2}\frac{s_i\cdotp(p_2+p_4)s^j_L\cdotp q}{s^i\cdotp s^j_T}.
\end{align}

There are some special cases worth discussing.
\begin{enumerate}
\item
If the lepton is longitudinally polarized,
\begin{align}
\frac{s_i\cdotp(p_2+p_4)s^j_L\cdotp q}{s^i\cdotp s^j_T}=\frac{|\mathbf{q}|}{\hat{\mathbf{p}}_i\cdotp\hat{\mathbf{s}}^j_T}\Big[-2m_N(1+\tau)\frac{|\mathbf{p}_i|}{E_i}+(\mathbf{p}_1-\mathbf{p}_3)\cdotp\hat{\mathbf{p}}_i\Big].
\end{align}

\item
If $\hat{\mathbf{s}}^j_T$ is in scattering plane,
\begin{align}
\hat{\mathbf{p}}_i\cdotp\hat{\mathbf{s}}^j_T=\frac{|\mathbf{p}_1||\mathbf{p}_3|}{|\mathbf{p}_i||\mathbf{q}|}\sin\theta.
\end{align}

\item
If lepton is massless,
\begin{align}
2m_N(1+\tau)\frac{|\mathbf{p}_i|}{E_i}-(\mathbf{p}_1-\mathbf{p}_3)\cdotp\hat{\mathbf{p}}_i\underset{m_l\rightarrow0}{=}\frac{m_N(E_1+E_3)}{E_i}
\end{align}

\item
Combining all the choices and limit above, (\ref{eq:ratio tech}) reduces to a familiar form\footnote{In other literature, $P_T$ and $P_L$ usually stand for spin polarization transfer which is the spin asymmetry part divided by spin symmetry part.}
\begin{align}
\frac{G_E}{G_M}=b^j\Bigg(\frac{P_T}{P_L}\Bigg)^{i,j}\frac{E_1+E_3}{2m_N}\tan\frac{\theta}{2},
\end{align}
which is widely used today to measure form factors in elastic electron-nucleon scattering \cite{Rock:2001zi,Arrington:2011kb,Punjabi:2014tna}.
\end{enumerate}
The commonly used cases are $(i,j)=(1,2)$ or $(1,4)$ because the polarization of outgoing lepton is harder to measure.

\subsection{Incoming and Outgoing Leptons Polarized\label{sec:form factor 13}}
\begin{figure}
\centering
\includegraphics[scale=2]{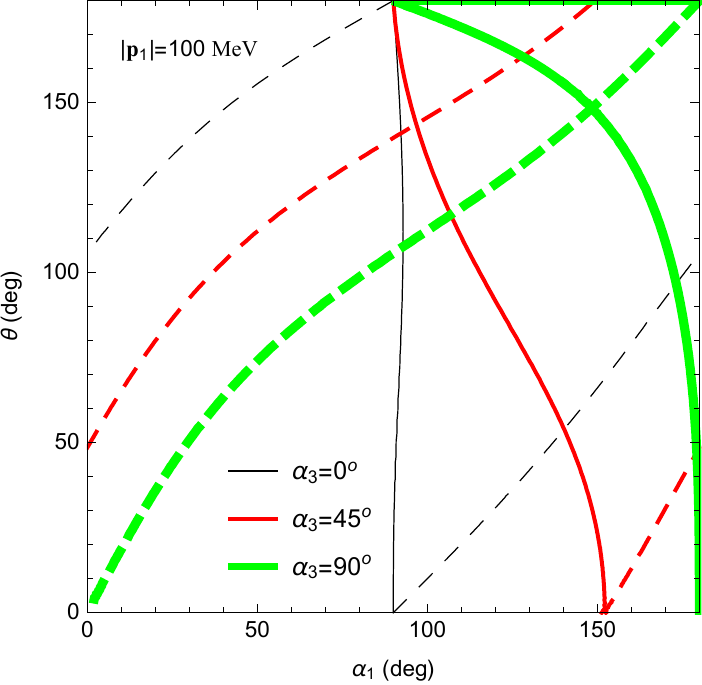}
\caption{\label{fig:P13 zeros} Elastic muon-nucleon scattering for $|\mathbf{p}_1|$=100 MeV with polarized incoming and outgoing muons in the scattering plane, $\beta_1=0$ and $\beta_3=0$, in the lab frame: One can find the zeroes of $G_M^2$ and $G_E^2$ terms in $P^{1,3}$ denoted as the solid and dashed lines in the figure,respectively; the thin black, red, and thick green lines correspond to $\alpha_3=0^\circ$, $45^\circ$, and $90^\circ$, respectively. The zeros for muon-proton and muon-neutron scattering are actually different, but too small to distinguish in the figure.}
\end{figure}

From (\ref{eq:I_gg^13 2nd}), one can choose kinematics conditions such that the spin asymmetry part
\begin{align}
P^{1,3}=I_{\gamma\gamma}^{1,3}-\frac{1}{2}I^{\rm u}_{\gamma\gamma}=\frac{1}{2}s^1\cdotp s^3I+s_\mu^1 s_\nu^3(J^{1,3})^{\mu\nu}
\end{align}
can selectively receive $G_M^2$ or $G_E^2$ contributions. We do not find a compact algebraic form for such condition, but one can numerically evaluate it. In figure \ref{fig:P13 zeros}, assuming $|\mathbf{p}_1|$=100 MeV and the spin polarizations are in scattering plane, $\beta_1=0$ and $\beta_3=0$, we find the zeros of $G_M^2$ and $G_E^2$ terms in $P^{1,3}$ of elastic muon-nucleon scattering.

\begin{figure}
\centering
\includegraphics[scale=1.8]{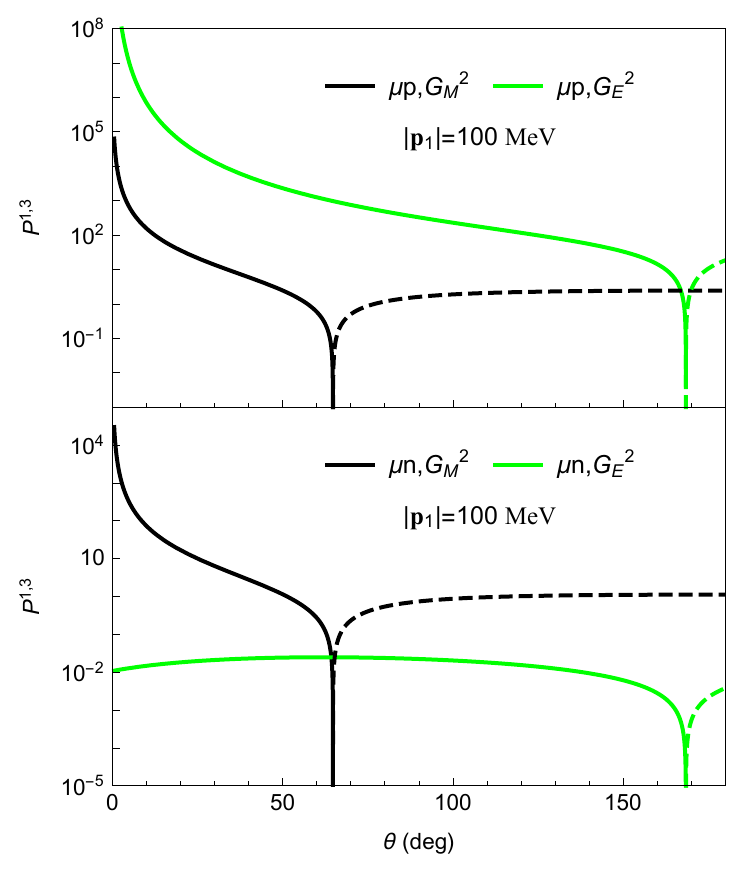}
\caption{\label{fig:P13} Elastic muon-nucleon scattering for $|\mathbf{p}_1|$=100 MeV with polarized incoming and outgoing muons, $(\alpha_1,\beta_1,\alpha_3,\beta_3)=(135^\circ,0^\circ,45^\circ,0^\circ)$: The solid and dashed lines represent the positive and negative values, respectively; the black and green lines correspond to $G_M^2$ and $G_E^2$ terms of $P^{1,3}$ of elastic muon-nucleon scattering cross section. The zeros of $G_M^2$ and $G_E^2$ terms of $P^{1,3}$ for elastic muon-proton and muon-neutron scattering are very close but different.} 
\end{figure}

As an example (see figure \ref{fig:P13}), consider $|\mathbf{p}_1|$=100 MeV and $(\alpha_1,\beta_1,\alpha_3,\beta_3)=(135^\circ,0^\circ,45^\circ,0^\circ)$, for elastic muon-proton scattering, $P^{1,3}$ receives contribution only from $G_E^2$ and $G_M^2$ term at $64.794^\circ$ and $168.508^\circ$, respectively; for elastic muon-neutron scattering, $P^{1,3}$ receives contribution only from $G_E^2$ and $G_M^2$ term at $64.798^\circ$ and $168.506^\circ$, respectively.

In practice, although the angular resolution is sufficient to perform the experiment, however, this method to measure the form factors could be hard due to the fact that to polarized and measure leptons to some specific non-longitudinal angle is not an easy task. 

\section{Discussion\label{sec:conclusions}}
In this chapter, we calculated the differential cross section of lepton-nucleon elastic scattering using the one-photon and one-scalar-boson exchange mechanism for all possible polarizations in a general reference frame. The expressions are shown in section \ref{sec:LN cross section}. There are some possible applications: finding a new scalar boson to resolve proton radius and muon $g-2$ puzzles; generalizing the current methods to measure the ratio of the nucleon form factors, $G_E/G_M$; providing an alternative way to measure $G_E^2$ and $G_M^2$ directly. 

The effects of new scalar boson are studied in section \ref{sec:searching for a scalar boson}. We conclude that there are two cases that the scalar boson is potentially observable. For elastic electron-neutron helicity flip forward scattering with incoming and outgoing electrons polarized, the effects of scalar boson are dominant and produce a bump, see figure \ref{fig:cross_section_13}. The optimistic estimates in section \ref{sec:searching for phi_leptons polarized} show that the experiment is possible for current technology but faces several obstacles. For elastic muon-neutron scattering at small scattering angles using polarized incoming and outgoing neutrons, the effects of scalar boson is significant and shifts the local minimum, see figure \ref{fig:cross_section_24}. However, the optimistic estimates in section \ref{sec:searching for phi_nucleons polarized} show that unless the intensity of muon flux can be dramatically increased, the experiment is impossible in the foreseeable future.

In section \ref{sec:form factor ij}, the current method to measure nucleon form factors, as known as ratio technique or polarization transfer method, is generalized and shown in (\ref{eq:ratio tech}). The current method is to polarize incoming electron and either polarize incoming or outgoing nucleons. We generalize it such that one of the incoming and outgoing leptons is polarized, and one of the incoming and outgoing nucelons is polarized. Also, in order to separate the contributions of $G_MG_E$ and $G_M^2$ terms, the current method requires polarization vector of the nucleon $\mathbf{s}^j$ to be either perpendicular or parallel to momentum transfer $\mathbf{q}$, and neglects the lepton mass and as a result the lepton polarization become longitudinal. In our expression, we only require the kinematic condition that the nucleon $\mathbf{s}^j$ is either perpendicular or parallel to momentum transfer $\mathbf{q}$. In conclusion, using our new expression, one can choose which leptons and nucleons to be polarized; the lepton mass is included; the polarization of leptons becomes two extra angular parameters in an experiment.

In section \ref{sec:form factor 13}, in studying elastic muon-nucleon scattering with incoming and outgoing muons polarized, we can separately measure the contributions of $G_E^2$ and $G_M^2$. Although we do not obtain a simple analytic expression, it is easy to evaluate the zeros of $G_E^2$ and $G_M^2$ terms of spin asymmetry part of the cross section numerically, see figure \ref{fig:P13 zeros}. As an example in figure \ref{fig:P13}, one can measure $G_E^2$ and $G_M^2$ directly, and the angle resolution of the current facility is sufficient to perform the experiment. However, measuring the polarization of the outgoing muon could be challenging.

\chapter{Beam Dump Experiments}\label{ch:beam dump}

\section{Introduction}
Beam dump experiments have been aimed at searching for new particles, such as dark photons and axions (see, e.g. \cite{Essig:2013lka} and references therein) that decay to lepton pairs and/or photons. Electron beam dumps in particular have received a large amount of theoretical attention in recent years~\cite{Bjorken:2009mm,Andreas:2012mt}. The typical setup of an electron beam dump experiment is to dump an electron beam into a target, in which the electrons are stopped. The new particles produced by the bremsstrahlung-like process pass through a shield region and decay. These new particles can be detected by their decay products, electron pairs and/or photons, measured by the detector downstream of the decay region. Previous work simplified the necessary phase space integral by using the Weizs\"{a}cker-Williams (WW) approximation \cite{vonWeizsacker:1934nji,Williams:1935dka} which, also known as method of virtual quanta, is a semiclassical approximation. The idea is that the electromagnetic field generated by a fast moving charged particle is nearly transverse which is like a plane wave and can be approximated by real photon. The use of the  WW approximation in bremsstrahlung processes was developed in Refs.~\cite{Kim:1973he,Tsai:1973py} and applied to beam dump experiments in Refs.~\cite{Bjorken:2009mm,Tsai:1986tx}. The WW approximation simplifies evaluation of the integral over phase space and approximates the 2 particle to 3 particle (2 to 3) cross section in terms of a 2 particle to 2 particle (2 to 2) cross section. For the WW approximation to work in a beam dump experiment, it needs the incoming beam energy to be much greater than the mass of the new particle, $m_\phi$, and electron mass $m_e$. 

The previous work \cite{Bjorken:2009mm} used the following three approximations:
\begin{enumerate}
\item WW approximation;
\item a further simplification of the phase space integral, see Eq. (\ref{eq:IWW tmin tmax});
\item $m_\phi\gg m_e$.
\end{enumerate}
The combination of the first two approximations has been denoted \cite{Kim:1973he} the improved WW (IWW) approximation. The name ``improved WW" might be somewhat misleading since the procedure reduces the computational time but not to improve accuracy). In this paper, we will focus on examining the validity of WW and IWW approximations. The third approximation used to simplify the calculation of amplitude, however, is not in our scope because it is merely a special case by cutting off our results when $m_\phi\lesssim 2m_e$. Nevertheless, we should point out that without using the third approximation we can use beam dump experiments to explore a larger parameter space.

As an example, we use the beam dump experiment E137 \cite{Bjorken:1988as} and the production of a new boson, which we denote $\phi$. To be general, we consider that the new boson could be scalar, pseudoscalar, vector, or axial-vector.

The outline of this chapter is as follows. In section \ref{sec:dynamics}, we calculate the squared amplitude for 2 to 3 and 2 to 2 processes. In section \ref{sec:cross section}, the cross sections for the 2 to 3 and 2 to 2 processes are calculated in the lab frame without any approximation. In section \ref{sec:WW approximation}, we introduce the WW approximation. In section \ref{sec:cross section comparison}, we derive and compare the cross sections with and without approximations. In section \ref{sec:particle production}, we compare the number of new particles produced in beam dump experiments with and without approximations. In section \ref{sec:data analysis}, we assume that this new boson is observed and measured in beam dump experiment, determine the mass and coupling constant, and compare the results with and without approximations. A discussion is presented in section \ref{sec:discussion}.

\section{Dynamics}\label{sec:dynamics}
For simplicity, we assume that there is only one new boson $\phi$, which only couples to electron by a Yukawa interaction, i.e. the boson does not couple to other standard model fermions other than electron. The Lagrangian contains either one of the following interactions
\begin{align}
\mathcal{L}_S&=e\epsilon\phi\bar\psi\psi\nonumber\\
\mathcal{L}_P&=i e\epsilon_P\phi\bar\psi\gamma_5\psi\nonumber\\
\mathcal{L}_V&=e\epsilon_V\phi_\mu\bar\psi\gamma^\mu\psi\\
\mathcal{L}_A&=e\epsilon_A\phi_\mu\bar\psi\gamma_5\gamma^\mu\psi\nonumber
\end{align}
where the subscripts $S$, $P$, $V$, and $A$ correspond to scalar, pseudoscalar, vector, and axial-vector, respectively; $\epsilon=g/e$, $g$ is the coupling of the new boson, and $e$ is the electric charge; $\psi$ is the electron field; $\gamma_5=-\frac{i}{4!}\epsilon_{\mu\nu\rho\sigma}\gamma^\mu\gamma^\nu\gamma^\rho\gamma^\sigma$; we choose the convention that there is an extra $i$ in $\mathcal{L}_P$, such that $\epsilon_P$ can be a non-negative number.

If $m_\phi>2m_e$, the dominant new boson decay is to electron pairs
\begin{align}\label{eq:decay to electrons}
\Gamma_{S}(\phi\to e^+e^-)&=\epsilon_S^2\frac{\alpha}{2}m_\phi\left(1-\frac{4m_e^2}{m_\phi^2}\right)^{3/2}\nonumber\\
\Gamma_{P}(\phi\to e^+e^-)&=\epsilon_P^2\frac{\alpha}{2}m_\phi\left(1-\frac{4m_e^2}{m_\phi^2}\right)^{1/2}\nonumber\\
\Gamma_{V}(\phi\to e^+e^-)&=\epsilon_V^2\frac{\alpha}{3}m_\phi\left(1+\frac{2m_e^2}{m_\phi^2}\right)\left(1-\frac{4m_e^2}{m_\phi^2}\right)^{1/2}\\
\Gamma_{A}(\phi\to e^+e^-)&=\epsilon_A^2\frac{\alpha}{3}m_\phi\left(1-\frac{4m_e^2}{m_\phi^2}\right)^{3/2}\nonumber,
\end{align}
where $\alpha$ is the fine structure constant.

If $m_\phi<2m_e$, the dominant decay channel involves photons produced through the electron loop. For spin-0 particles, they decay to two photons 
\begin{align}\label{eq:P decay to photons}
\Gamma_{S,P}(\phi\to\gamma\gamma)&=\epsilon_{S,P}^2\frac{\alpha^3}{4\pi^2}\frac{m_\phi^3}{m_e^2}f_{S,P}\left(\frac{m_\phi^2}{4m_e^2}\right)
\end{align}
where $f_P(\tau)=\frac{1}{64\tau^2}\left|\ln\left[1-2\left(\tau+\sqrt{\tau^2-\tau}\right)\right]^2 \right|^2$ and $f_S(\tau)=\frac{1}{4\tau^2}\left|1+(1-\frac{1}{\tau})\left(\sin^{-1}\sqrt{\tau}\right)^2\right|^2$. For spin-1 particles, however, the two photon decay channel is forbidden by Landau--Yang theorem \cite{Landau:1948kw,Yang:1950rg,Zhemchugov:2014dza}. Therefore, the dominant decay channel of the vector boson is 3 photon decay 
\begin{align}
\Gamma(\phi\to\gamma_1+\gamma_2+\gamma_3)=\frac{1}{64S\pi^3 m_\phi}\int^\frac{m_\phi}{2}_0dE_1\int^\frac{m_\phi}{2}_{\frac{m_\phi}{2}-E1}dE_2|\mathcal{M}|^2
\end{align}
where $S$ is the symmetry factor accounting for identical particles in the final state and in this case $S=3!$; $E_1$ and $E_2$ are energy of $\gamma_1$ and $\gamma_2$, respectively; $\mathcal{M}$ is the amplitude containing 6 diagrams. We express the decay rate in term of $\frac{m_\phi}{m_e}$ expansion
\begin{align}\label{eq:V decay to photons}
\Gamma_{V}(\phi\to3\gamma)=\epsilon_V^2\frac{\alpha^4}{2^7 3^6 5^2 \pi^3}\frac{m_\phi^9}{m_e^8}\left[\frac{17}{5}+\frac{67}{42}\frac{m_\phi^2}{m_e^2}+\frac{128941}{246960}\frac{m_\phi^4}{m_e^4}+\mathcal{O}\left(\frac{m_\phi^6}{m_e^6}\right)\right].
\end{align}
The leading term of this result agrees with \cite{Pospelov:2008jk}, which used effective field theory.

For axial-vector, the 3 photon decay channel is further forbidden by the charge conjugation symmetry (similar with the argument of Furry's theorem). Thus the dominant decay channel of the axial-vector boson is 4 photon decay. There are 24 diagrams and the 4 body phase space integral of the decay rate is done in Refs. \cite{Anastasiou:2003gr,Asatrian:2012tp}. We express the result in term of $\frac{m_\phi}{m_e}$ expansion
\begin{align}\label{eq:A decay to photons}
\Gamma_{A}(\phi\to 4\gamma)=\epsilon_A^2\frac{127\alpha^5}{2^{11} 3^8 5^4 7^2 \pi^4}\frac{m_\phi^{13}}{m_e^{12}}+\mathcal{O}\left(\frac{m_\phi^{15}}{m_e^{14}}\right).
\end{align}

\subsection{2 to 3 production}
\begin{figure}
\centering
\includegraphics[scale=1]{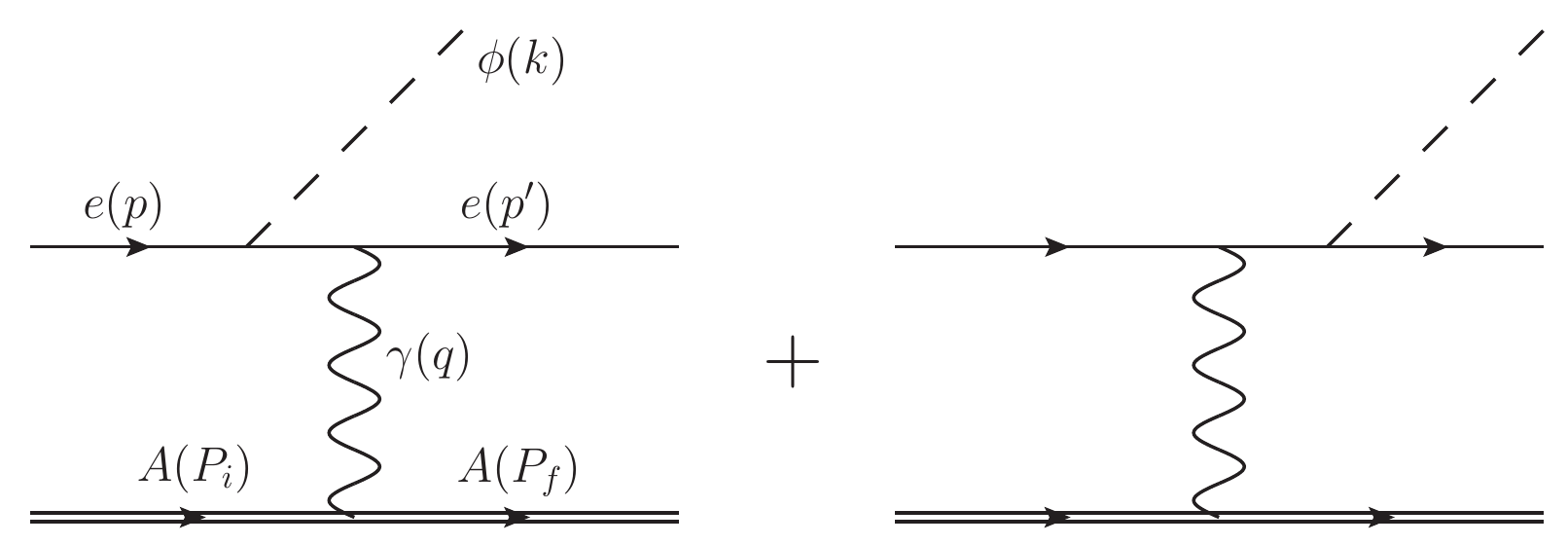}
\caption{\label{fig:2 to 3} Lowest order 2 to 3 production process: $e(p)+A(P_i)\rightarrow e(p')+A(P_f)+\phi(k)$. $A$, $\gamma$, $e$, and $\phi$ stand for the target atom, photon, electron, and the new boson.}
\end{figure}

The leading production process is the bremsstrahlung-like radiation of the new boson from the electron, shown in figure \ref{fig:2 to 3},
\begin{align}\label{eq:2 to 3 production process}
e(p)+A(P_i)\rightarrow e(p')+A(P_f)+\phi(k)
\end{align}
where $e$, $A$, and $\phi$ stand for electron, target atom, and the new boson, respectively. We define the following quantities using the mostly-plus metric
\begin{align}\label{eq:2 to 3 variables}
\tilde{s}&=-(p'+k)^2-m_e^2=-2p'\cdotp k+m_\phi^2\nonumber\\
\tilde{u}&=-(p-k)^2-m_e^2=2p\cdotp k+m_\phi^2\nonumber\\
t_2&=-(p'-p)^2=2p'\cdotp p+2m_e^2\\
q&=P_i-P_f\nonumber\\
t&=q^2\nonumber
\end{align}
which satisfy
\begin{align}
\tilde{s}+t_2+\tilde{u}+t=m_\phi^2.
\end{align}

For definiteness, we assume the atom is a scalar boson (its spin is not consequential here) so that the Feynman rule for the photon-atom vertex is 
\begin{align}
ieF(q^2)(P_i+P_f)_\mu\equiv ieF(q^2)P_\mu
\end{align}
where $F(q^2)$ is the form factor which accounts for the nuclear form factor \cite{DeJager:1987qc} and the atomic form factor \cite{atomic_form_factor}. Here, we only include the elastic form factor since the contribution of the inelastic one is much smaller and can be neglected in computing the cross section. The amplitude of the process in figure \ref{fig:2 to 3} using the mostly-plus metric is
\begin{align}
\mathcal{M}^{23}_S&=e^2g_S\frac{F(q^2)}{q^2}\bar{u}_{p',s'}\left[\slashed{P}\frac{-(\slashed{p}-\slashed{k})+m_e}{-\tilde{u}}+\frac{-(\slashed{p'}+\slashed{k})+m_e}{-\tilde{s}}\slashed{P}\right]u_{p,s}\nonumber\\
\mathcal{M}^{23}_P&=ie^2g_P\frac{F(q^2)}{q^2}\bar{u}_{p',s'}\left[\slashed{P}\frac{(\slashed{p}-\slashed{k})-m_e}{\tilde{u}}\gamma_5+\gamma_5\frac{(\slashed{p'}+\slashed{k})-m_e}{\tilde{s}}\slashed{P}\right]u_{p,s}\nonumber\\
\mathcal{M}^{23}_V&=e^2g_V\frac{F(q^2)}{q^2}\tilde{\epsilon}^\mu_{k,\lambda}\bar{u}_{p',s'}\left[\slashed{P}\frac{(\slashed{p}-\slashed{k})-m_e}{\tilde{u}}\gamma_\mu+\gamma_\mu\frac{(\slashed{p'}+\slashed{k})-m_e}{\tilde{s}}\slashed{P}\right]u_{p,s}\\
\mathcal{M}^{23}_A&=e^2g_A\frac{F(q^2)}{q^2}\tilde{\epsilon}^\mu_{k,\lambda}\bar{u}_{p',s'}\left[\slashed{P}\frac{(\slashed{p}-\slashed{k})-m_e}{\tilde{u}}\gamma_5\gamma_\mu+\gamma_5\gamma_\mu\frac{(\slashed{p'}+\slashed{k})-m_e}{\tilde{s}}\slashed{P}\right]u_{p,s}\nonumber
\end{align}
where $S$, $P$, $V$, and $A$ stand for scalar, pseudoscalar, vector, and axial-vector, respectively; $u_{p,s}$ is the electron spinor and $s=\pm 1$; $\tilde{\epsilon}$ is the polarization of the new spin-1 particle and $\lambda=0,\,\pm 1$. The polarization sum for the new massive spin-1 particle is
\begin{align}
\sum_\lambda\tilde{\epsilon}^\mu_{k,\lambda}\tilde{\epsilon}^{\nu*}_{k,\lambda}=g^{\mu\nu}+\frac{k^\mu k^\nu}{m_\phi^2}.
\end{align}
After averaging and summing over initial and final spins, we have
\begin{align}
\overline{|\mathcal{M}^{23}_{S,P}|^2}&=\left(\frac{1}{2}\sum_s\right)\sum_{s'}|\mathcal{M}_{S,P}^{23}|^2=e^4g_{S,P}^2\frac{F(q^2)^2}{q^4}\mathcal{A}^{23}_{S,P}\nonumber\\
\overline{|\mathcal{M}^{23}_{V,A}|^2}&=\left(\frac{1}{2}\sum_s\right)\sum_{s'}\sum_{\lambda}|\mathcal{M}_{V,A}^{23}|^2=e^4g_{V,A}^2\frac{F(q^2)^2}{q^4}\mathcal{A}^{23}_{V,A}
\end{align}
where
\begin{align}
\mathcal{A}^{23}_S=&-\frac{(\tilde{s}+\tilde{u})^2}{\tilde{s}\tilde{u}}P^2-4\frac{t}{\tilde{s}\tilde{u}}(P\cdotp k)^2-\frac{(\tilde{s}+\tilde{u})^2}{\tilde{s}^2\tilde{u}^2}(m_\phi^2-4m_e^2)\left[P^2 t+4\left(\frac{\tilde{u}P\cdotp p+\tilde{s}P\cdotp p'}{\tilde{s}+\tilde{u}}\right)^2\right]\nonumber\\
\mathcal{A}^{23}_P=&-\frac{(\tilde{s}+\tilde{u})^2}{\tilde{s}\tilde{u}}P^2-\frac{4t}{\tilde{s}\tilde{u}}(P\cdotp k)^2-\frac{(\tilde{s}+\tilde{u})^2}{\tilde{s}^2\tilde{u}^2}m_\phi^2\left[P^2 t+4\left(\frac{\tilde{u}P\cdotp p+\tilde{s}P\cdotp p'}{\tilde{s}+\tilde{u}}\right)^2\right]\nonumber\\
\mathcal{A}^{23}_V=&-2\frac{\tilde{s}^2+\tilde{u}^2}{\tilde{s}\tilde{u}}P^2-\frac{8t}{\tilde{s}\tilde{u}}\left[(P\cdotp p)^2+(P\cdotp p')^2-\frac{t_2+m_\phi^2}{2}P^2\right]\nonumber\\
&-2\frac{(\tilde{s}+\tilde{u})^2}{\tilde{s}^2\tilde{u}^2}(m_\phi^2+2m_e^2)\left[P^2 t+4\left(\frac{\tilde{u}P\cdotp p+\tilde{s}P\cdotp p'}{\tilde{s}+\tilde{u}}\right)^2\right]\\
\mathcal{A}^{23}_A=&-2\frac{\tilde{s}^2+\tilde{u}^2}{\tilde{s}\tilde{u}}P^2-\frac{8t}{\tilde{s}\tilde{u}}\left[(P\cdotp p)^2+(P\cdotp p')^2-\frac{t_2-m_\phi^2}{2}P^2\right]-4m_e^2\frac{(\tilde{s}+\tilde{u})^2 P^2+4t(P\cdotp k)^2}{m_\phi^2\tilde{s}\tilde{u}}\nonumber\\
&-2\frac{(\tilde{s}-\tilde{u})^2}{\tilde{s}^2\tilde{u}^2}(m_\phi^2-4m_e^2)\left[P^2 t+4\left(\frac{\tilde{u}P\cdotp p+\tilde{s}P\cdotp p'}{\tilde{s}-\tilde{u}}\right)^2\right]\nonumber.
\end{align}

\subsection{2 to 2 production}
\begin{figure}
\centering
\includegraphics[scale=1]{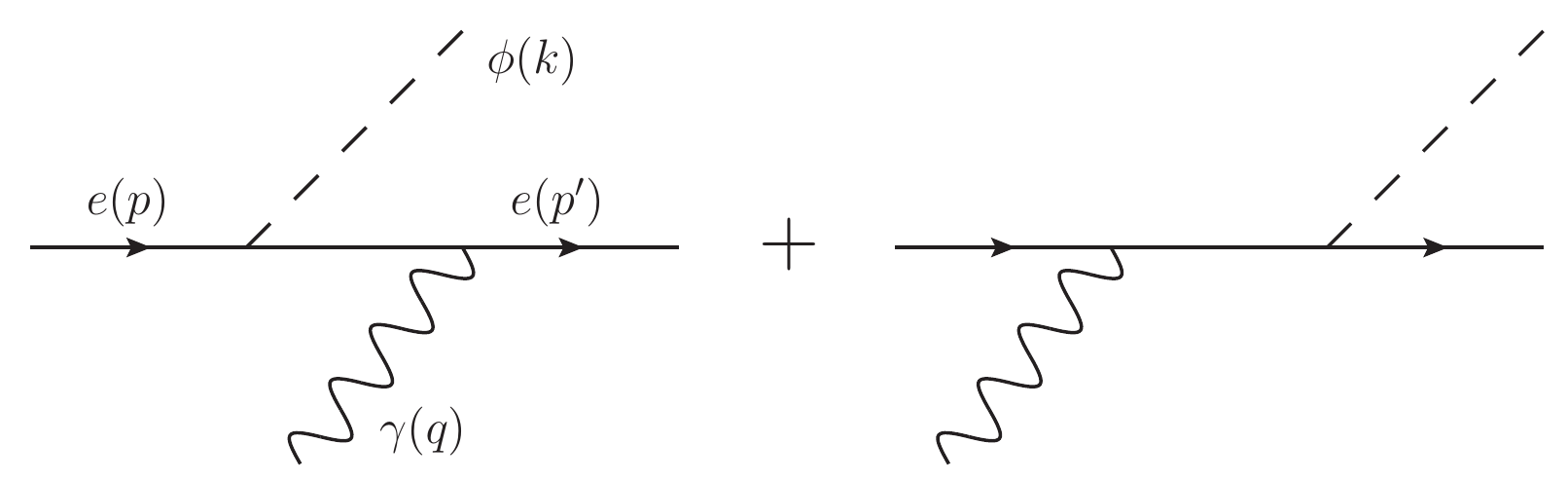}
\caption{\label{fig:2 to 2}  Lowest order 2 to 2 production process: $e(p)+\gamma(q)\rightarrow e(p')+\phi(k)$. $\gamma$, $e$, and $\phi$ stand for photon, electron, and the new boson.}
\end{figure}

For the 2 to 2 process in figure \ref{fig:2 to 2}, a ``subprocess'' of the full 2 to 3 interaction,
\begin{align}\label{eq:2 to 2 production process}
e(p)+\gamma(q)\rightarrow e(p')+\phi(k).
\end{align} 
With the same definition in Eq. (\ref{eq:2 to 3 variables}), $\tilde{s}$, $\tilde{u}$, and $t_2$ satisfy
\begin{align}
\tilde{s}+t_2+\tilde{u}&=m_\phi^2
\end{align}
and the amplitude in figure \ref{fig:2 to 2} is
\begin{align}
\mathcal{M}^{22}_S=&eg_S\epsilon^\mu_{q,\lambda}\bar{u}_{p',s'}\left[\gamma_\mu\frac{(\slashed{p}-\slashed{k})-m_e}{\tilde{u}}+\frac{(\slashed{p'}+\slashed{k})-m_e}{\tilde{s}}\gamma_\mu\right]u_{p,s}\nonumber\\
\mathcal{M}^{22}_P=&ieg_P\epsilon^\mu_{q,\lambda}\bar{u}_{p',s'}\left[\gamma_\mu\frac{(\slashed{p}-\slashed{k})-m_e}{\tilde{u}}\gamma_5+\gamma_5\frac{(\slashed{p'}+\slashed{k})-m_e}{\tilde{s}}\gamma_\mu\right]u_{p,s}\nonumber\\
\mathcal{M}^{22}_V=&eg_V\epsilon^\mu_{q,\lambda}\tilde{\epsilon}^\nu_{k,\lambda'}\bar{u}_{p',s'}\left[\gamma_\mu\frac{(\slashed{p}-\slashed{k})-m_e}{\tilde{u}}\gamma_\nu+\gamma_\nu\frac{(\slashed{p'}+\slashed{k})-m_e}{\tilde{s}}\gamma_\mu\right]u_{p,s}\\
\mathcal{M}^{22}_A=&eg_A\epsilon^\mu_{q,\lambda}\tilde{\epsilon}^\nu_{k,\lambda'}\bar{u}_{p',s'}\left[\gamma_\mu\frac{(\slashed{p}-\slashed{k})-m_e}{\tilde{u}}\gamma_5\gamma_\nu+\gamma_5\gamma_\nu\frac{(\slashed{p'}+\slashed{k})-m_e}{\tilde{s}}\gamma_\mu\right]u_{p,s}\nonumber
\end{align}
where $\epsilon$ is the photon polarization vector and $\lambda=\pm 1$. The polarization sum for photon is
\begin{align}
\sum_\lambda \epsilon^\mu_{q,\lambda}\epsilon^{\nu*}_{q,\lambda}=g^{\mu\nu}.
\end{align}
After averaging and summing over the initial and final spins and polarization,
\begin{align}\label{eq:2 to 2 M}
\overline{|\mathcal{M}^{22}|^2}_{S,P}&=\left(\frac{1}{2}\sum_s\right)\sum_{s'}\left(\frac{1}{2}\sum_\lambda\right)|\mathcal{M}^{22}_{S,P}|^2=e^2g_{S,P}^2\mathcal{A}^{22}_{S,P}\nonumber\\
\overline{|\mathcal{M}^{22}|^2}_{V,A}&=\left(\frac{1}{2}\sum_s\right)\sum_{s'}\left(\frac{1}{2}\sum_\lambda\right)\sum_{\lambda'}|\mathcal{M}^{22}_{V,A}|^2=e^2g_{V,A}^2\mathcal{A}^{22}_{V,A}
\end{align}
where
\begin{align}\label{eq:2 to 2 A}
\mathcal{A}^{22}_S=&-\frac{(\tilde{s}+\tilde{u})^2}{\tilde{s}\tilde{u}}+2(m_\phi^2-4m_e^2)\left[\left(\frac{\tilde{s}+\tilde{u}}{\tilde{s}\tilde{u}}\right)^2m_e^2-\frac{t_2}{\tilde{s}\tilde{u}}\right]\nonumber\\
\mathcal{A}^{22}_P=&-\frac{(\tilde{s}+\tilde{u})^2}{\tilde{s}\tilde{u}}+2m_\phi^2\left[\left(\frac{\tilde{s}+\tilde{u}}{\tilde{s}\tilde{u}}\right)^2m_e^2-\frac{t_2}{\tilde{s}\tilde{u}}\right]\nonumber\\
\mathcal{A}^{22}_V=&4-2\frac{(\tilde{s}+\tilde{u})^2}{\tilde{s}\tilde{u}}+4(m_\phi^2+2m_e^2)\left[\left(\frac{\tilde{s}+\tilde{u}}{\tilde{s}\tilde{u}}\right)^2m_e^2-\frac{t_2}{\tilde{s}\tilde{u}}\right]\\
\mathcal{A}^{22}_A=&4-\left(2+\frac{4m_e^2}{m_\phi^2}\right)\frac{(\tilde{s}+\tilde{u})^2}{\tilde{s}\tilde{u}}+4(m_\phi^2-4m_e^2)\left[\left(\frac{\tilde{s}+\tilde{u}}{\tilde{s}\tilde{u}}\right)^2m_e^2-\frac{t_2}{\tilde{s}\tilde{u}}\right]\nonumber.
\end{align}

\section{Cross Section}\label{sec:cross section}
\subsection{2 to 3}
The cross section for the 2 to 3 process, see Eq. (\ref{eq:2 to 3 production process}) and figure \ref{fig:2 to 3}, in the lab frame is given by
\begin{align}
d\sigma=\frac{1}{4|\textbf{p}|M}\overline{|\mathcal{M}^{2\to3}|^2}(2\pi)^4\delta^4(p'+k-p-q)\frac{d^3\textbf{p}'}{(2\pi)^3 2E'}\frac{d^3\textbf{P}_f}{(2\pi)^3 2E_f}\frac{d^3\textbf{k}}{(2\pi)^3 2E_k}
\end{align}
where $M$ is the mass of the target atom. Integrating over $\mathbf{p}'$ and changing the variable from $\mathbf{P}_f$ to $\mathbf{q}$, we have
\begin{align}
d\sigma=\frac{\overline{|\mathcal{M}^{2\to3}|^2}}{1024\pi^5|\textbf{p}|M E_f E' E_k}\delta(E'+E_k-E-q_0)d^3\textbf{q}d^3\textbf{k}.
\end{align}
In order to integrate over $\mathbf{q}$, we choose the spherical coordinate $(Q,\theta_q,\phi_q)$ where $Q=|\mathbf{q}|$, and $\theta_q$ and $\phi_q$ are the polar and azimuthal angles of \textbf{q} in the direction of $\mathbf{V}=\mathbf{k}-\mathbf{p}$. First, we use the remaining $\delta$-function to integrate out $Q$, and then change variables from $\theta_q$ to $t$. We obtain
\begin{align}
d\sigma=\frac{d^3\textbf{k}}{128\pi^4|\textbf{p}|V E_k}\int^{t_{max}}_{t_{min}}dt\left(\frac{1}{8M^2}\int_0^{2\pi}\frac{d\phi_q}{2\pi}\overline{|\mathcal{M}^{2\to3}|^2}\right)
\end{align}
where $V=|\textbf{V}|$, $t(Q)=q^2=2M(\sqrt{M^2+Q^2}-M)$,
\begin{align}\label{eq:tmin tmax}
t_{max}=t(Q_+),\;\, t_{min}=t(Q_-), 
\end{align}
and
\begin{align}
Q_\pm=\frac{V[\tilde{u}+2M(E'+E_f)]\pm(E'+E_f)\sqrt{\tilde{u}^2+4M\tilde{u}(E'+E_f)+4M^2V^2}}{2(E'+E_f)^2-2V^2}.
\end{align}
Integrate over the polar angle, $\theta$, and azimuthal angle of \textbf{k} in the direction of \textbf{p}, and then change the variable from $|\mathbf{k}|$ to $x$ where $x\equiv E_k/E$. We have
\begin{align}\label{eq:2 to 3}
\frac{d\sigma}{dx d\cos\theta}&=\frac{|\textbf{k}|E}{64\pi^3|\textbf{p}|V}\int^{t_{max}}_{t_{min}}dt\left(\frac{1}{8M^2}\int_0^{2\pi}\frac{d\phi_q}{2\pi}\overline{|\mathcal{M}^{2\to3}|^2}\right)\nonumber\\
&=\epsilon^2\alpha^3\frac{|\textbf{k}|E}{|\textbf{p}|V}\int^{t_{max}}_{t_{min}}dt\frac{F(t)^2}{t^2}\left(\frac{1}{8M^2}\int_0^{2\pi}\frac{d\phi_q}{2\pi}\mathcal{A}^{2\to3}\right).
\end{align}

\subsection{2 to 2}
The 2 to 2 cross section, see Eq. (\ref{eq:2 to 2 production process}) and figure \ref{fig:2 to 2}, in the lab frame is straightforwardly expressed in terms of the amplitude,
\begin{align}\label{eq:2 to 2}
\frac{d\sigma}{d(p\cdotp k)}=2\frac{d\sigma}{dt_2}=\frac{\overline{|\mathcal{M}^{2\to2}|^2}}{8\pi\tilde{s}^2}=\epsilon^2\alpha^2\frac{2\pi}{\tilde{s}^2}\mathcal{A}^{2\to2}.
\end{align}

\section{Weizs\"{a}cker-Williams Approximation}\label{sec:WW approximation}
It is explained in Ref.~\cite{Kim:1973he} that the WW approximation relies on the incoming electron energy being much greater than $m_\phi$ and $m_e$, such that the final state electron and the new boson are highly collinear. In that case the phase space integral can be approximated by
\begin{align}\label{eq:WW}
\frac{1}{8M^2}\int\frac{d\phi_q}{2\pi}\mathcal{A}^{2\to3}\approx\frac{t-t_{min}}{2t_{min}}\mathcal{A}^{2\to2}_{t=t_{min}}.
\end{align}
With the WW approximation, Eq.~(\ref{eq:2 to 3}) can be approximated to be
\begin{align}
\frac{d\sigma}{dx d\cos\theta}\approx\epsilon^2\alpha^3\frac{|\textbf{k}|E}{|\textbf{p}|V}\frac{\mathcal{A}^{2\to2}_{t=t_{min}}}{2t_{min}}\chi
,\end{align}
where
\begin{align}\label{eq:chi}
\chi=\int^{t_{max}}_{t_{min}}dt\frac{t-t_{min}}{t^2}F(t)^2.
\end{align}
Using Eq. (\ref{eq:2 to 2}), we have
\begin{align}
\frac{d\sigma}{dx d\cos\theta}\approx\frac{\alpha\chi}{4\pi}\frac{|\textbf{k}|E}{|\textbf{p}|V}\frac{\tilde{s}^2}{t_{min}}\left.\frac{d\sigma}{d(p\cdotp k)}\right|_{t=t_{min}}.
\end{align}
Following the discussion in Refs.~\cite{Bjorken:2009mm,Tsai:1986tx}, near $t=t_{min}$ (when $\mathbf{q}$ and $\mathbf{V}=\mathbf{k}-\mathbf{p}$ are collinear), we can approximate the following quantities
\begin{align}\label{eq:WW variables}
\tilde{s}&\approx-\frac{\tilde{u}}{1-x}\nonumber\\
\tilde{u}&\approx-xE^2\theta^2-m_\phi^2\frac{1-x}{x}-m_e^2 x\nonumber\\
t_2&\approx\frac{\tilde{u}x}{1-x}+m_\phi^2\\
V&\approx E(1-x)\nonumber\\
t_{min}&\approx\frac{\tilde{s}^2}{4E^2}\nonumber
\end{align}
Using Eq. (\ref{eq:WW variables}), we arrive at the well-known equation \cite{Bjorken:2009mm,Tsai:1986tx}
\begin{align}\label{eq:WW Tsai}
\frac{d\sigma}{dx d\cos\theta}\approx\frac{\alpha\chi}{\pi}\frac{xE^2\beta}{1-x}\left.\frac{d\sigma}{d(p\cdotp k)}\right|_{t=t_{min}}
\end{align}
where $\beta=\sqrt{1-m_\phi^2/E_k^2}$. Note that in Eq. (\ref{eq:WW Tsai}) $d\sigma/d(p\cdotp k)$ is evaluated at $t=t_{min}$. So the amplitude $\mathcal{A}^{2\to2}$ in Eq. (\ref{eq:2 to 2}) evaluated at $t=t_{min}$ using Eq. (\ref{eq:WW variables}) is
\begin{align}\label{eq:2 to 2 A tmin}
\mathcal{A}^{22}_{S,t=t_{min}}\approx&\frac{x^2}{1-x}+2(m_\phi^2-4m_e^2)\frac{\tilde{u}x+m_\phi^2(1-x)+m_e^2x^2}{\tilde{u}^2}\nonumber\\
\mathcal{A}^{22}_{P,t=t_{min}}\approx&\frac{x^2}{1-x}+2m_\phi^2\frac{\tilde{u}x+m_\phi^2(1-x)+m_e^2x^2}{\tilde{u}^2}\nonumber\\
\mathcal{A}^{22}_{V,t=t_{min}}\approx&2\frac{2-2x+x^2}{1-x}+4(m_\phi^2+2m_e^2)\frac{\tilde{u}x+m_\phi^2(1-x)+m_e^2x^2}{\tilde{u}^2}\\
\mathcal{A}^{22}_{A,t=t_{min}}\approx&\frac{4m^2x^2}{m_\phi^2(1-x)}+2\frac{2-2x+x^2}{1-x}+4(m_\phi^2-4m_e^2)\frac{\tilde{u}x+m_\phi^2(1-x)+m_e^2x^2}{\tilde{u}^2}\nonumber.
\end{align}

\section{Cross Section Comparison}\label{sec:cross section comparison}

\begin{figure}
\centering
\subfigure[\;$d\sigma_S/(\epsilon_S^2dx)$]{\includegraphics[scale=0.9]{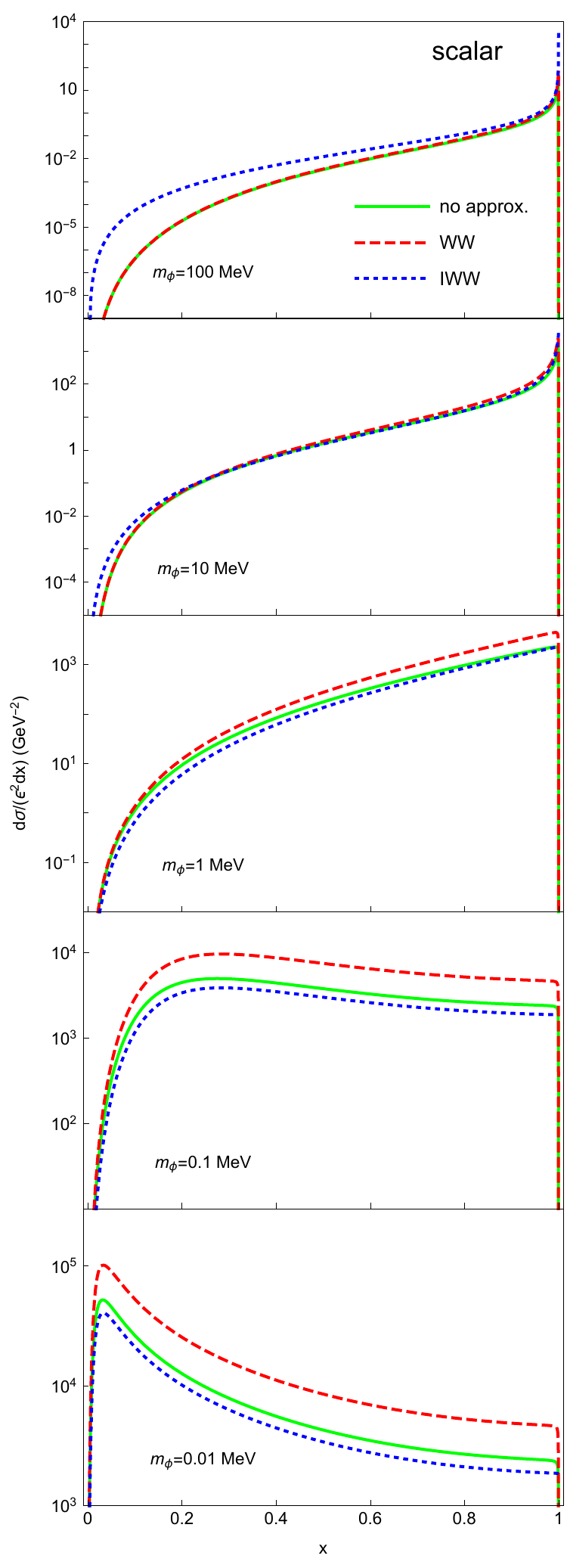}}
\subfigure[\;relative error of $d\sigma_S/(\epsilon_S^2 dx)$]{\includegraphics[scale=0.9]{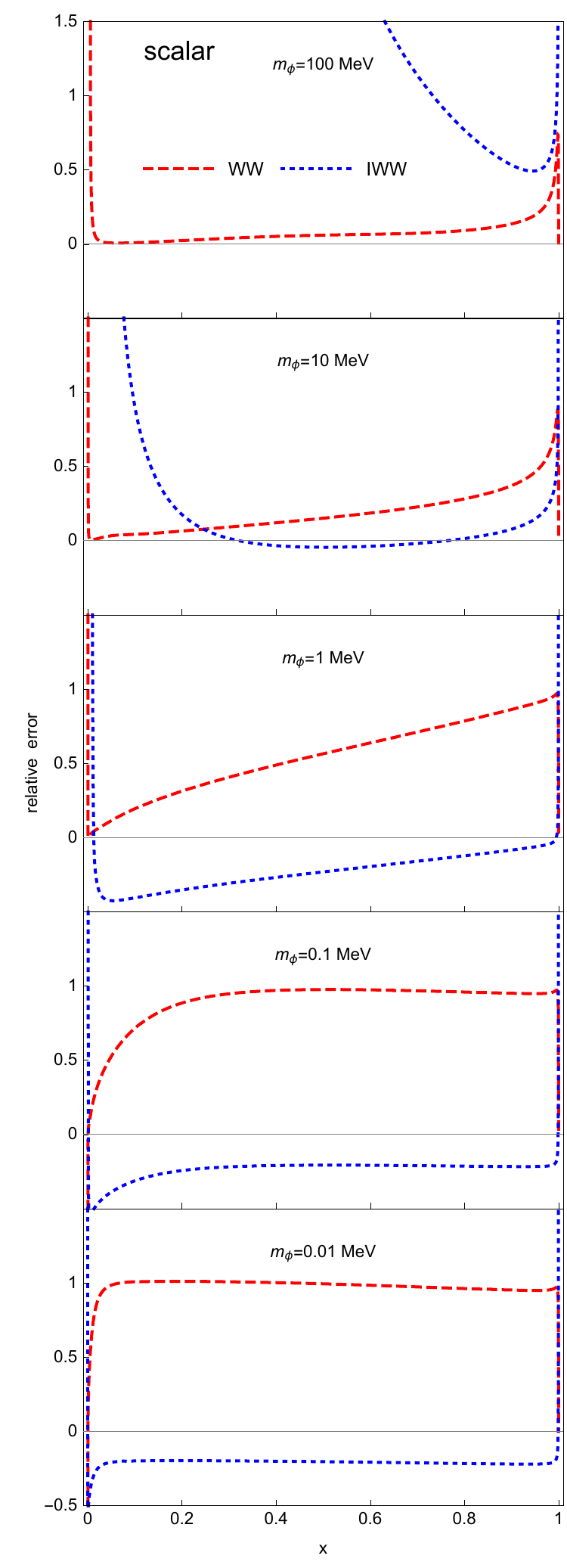}}
\caption{\label{fig:cross_section} Cross section of a scalar boson production: The solid green, dashed red, and dotted blue lines correspond to the differential cross section with no, WW, and IWW approximation. The relative error of $\mathcal{O}$ is defined by $(\mathcal{O}_{\rm approx.}-\mathcal{O}_{\rm exact})/\mathcal{O}_{\rm exact}$.}
\end{figure}

\begin{figure}
\centering
\subfigure[\;$d\sigma_P/(\epsilon_P^2 dx)$]{\includegraphics[scale=0.9]{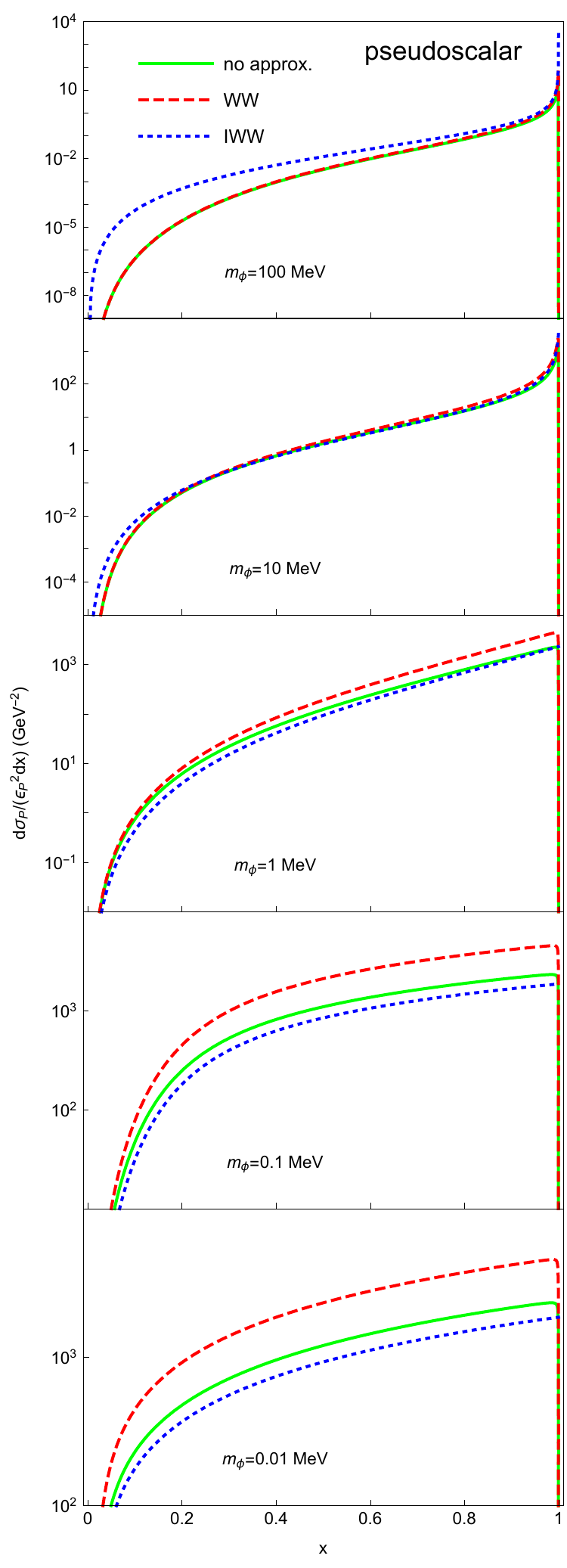}}
\subfigure[\;relative error of $d\sigma_P/(\epsilon_P^2 dx)$]{\includegraphics[scale=0.9]{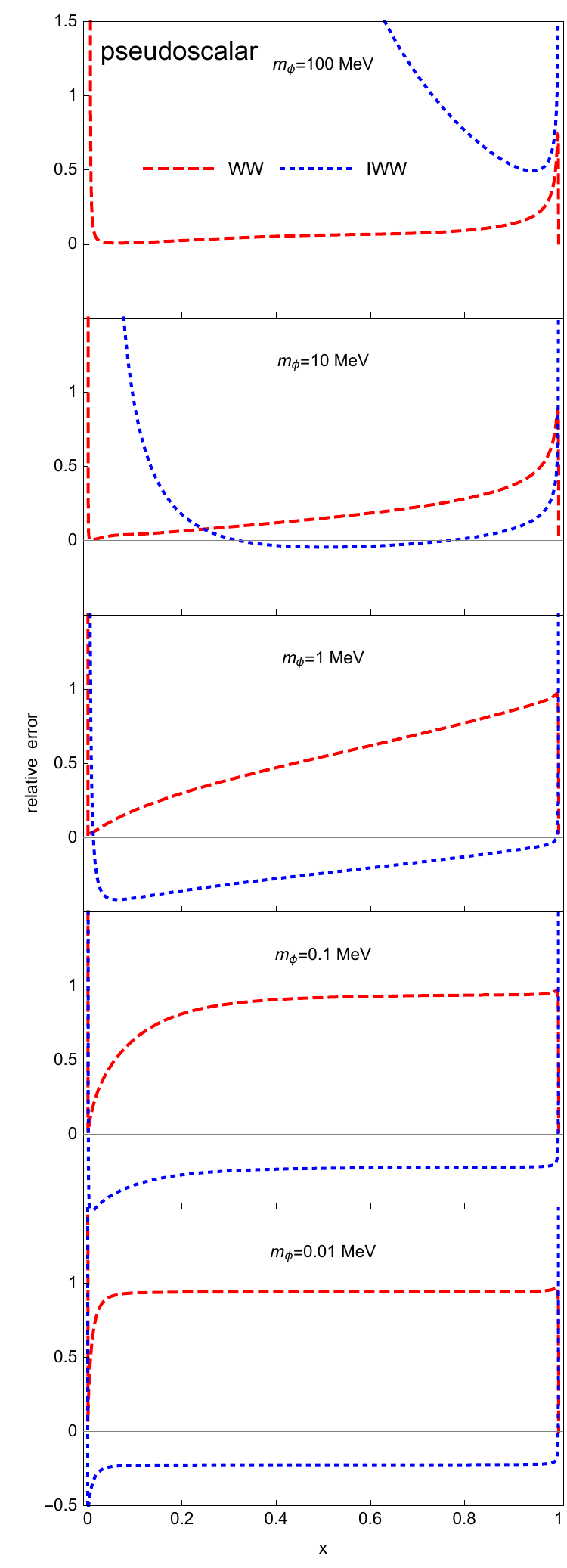}}
\caption{\label{fig:pseudoscalar_cross_section} Cross section of a pseudoscalar boson production: See caption of figure \ref{fig:cross_section} for detail.}
\end{figure}

\begin{figure}
\centering
\subfigure[\;$d\sigma_V/(\epsilon_V^2 dx)$]{\includegraphics[scale=0.9]{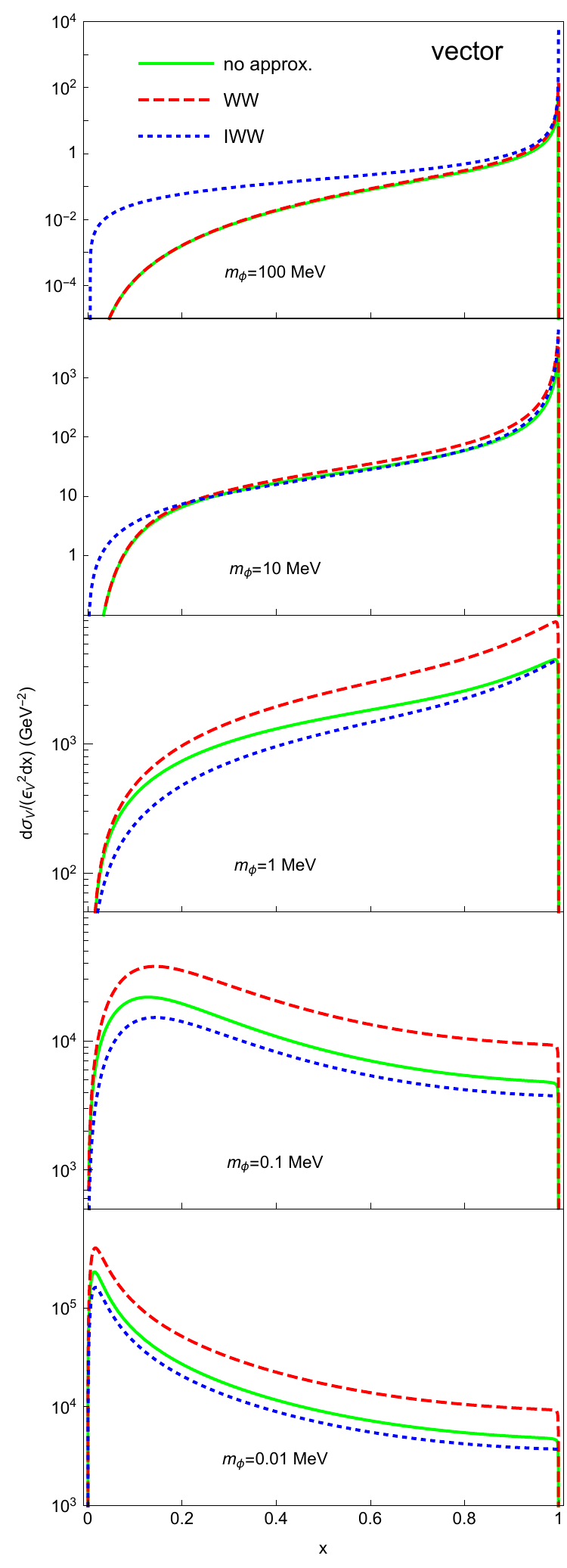}}
\subfigure[\;relative error of $d\sigma_V/(\epsilon_V^2 dx)$]{\includegraphics[scale=0.9]{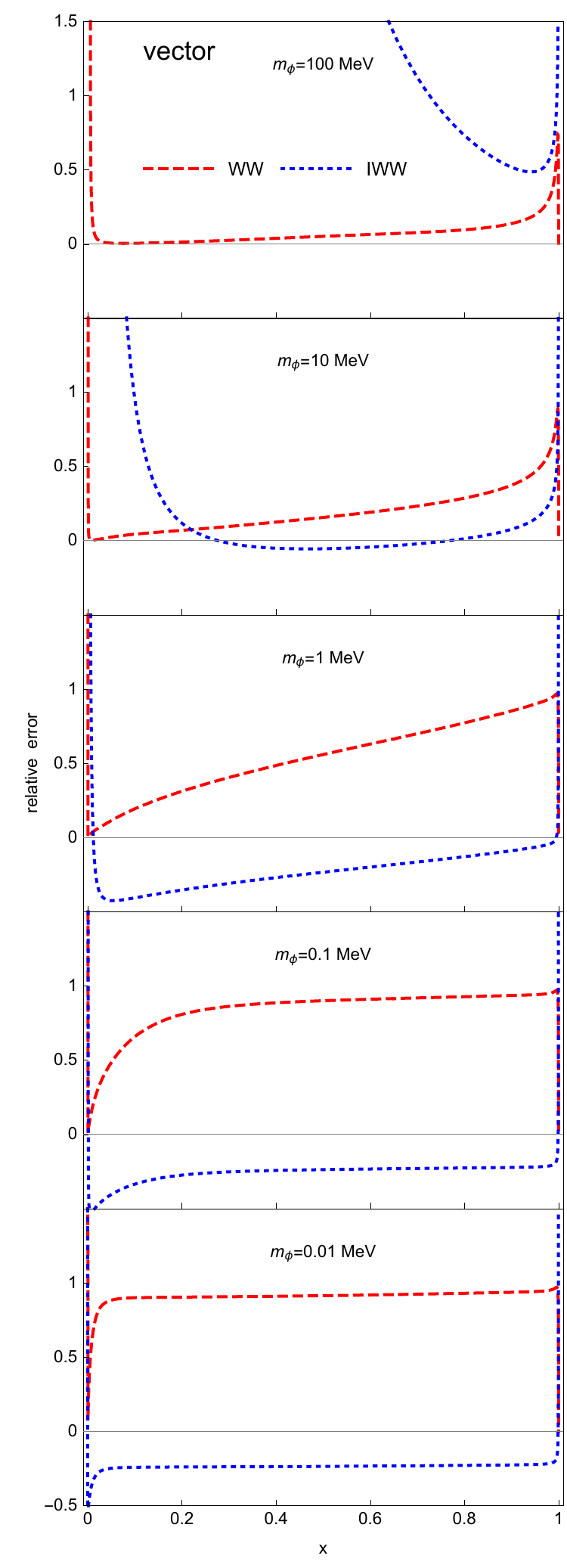}}
\caption{\label{fig:vector_cross_section} Cross section of a vector boson production: See caption of figure \ref{fig:cross_section} for detail.}
\end{figure}

\begin{figure}
\centering
\subfigure[\;$d\sigma_A/(\epsilon_A^2 dx)$]{\includegraphics[scale=0.9]{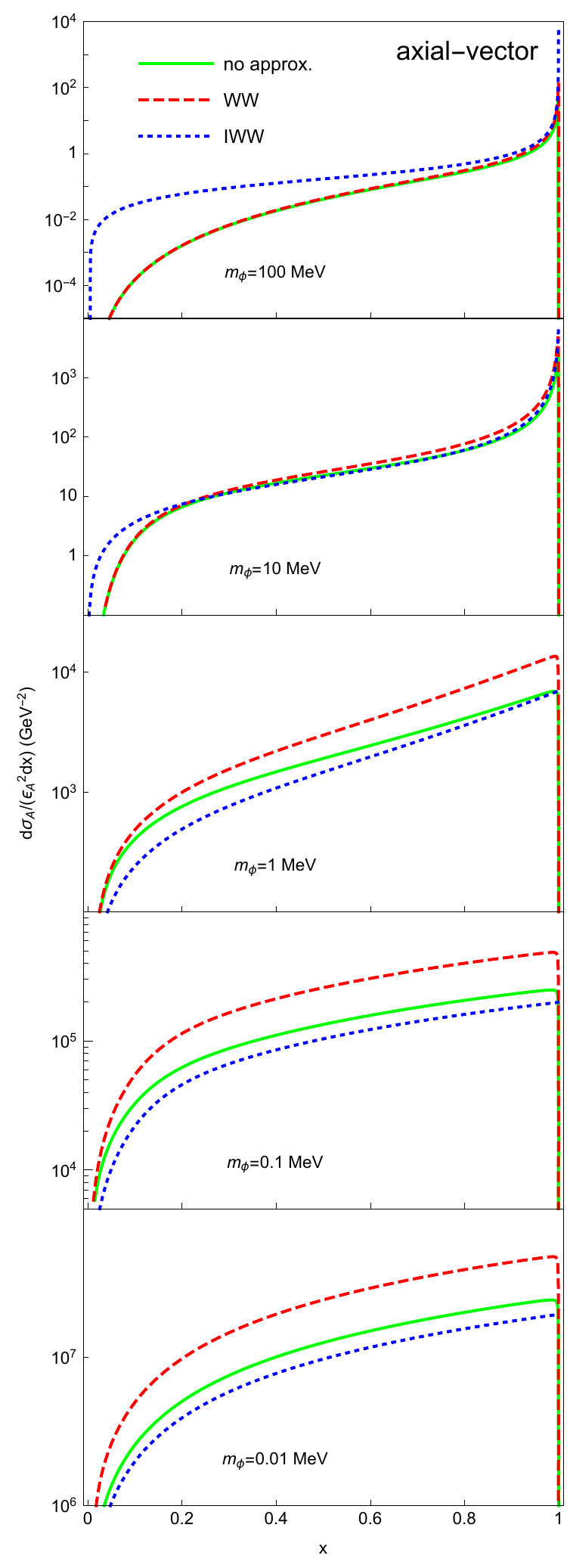}}
\subfigure[\;relative error of $d\sigma_A/(\epsilon_A^2 dx)$]{\includegraphics[scale=0.9]{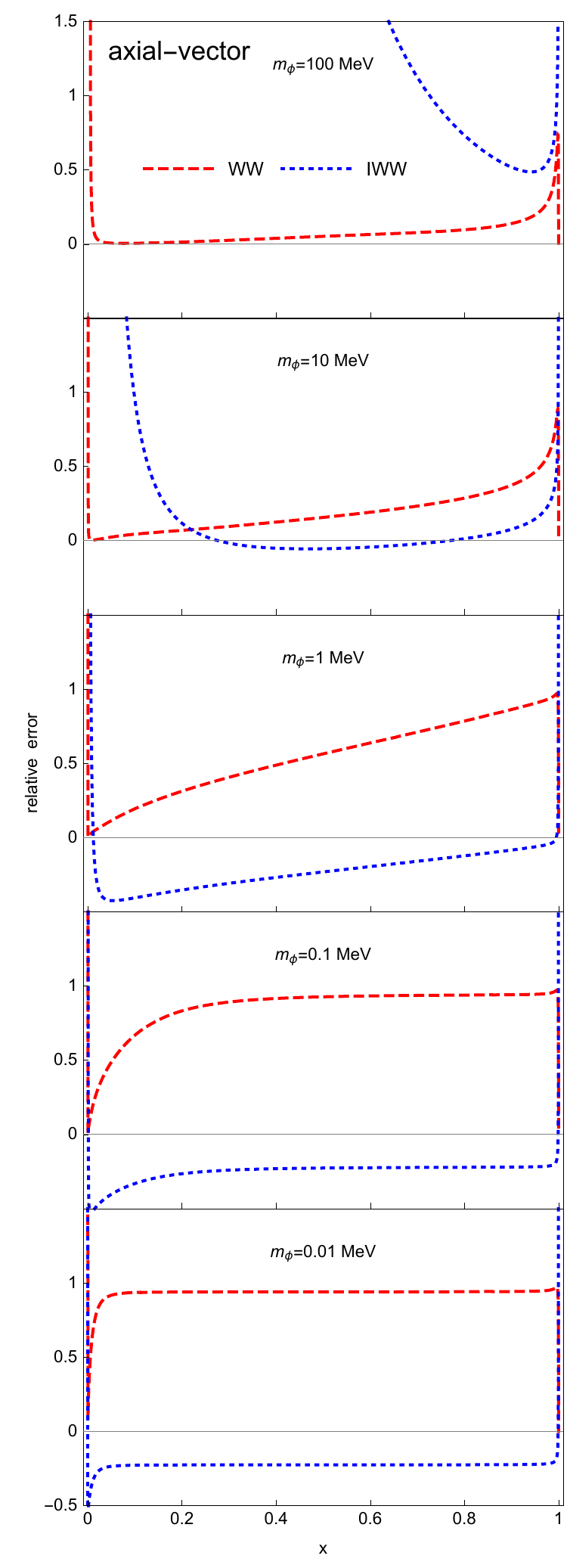}}
\caption{\label{fig:axial_vector_cross_section} Cross section of a axial-vector boson production: See caption of figure \ref{fig:cross_section} for detail.}
\end{figure}

To test approximations of the cross section for $\phi$ production, we examine three cases. 
\begin{enumerate}
\item The complete calculation, Eq. (\ref{eq:2 to 3}),
\begin{align}\label{eq:d sigma dx 1}
\frac{d\sigma}{dx}=\epsilon^2\alpha^3\frac{|\textbf{k}|E}{|\textbf{p}|}\int_0^{\theta_{max}} d\cos\theta\frac{1}{V}\int^{t_{max}}_{t_{min}}dt\frac{F(t)^2}{t^2}\left(\frac{1}{8M^2}\int_0^{2\pi}\frac{d\phi_q}{2\pi}\mathcal{A}^{2\to3}\right)
\end{align}
where $\theta_{max}$ depends on the configuration of the detector. For beam dump E137, $\theta_{max}\approx 4.4\times10^{-3}$.

\item WW: using the WW approximation, Eq. (\ref{eq:WW}),
\begin{align}\label{eq:d sigma dx 2}
\left(\frac{d\sigma}{dx}\right)_{WW}=2\epsilon^2\alpha^3|\textbf{k}|E(1-x)\int_0^{\theta_{max}} d\cos\theta\frac{\mathcal{A}^{2\to2}_{t=t_{min}}}{\tilde{u}^2}\chi
\end{align}
where $\theta_{max}$ is the same as the first case and $\chi$ is defined in Eq. (\ref{eq:chi}). Note that the upper and lower limits of $\chi$ depend on $x$ and $\theta$.

\item Improved WW (IWW): If the upper and lower limits of the $t$-integral in $\chi$ in Eq. (\ref{eq:d sigma dx 2}) are not sensitive to $x$ and $\theta$; i.e., the integration limit can be set to be independent of $x$ and $\theta$, we can further approximate the integration limits of $t$. Similar to the argument in Ref.~\cite{Bjorken:2009mm}, we set
\begin{align}\label{eq:IWW tmin tmax}
t_{min}=\left(\frac{m_\phi^2}{2E}\right)^2 {\rm\; and\;\;} t_{max}=m_\phi^2+m_e^2
\end{align}
which is valid when the production cross section is dominantly collinear with $x$ close to 1. The difference in $t_{max}$ between \cite{Bjorken:2009mm} and our approach occurs because we do not assume $m_\phi\gg m_e$. Therefore, we can pull $\chi$ out of the integral over $\cos\theta$. Then, changing variables from $\cos\theta$ to $\tilde{u}$ and extending the lower limit of $\tilde{u}$ to $-\infty$,
\begin{align}\label{eq:d sigma dx 3-1}
\left(\frac{d\sigma}{dx}\right)_{IWW}&=\epsilon^2\alpha^3\chi\frac{|\textbf{k}|}{E}\frac{1-x}{x}\int^{\tilde{u}_{max}}_{-\infty}d\tilde{u}\frac{\mathcal{A}^{22}_{t=t_{min}}}{\tilde{u}^2}
\end{align}
using Eq. (\ref{eq:2 to 2 A tmin}) we have
\begin{align}\label{eq:d sigma dx 3-2}
\left(\frac{d\sigma_S}{dx}\right)_{IWW}=&\epsilon_S^2\alpha^3\chi\frac{|\textbf{k}|}{E}\frac{m_e^2(2-x)^2-2x \tilde{u}_{max}}{3\tilde{u}_{max}^2}\nonumber\\
\left(\frac{d\sigma_P}{dx}\right)_{IWW}=&\epsilon_P^2\alpha^3\chi\frac{|\textbf{k}|}{E}\frac{m_e^2x^2-2x \tilde{u}_{max}}{3\tilde{u}_{max}^2}\nonumber\\
\left(\frac{d\sigma_V}{dx}\right)_{IWW}=&2\epsilon_V^2\alpha^3\chi\frac{|\textbf{k}|}{E}\frac{m_e^2x(-2+2x+x^2)-2(3-3x+x^2)\tilde{u}_{max}}{3x\tilde{u}_{max}^2}\\
\left(\frac{d\sigma_A}{dx}\right)_{IWW}=&2\epsilon_A^2\alpha^3\chi\frac{|\textbf{k}|}{E}\left[\frac{m_e^2x(2-x)^2-2(3-3x+x^2)\tilde{u}_{max}}{3x\tilde{u}_{max}^2}+\frac{2m_e^2(1-x)}{\tilde{u}_{max}(\tilde{u}_{max}+m_e^2x)}\right]\nonumber
\end{align}
where $\tilde{u}_{max}=-m_\phi^2\frac{1-x}{x}-m_e^2 x$. We emphasize that the name ``improved" means reducing the computational time (because of one fewer integral than in the WW approximation above) and does not imply more accuracy.

\end{enumerate}

In figures \ref{fig:cross_section}--\ref{fig:axial_vector_cross_section}, we show the cross sections in each of the above three cases for five values of the new boson mass, setting the incoming electron beam energy to 20 GeV and the target to be aluminum. The cross sections for different bosons are different, as expected, because they have different dynamics; the relative errors with the same approximation between different bosons are similar, also as expected, because the approximation deals with phase space integral and the kinematics between different bosons are similar.

In both approximations, the cross section is of the same order of magnitude as that using the complete calculation. However, there are regions where there are ${\cal O}\left(1\right)$ relative errors. The WW approximation (dashed red lines in figures \ref{fig:cross_section}--\ref{fig:axial_vector_cross_section}) can differ from the complete calculation by 100\% when $m_\phi\lesssim 1$ MeV; in the IWW case (dotted blue lines in figures \ref{fig:cross_section}--\ref{fig:axial_vector_cross_section}), the approximation starts to fail when $m_\phi\gtrsim 100$ MeV.

\begin{figure}
\centering
\includegraphics[scale=1.7]{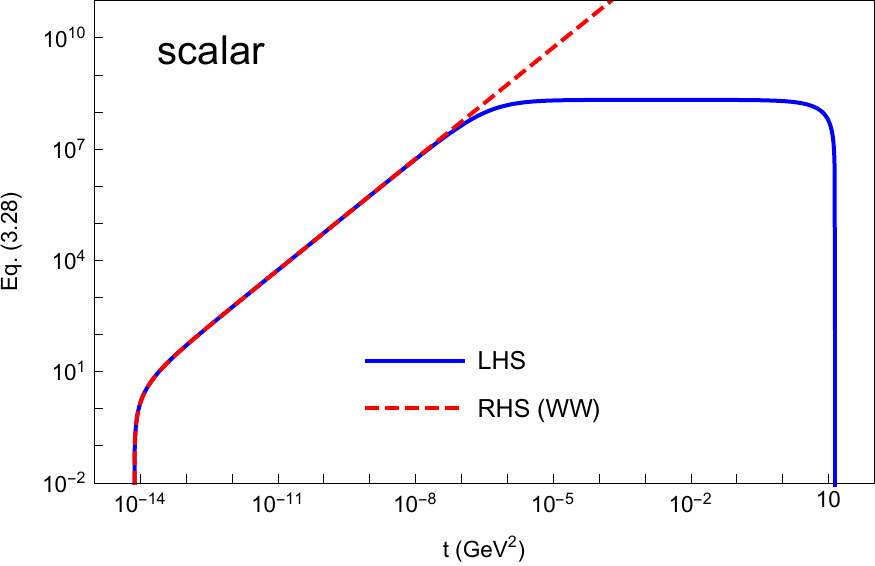}
\caption{\label{fig:WW} The solid blue (dashed red) line is the right-hand (left-hand) side of Eq. (\ref{eq:WW}) evaluated (using scalar boson) at $m_\phi=1$ MeV, $E=20$ GeV, $x=0.9$, and $\theta=0$. The WW approximation works well at low $t$ region but starts to fail at higher $t$.}
\end{figure}

\begin{figure}
\centering
\includegraphics[scale=1.7]{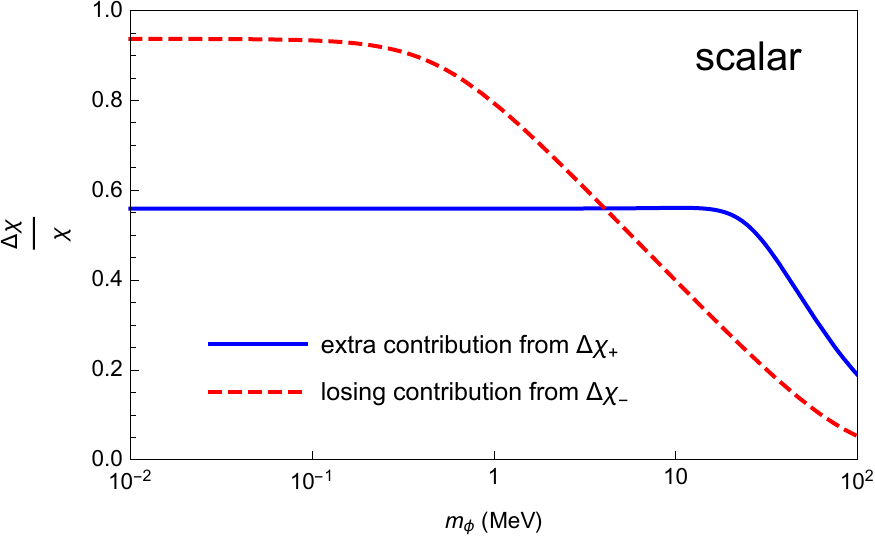}
\caption{\label{fig:IWW}  The solid blue (dashed red) line is the extra (losing) contribution when calculating the factor $\chi$ in Eq. (\ref{eq:chi}). We
evaluate the ratio of $\Delta\chi_{+}$ Eq. (\ref{eq:chi+}) and $\Delta\chi_{-}$ Eq. (\ref{eq:chi-}) to $\chi$ Eq. (\ref{eq:chi}) (using scalar boson) at $E=20$ GeV, $x=0.9$, and $\theta=2\times 10^{-2}$. We see that the total $chi$ using IWW approximation is smaller (bigger) in lower (higher) $m_\phi$ than $chi$ with complete calculation. This also reflex to the differential cross section using IWW approximation in figures \ref{fig:cross_section}--\ref{fig:axial_vector_cross_section}.}
\end{figure}

It is worth noting that in figures \ref{fig:cross_section}--\ref{fig:axial_vector_cross_section} the differential cross section using WW approximation is always greater than the complete calculation. This can be further understood by Eq. (\ref{eq:WW}) and an example in figure \ref{fig:WW}. The WW approximation works well at low $t$ region but starts to fail at higher $t$. The bigger cross sections obtained using the WW approximation indicate that the integral over $t$ picks up an extra contribution at large $t$ where the WW approximation using real photons fails.

On the other hand, the behavior of IWW approximation is more bizarre. In figures \ref{fig:cross_section}--\ref{fig:axial_vector_cross_section}, the differential cross section using IWW approximation can either be bigger or smaller than the complete calculation. To explain this, first we notice that both $t_{min,IWW}$ and  $t_{max,IWW}$ in Eq. (\ref{eq:IWW tmin tmax}) are smaller than actual $t_{min}$ and  $t_{max}$ in Eq. (\ref{eq:tmin tmax}), therefore when we calculate the factor $\chi$ in Eq. (\ref{eq:chi}), there will be an extra contribution from the modified lower bound
\begin{align}\label{eq:chi+}
\Delta\chi_{+}=\int^{t_{min}}_{t_{min,IWW}}dt\frac{t-t_{min}}{t^2}F(t)^2
\end{align}
and lose a  contribution from the modified upper bound
\begin{align}\label{eq:chi-}
\Delta\chi_{-}=\int^{t_{max,IWW}}_{t_{max}}dt\frac{t-t_{min}}{t^2}F(t)^2.
\end{align}
We show an example in figure \ref{fig:IWW} that the total $\chi$ using IWW approximation lose more contribution from $\Delta\chi_{-}$ than $\Delta\chi_{+}$ gained in lower $m_\phi$ region, and vice versa in higher $m_\phi$ region. Therefore we see in figures \ref{fig:cross_section}--\ref{fig:axial_vector_cross_section} that the differential cross section is smaller (bigger) in lower (higher) $m_\phi$ region than the differential cross section using complete calculation.

\section{Particle Production}\label{sec:particle production}

There are two characteristic lengths which are crucial in beam dump experiments. The first is the decay length of the new particle in the lab frame,
\begin{align}
l_\phi=\frac{|\mathbf{k}|}{m_\phi}\frac{1}{\Gamma_\phi},
\end{align}
where $\Gamma_\phi=\Gamma(\phi\to e^+e^-)+\Gamma(\phi\to{\rm photons})$, see Eqs. (\ref{eq:decay to electrons},\ref{eq:P decay to photons},\ref{eq:V decay to photons},\ref{eq:A decay to photons}). The new particle, after production, must decay after going through the target and shielding and before going through the detector in order to be observed. If the target is thick (much greater than a radiation length), most of the new particles will be produced in the first few radiation lengths. The production rate is approximately proportional to the probability $e^{-L_{sh}/l_\phi}(1-e^{-L_{dec}/l_\phi})$, where $L_{sh}$ is length of the target and shield and $L_{dec}$ is length for the new particle to decay into electron or photon pairs after the shield and before the detector.

The second characteristic length is the absorption length
\begin{align}
\lambda=\frac{1}{n_e\sigma_{abs}},
\end{align}
where $n_e$ is the number density of the target electrons and $\sigma_{abs}$ is the cross section of absorption process. The leading process of absorption is
\begin{align}\label{eq:absorption}
e(p)+\phi(k)\rightarrow e(p')+\gamma(q),
\end{align} 
which is related to the 2 to 2 production process Eq. (\ref{eq:2 to 2 production process}) via crossing symmetry $\tilde{s}\leftrightarrow\tilde{u}$. Since Eq. (\ref{eq:2 to 2 A}) is symmetric in $\tilde{s}\leftrightarrow\tilde{u}$, the algebraic form of amplitude squared of absorption process is the same as Eq. (\ref{eq:2 to 2 A}) but differs by a factor 2 from summing over final state instead of averaging over initial state in Eq. (\ref{eq:2 to 2 M})
\begin{align}
\mathcal{A}^{22}_{abs}=c\mathcal{A}^{22}
\end{align}
where $c=2$ for spin-0 and $c=\frac{2}{3}$ for spin-1 particles.

The cross section of the process (\ref{eq:absorption}) is 
\begin{align}
\frac{d\sigma}{d\Omega}&=\frac{1}{64\pi^2 m_e}\frac{|\mathbf{q}|}{|\mathbf{k}|}\frac{\overline{|\mathcal{M}^{2\to2}_{abs}|^2}}{E_k+m_e-|\mathbf{k}|\cos\theta_\gamma}\\
\sigma_{abs}&=\frac{\pi\epsilon^2\alpha^2}{m_e|\mathbf{k}|}\int_{-1}^1 d\cos\theta_\gamma\frac{|\mathbf{q}|\mathcal{A}^{2\to2}}{E_k+m_e-|\mathbf{k}|\cos\theta_\gamma},
\end{align}
where $\theta_\gamma$ is the angle between outgoing photon and incoming new particle. The new particle, after produced, must not be absorbed by the target and shield to be detected. If the target is thick (much greater than absorption length), the production rate will be approximately proportional to the probability $e^{-L_{sh}/\lambda}$.

The number of the new particles produced in terms of the cross section (without considering the absorption process) can be found in, {\it e.g.}, Refs.~\cite{Bjorken:2009mm,Tsai:1986tx,Andreas:2012mt}. Using the thick target approximation and including the absorption process, we find
\begin{align}
N_\phi\approx\frac{N_eX}{M}\int_{E_{min}}^{E_0}dE\int_{x_{min}}^{x_{max}}dx\int_0^TdtI_e(E_0,E,t)\frac{d\sigma}{dx}e^{-L_{sh}\left(\frac{1}{l_\phi}+\frac{1}{\lambda}\right)}(1-e^{-L_{dec}/l_\phi}),
\end{align}
where $M$ is the mass of the target atom (aluminium); $N_e$ is the number of incident electrons; $X$ is the unit radiation length of the target; $E_0$ is the incoming electron beam energy, $E_{min}=m_e+\max(m_\phi,E_{cut})$ and $x_{min}=\frac{\max(m_\phi,E_{cut})}{E}$ where $E_{cut}$ is the measured energy cutoff depending on the detectors; $x_{max}$, which is smaller  but very close to 1 ($x_{max}$ can be approximated to be $1-\frac{m_e}{E}$ if the new particle and electron initial and final state are collinear); $T=\rho L_{sh}/X$ where $\rho$ is the density of the target; $l_\phi$ is the decay length of the new particle in lab frame; $\lambda$ is the absorption length of the new particle passing through the target and shield; $I_e$, derived in Ref.~\cite{Tsai:1966js}, is the energy distribution of the electrons after passing through a medium of $t$ radiation length  
\begin{align}
I_e(E_0,E,t)=\frac{\left(\ln\frac{E_0}{E}\right)^{bt-1}}{E_0\Gamma(bt)},
\end{align}
where $\Gamma$ is the gamma function and $b=4/3$. For beam dump E137 which we take as our prototypical setup, $E_0=20$ GeV and $E_{cut}=2$ GeV; $N_e=1.87\times 10^{{20}}$; $L_{sh}=179$ m and $L_{dec}=204$ m. The experiment has a null result which translates to 95\% C.L. of $N_\phi$ to be 3 events.

\section{Exclusion Plots}

\begin{figure}
\centering
\includegraphics[scale=2.1]{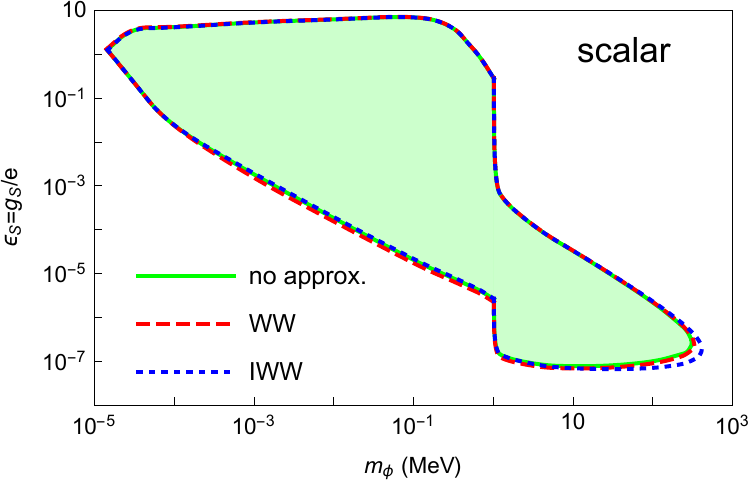}
\caption{\label{fig:E137} Exclusion (shaded region) plot for scalar using the beam dump experiment E137. The solid green, dashed red, and dotted blue lines correspond to the differential cross section with no, WW, and IWW approximation.}
\end{figure}

\begin{figure}
\centering
\includegraphics[scale=2.1]{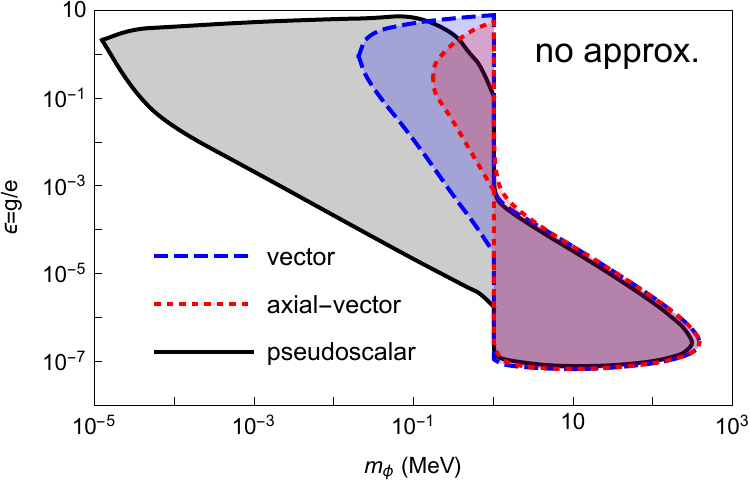}
\caption{\label{fig:E137all} Exclusion (shaded region) plot for pseudoscalar (black), vector (dashed blue), and axial-vector (dotted red) bosons without approximation using the beam dump experiment E137.}
\end{figure}

\begin{figure}
\centering
\subfigure[\;exclusion plot (mass in linear scale)]{\includegraphics[scale=1.5]{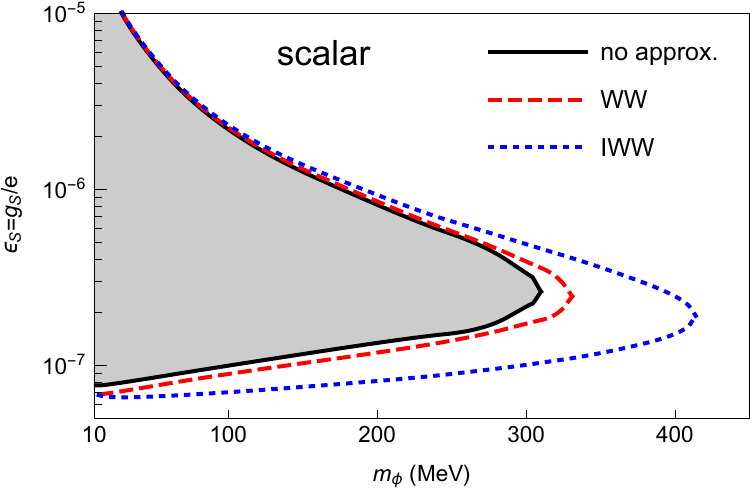}}
\subfigure[\;relative error of exclusion boundary]{\includegraphics[scale=1.5]{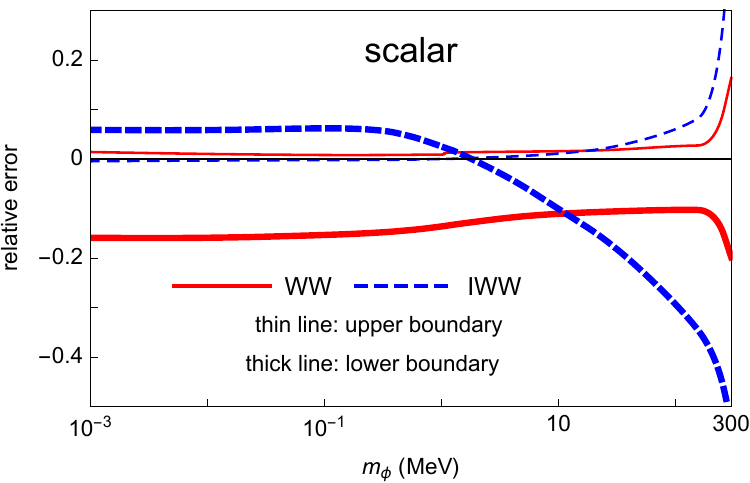}\label{fig:E137Sb}}
\caption{\label{fig:E137S} Exclusion (shaded region) plot for $\epsilon_S$ using the beam dump experiment E137: (a) The solid black, dashed red, and dotted blue lines correspond to using the differential cross section with no, WW, and IWW approximation. (b) The solid red and dashed blue lines correspond to the relative error of the exclusion boundary for a fixed value of $m_\phi$ with WW and IWW approximation. The thin and thick lines correspond to the upper and lower boundaries of the exclusion plot. The relative error is defined in figure \ref{fig:cross_section}.}
\end{figure}

\begin{figure}
\centering
\subfigure[\;exclusion plot (mass in linear scale)]{\includegraphics[scale=1.5]{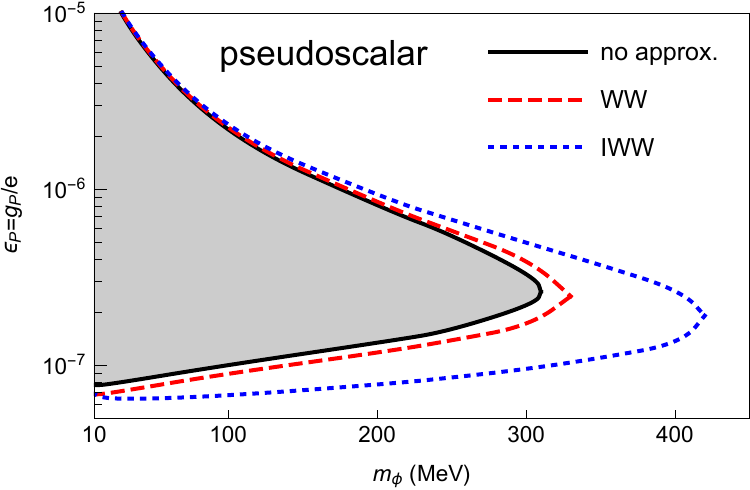}}
\subfigure[\;relative error of exclusion boundary]{\includegraphics[scale=1.5]{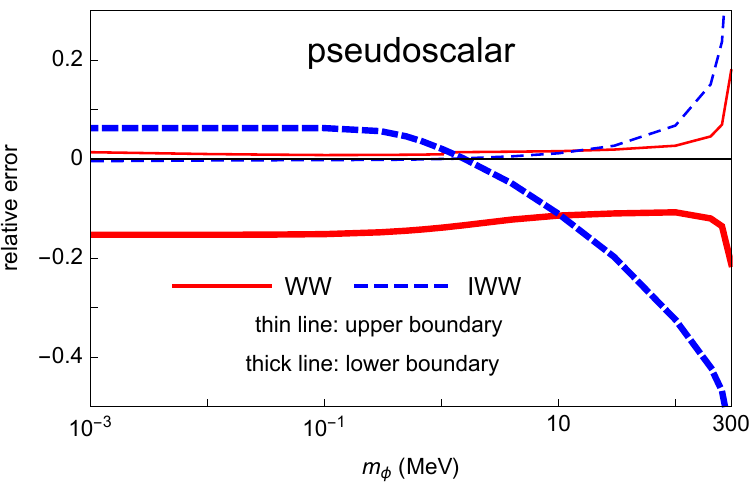}}
\caption{\label{fig:E137P} Exclusion (shaded region) plot for $\epsilon_P$: See caption of figure \ref{fig:E137S} for detail.}
\end{figure}

\begin{figure}
\centering
\subfigure[\;exclusion plot (mass in linear scale)]{\includegraphics[scale=1.5]{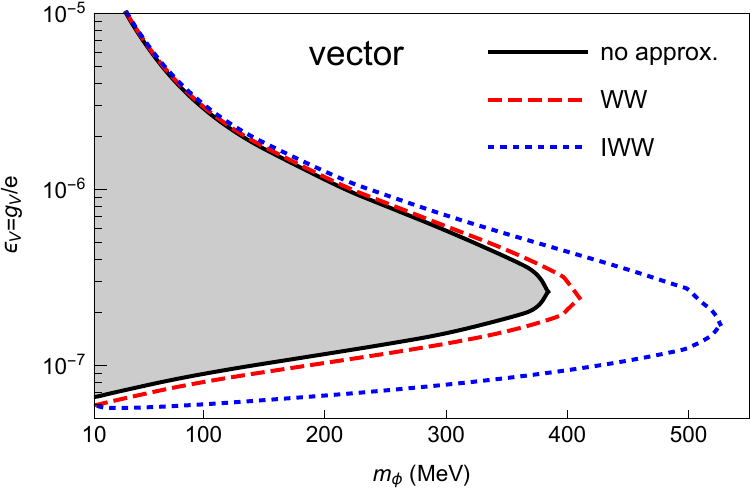}}
\subfigure[\;relative error of exclusion boundary]{\includegraphics[scale=1.5]{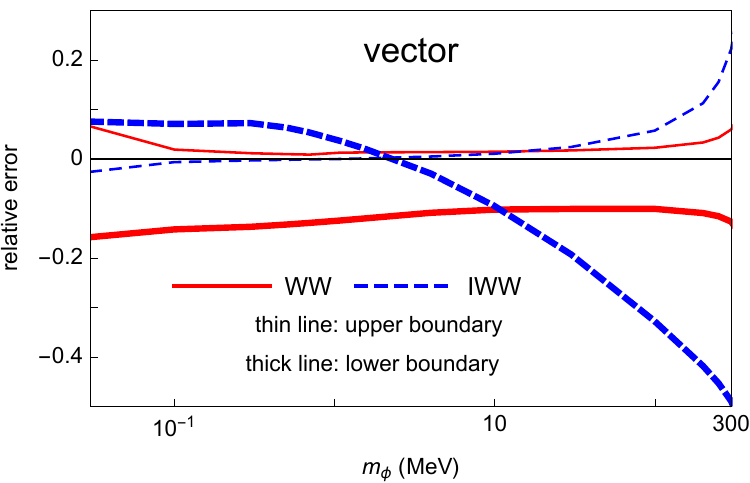}}
\caption{\label{fig:E137V} Exclusion (shaded region) plot for $\epsilon_V$: See caption of figure \ref{fig:E137S} for detail.}
\end{figure}

\begin{figure}
\centering
\subfigure[\;exclusion plot (mass in linear scale)]{\includegraphics[scale=1.5]{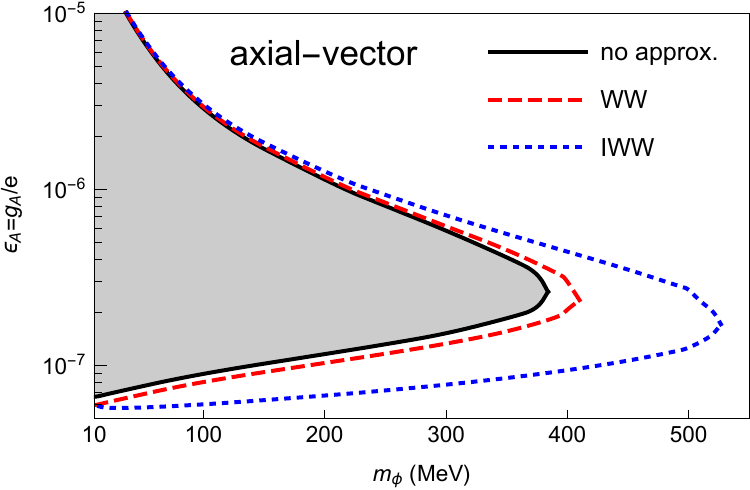}}
\subfigure[\;relative error of exclusion boundary]{\includegraphics[scale=1.5]{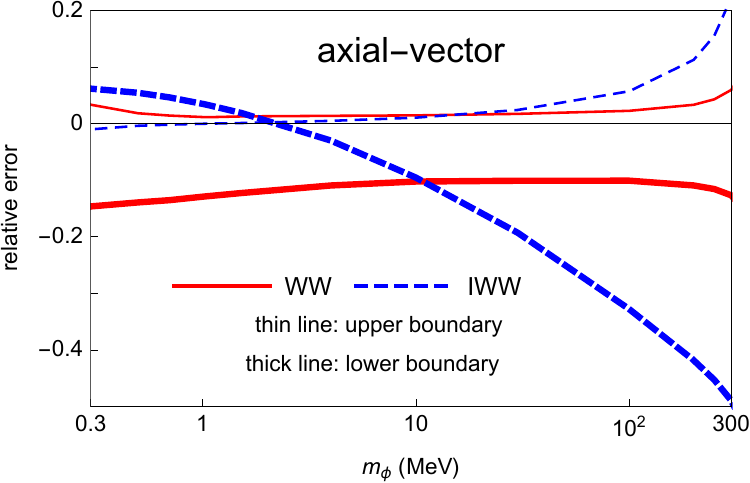}}
\caption{\label{fig:E137A} Exclusion (shaded region) plot for $\epsilon_A$: See caption of figure \ref{fig:E137S} for detail.}
\end{figure}

In figure \ref{fig:E137}, we show regions of coupling and mass excluded by the lack of a signal at E137 for scalar boson, using the three different ways, Eq. (\ref{eq:d sigma dx 1},\ref{eq:d sigma dx 2},\ref{eq:d sigma dx 3-2}), to calculate the differential cross section, $d\sigma/dx$. The error caused by approximation is almost indistinguishable in this log-log plot across several order of magnitude.

In figure \ref{fig:E137all}, using Eq. (\ref{eq:d sigma dx 1}), we show regions of coupling and mass excluded by the lack of a signal at E137 for other bosons. In the region where $m_\phi>2m_e$ the exclusion plots are similar with each other, however, in the region where $m_\phi<2m_e$ the exclusion plots are very different because the the decay widths for different bosons are fundamentally different. 

In figures \ref{fig:E137S}--\ref{fig:E137A}, using Eq. (\ref{eq:d sigma dx 1},\ref{eq:d sigma dx 2},\ref{eq:d sigma dx 3-2}), we show the exclusion regions using the three different ways to calculate the differential cross section. Because of the exponential factor from decay and absorption lengths, the error in the exclusion plot due to making approximations to the cross section is smaller along the upper boundary, which is mainly determined by whether $\phi$ lives long enough to make it to the detector. With the WW approximation, the 100\% error in cross section causes an error of less than 20\% along the lower boundary, and in a log-log plot across several scales, a 20\% error is almost indistinguishable by eyesight. On the other hand, with the IWW approximation, the difference is clearly visible when $m_\phi\gtrsim 100$ MeV. In the region where $m_\phi>2m_e$, the relative errors of the exclusion plots boundary for different bosons are similar based on the same reason which causes the similar relative errors of the cross section: the approximations deal with the phase space integral and the kinematics for different bosons are similar.

In figures \ref{fig:E137} and \ref{fig:E137all}, we see that the absorption process, Eq. (\ref{eq:absorption}), cuts off the exclusion plot around $\epsilon\sim\mathcal{O}(1)$ where the coupling of $\phi$ to electrons is of same order of the electromagnetic coupling. Therefore, in this region, there is another significant process to consider for beam dump experiments. This is the trapping process due to the rescattering
\begin{align}
e(p)+\phi(k)\rightarrow e(p')+\phi(k').
\end{align} 
The trapping process is expected to be as important as the absorption process in this example (new bosons and beam dump E137), and also cuts off the exclusion plot around $\epsilon\sim\mathcal{O}(1)$. However, in figures \ref{fig:E137} and \ref{fig:E137all} the region where $\epsilon\sim\mathcal{O}(1)$ has been excluded by other experiments, such as electron $g-2$ \cite{Pospelov:2008zw,Bouchendira:2010es} and hydrogen Lamb shift \cite{Eides:2000xc}, which are discussed in chapter \ref{ch:boson exclusion}. Therefore we do not include the trapping process, but it might be crucial for other experiments.

\section{A Positive Signal: Production of a New Scalar Boson}\label{sec:data analysis}
To further explore the accuracy of the approximations to the cross section, let us imagine that there is a signal of a new scalar boson being produced at a beam dump experiment. In such a case, the mass and the coupling of this particle can be determined by examining the data, {\it i.e.}, the distribution of events as a function of energy deposited in the detector. We perform 3 sets of pseudoexperiment by using the setup of E137; assume that the scalar boson exists with $(m_\phi,\epsilon)=(110{\rm\;MeV,10^{-7}})$, $(m_\phi,\epsilon)=(200{\rm\;MeV,1.3\times10^{-7}})$, and $(m_\phi,\epsilon)=(0.3{\rm\;MeV,8\times 10^{-6}})$, which are outside of the current exclusion in figure \ref{fig:E137}. We increase the incoming beam luminosities by 36, 36, and 137 times (increasing the total number of electrons dumped into the target), so that the expected total number of events is around 100, 100, and 400. We assume that the resolution of the detector is 1 GeV (which means that there are 18 bins) and generate the ``observed" number of events using a Poisson distribution with the mean value from the complete calculation for each bin. Finally, we can fit the ``observed" data with the calculation with no, WW, and IWW approximation using $\chi^2$ test, and we assume that the variance of the calculated value also satisfies Poisson distribution ({\it i.e.} we ignore systematic errors on the observed numbers of events for simplicity). Therefore, the definition of $\chi^2$ becomes
\begin{align}
\chi^2=\sum_i\frac{(N_{cal,i}-N_{obs,i})^2}{\sigma^2_i}=\frac{(N_{cal,i}-N_{obs,i})^2}{N_{cal,i}+N_{obs,i}}
\end{align}
where $N_{cal}$ and $N_{obs}$ are calculated and ``observed" number of events; the subscript $i$ is for the bins. Since there are two independent parameters (mass and coupling) to fit, the $1\sigma$ and $2\sigma$ range correspond to $\Delta\chi^2=2.30$ and $\Delta\chi^2=6.18$, where $\Delta\chi^2=\chi^2-\chi^2_{min}$.

\begin{figure}
\centering
\subfigure[\;generated data]{\includegraphics[scale=1]{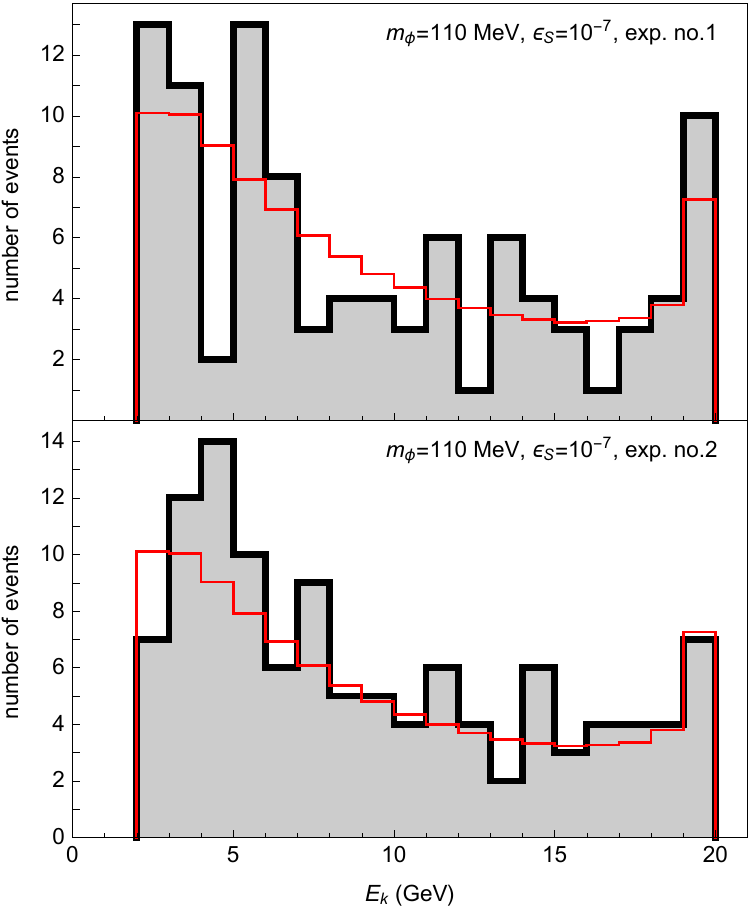}}
\subfigure[\;$\chi^2$ fit]{\includegraphics[scale=1.03]{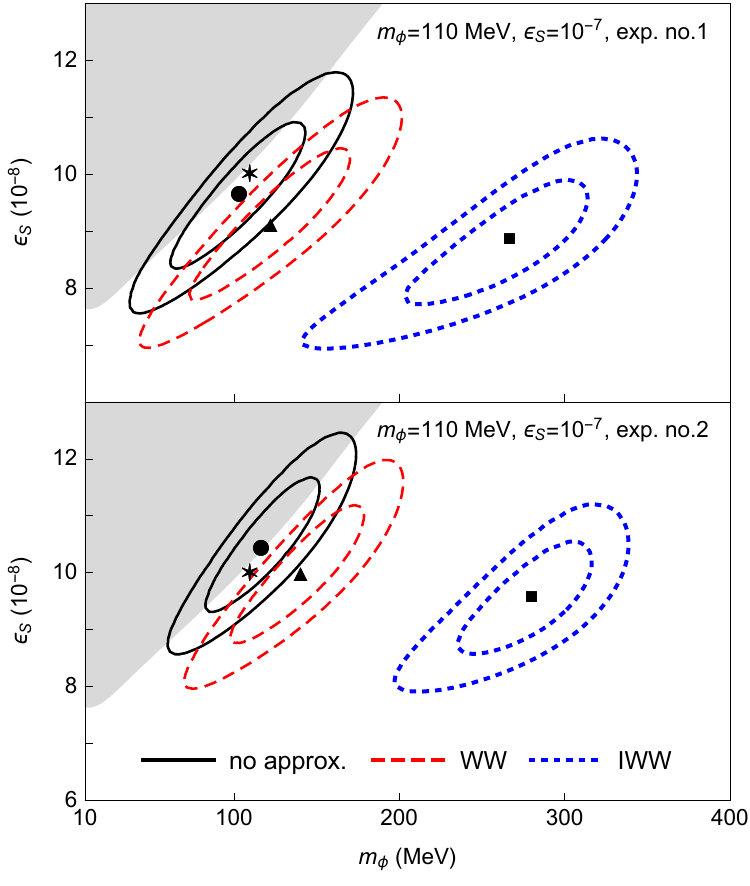}}
\caption{\label{fig:pseudo_exp_1} Assuming the scalar boson exists with $(m_\phi,\epsilon)=(110{\rm\;MeV,10^{-7}})$ and is observed in E137 with 36 times luminosity. (a) The number of events distribution with respect to the energy of the scalar boson: the thin red line is obtained by the complete calculation (no approximation), and the thick black lines is the ``data" generated by Poisson distribution with mean value given by the complete calculation. (b) The best fit point, $1\sigma$ range, and $2\sigma$ range with no, WW, and IWW approximation: the star is the ``true" value; the circle, triangle, and squares are the best fit parameters with no, WW, and IWW approximation, respectively; the black, dashed red, and dotted blue inner (outer) loop correspond to the $1\sigma$ ($2\sigma$) range with no, WW, and IWW approximation, respectively; the shaded area is the excluded region with no approximation from figure \ref{fig:E137}. The top and bottom rows correspond to the results of two separate pseudoexperiments.}
\end{figure}

\begin{figure}
\centering
\subfigure[\;generated data]{\includegraphics[scale=1]{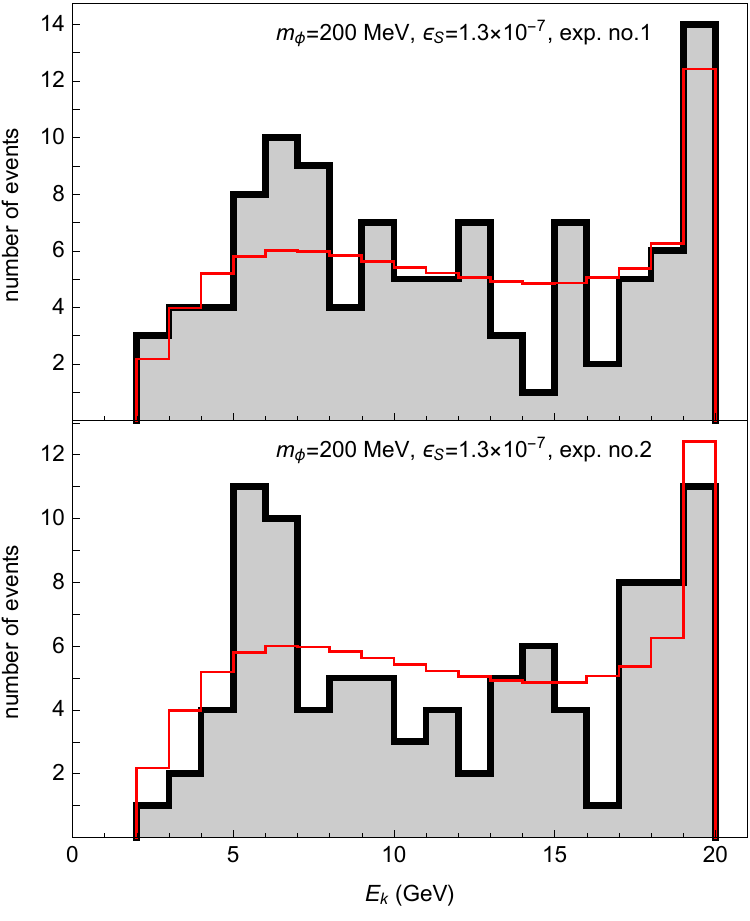}}
\subfigure[\;$\chi^2$ fit]{\includegraphics[scale=1]{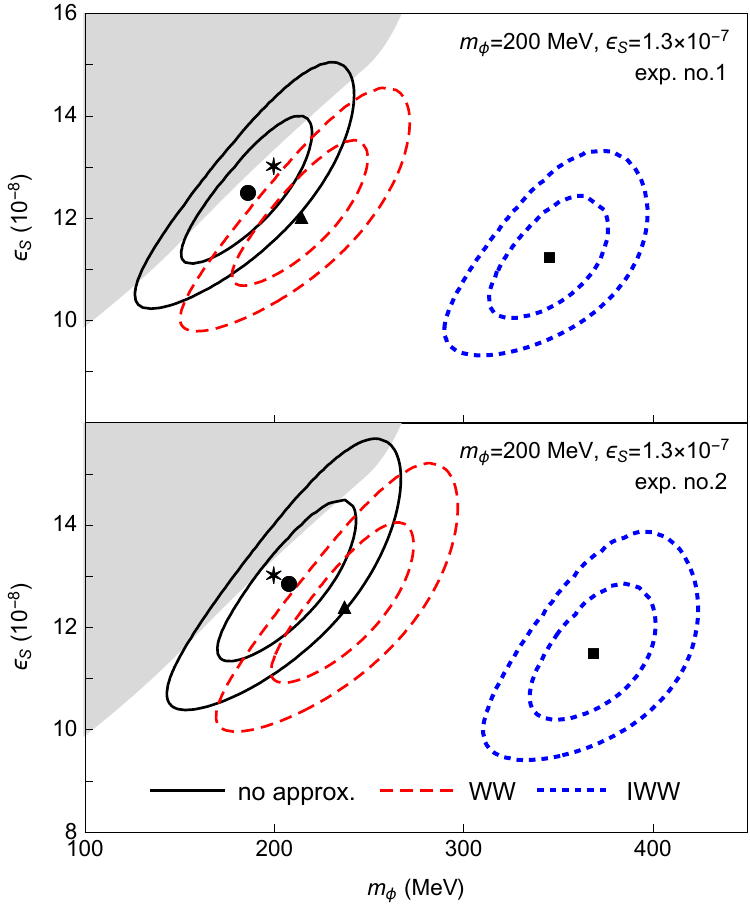}}
\caption{\label{fig:pseudo_exp_2} Assuming the scalar boson exists with $(m_\phi,\epsilon)=(200{\rm\;MeV,1.3\times10^{-7}})$ and is observed in E137 with 36 times luminosity. See the caption in figure \ref{fig:pseudo_exp_1} for details.}
\end{figure}

\begin{figure}
\centering
\subfigure[\;generated data]{\includegraphics[scale=0.75]{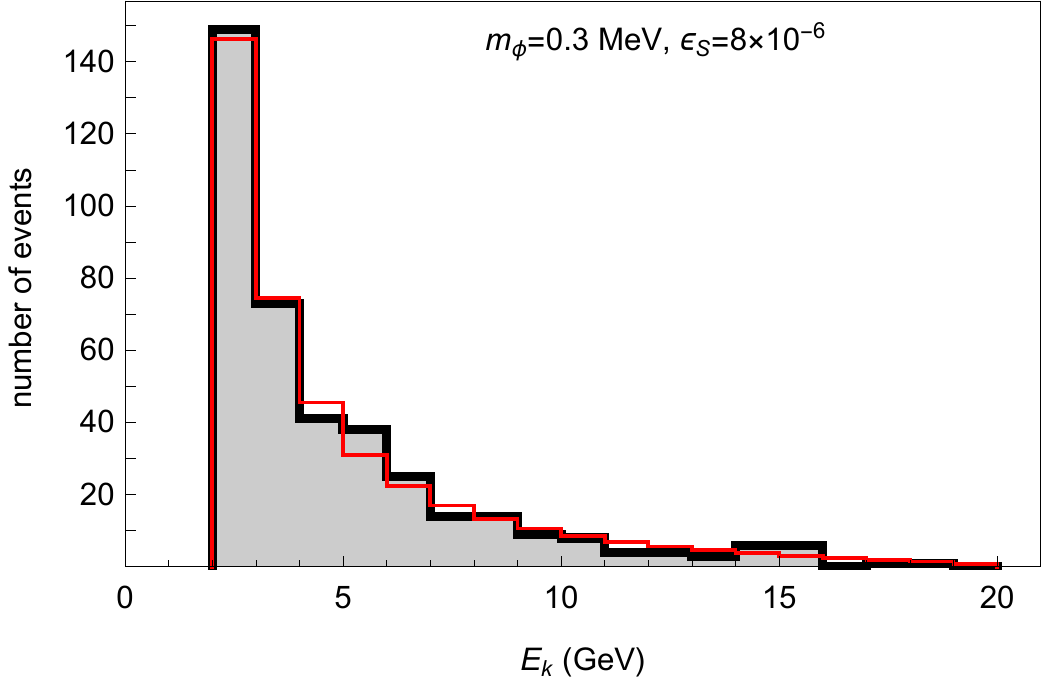}}
\subfigure[\;$\chi^2$ fit]{\includegraphics[scale=0.75]{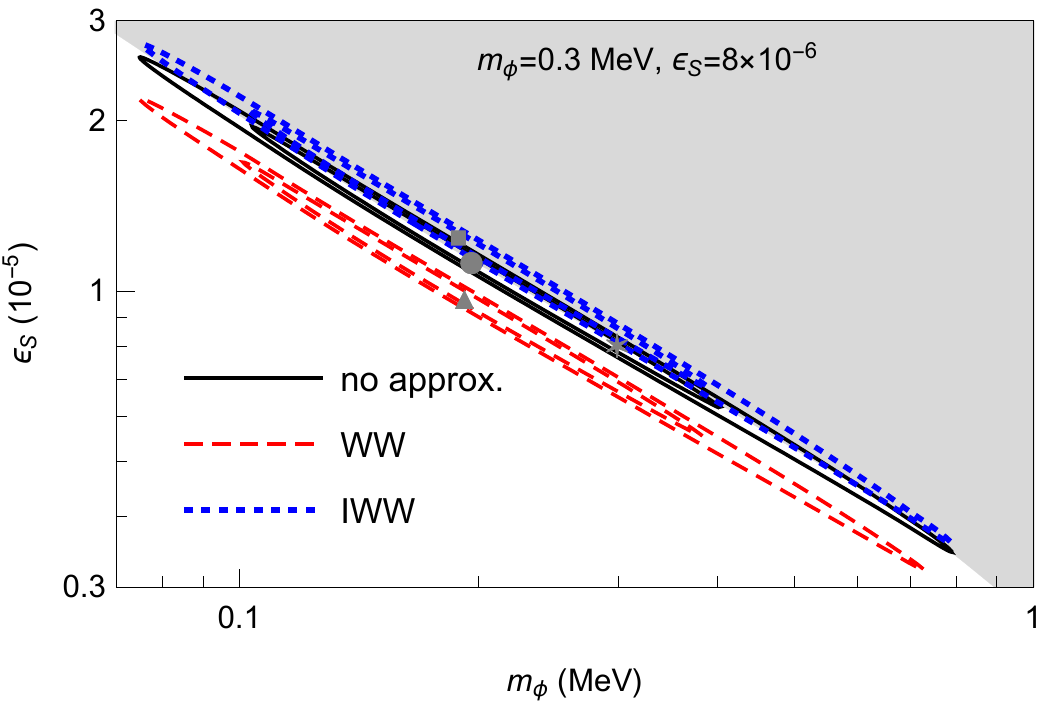}}
\subfigure[\;$\chi^2$ fit (zoomed in)]{\includegraphics[scale=0.75]{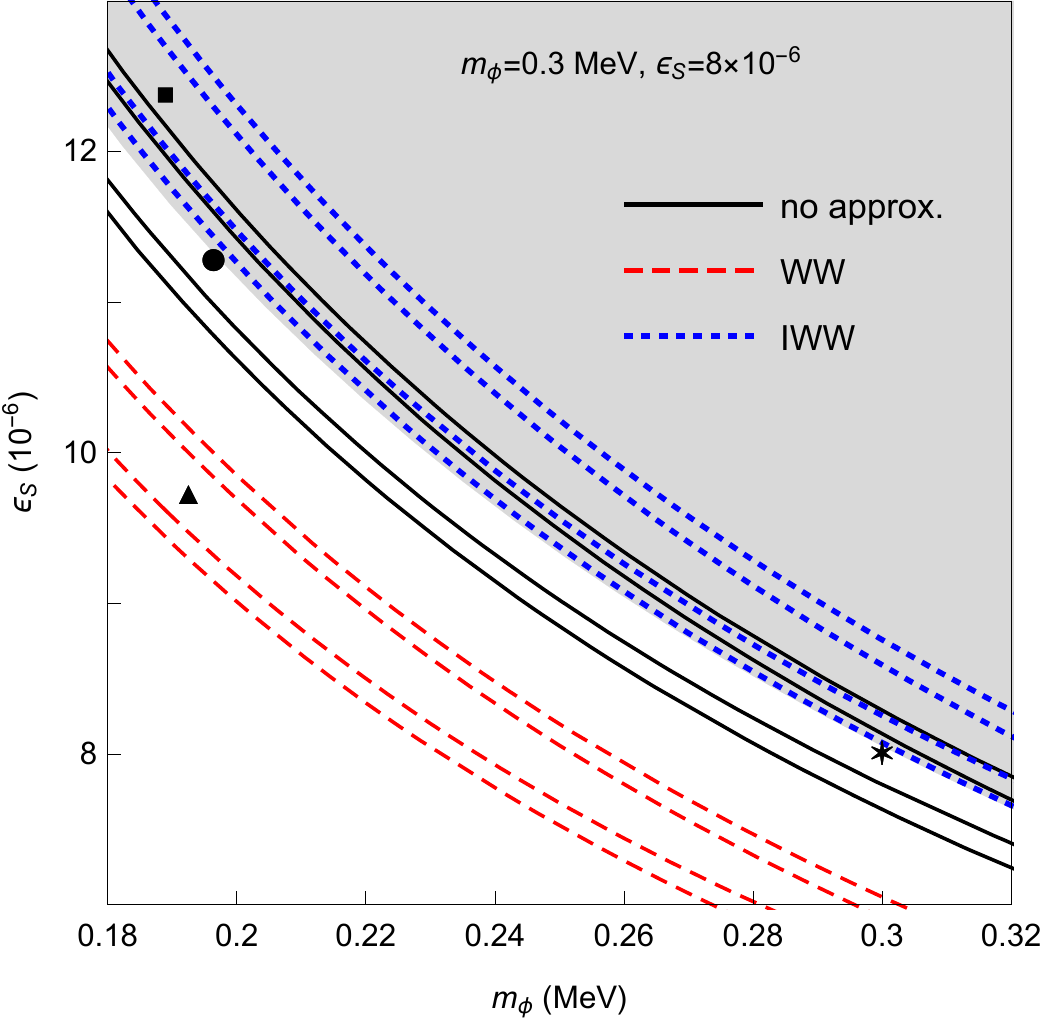}}
\subfigure[\;$\chi^2$ fit (change of coordinate)]{\includegraphics[scale=0.76]{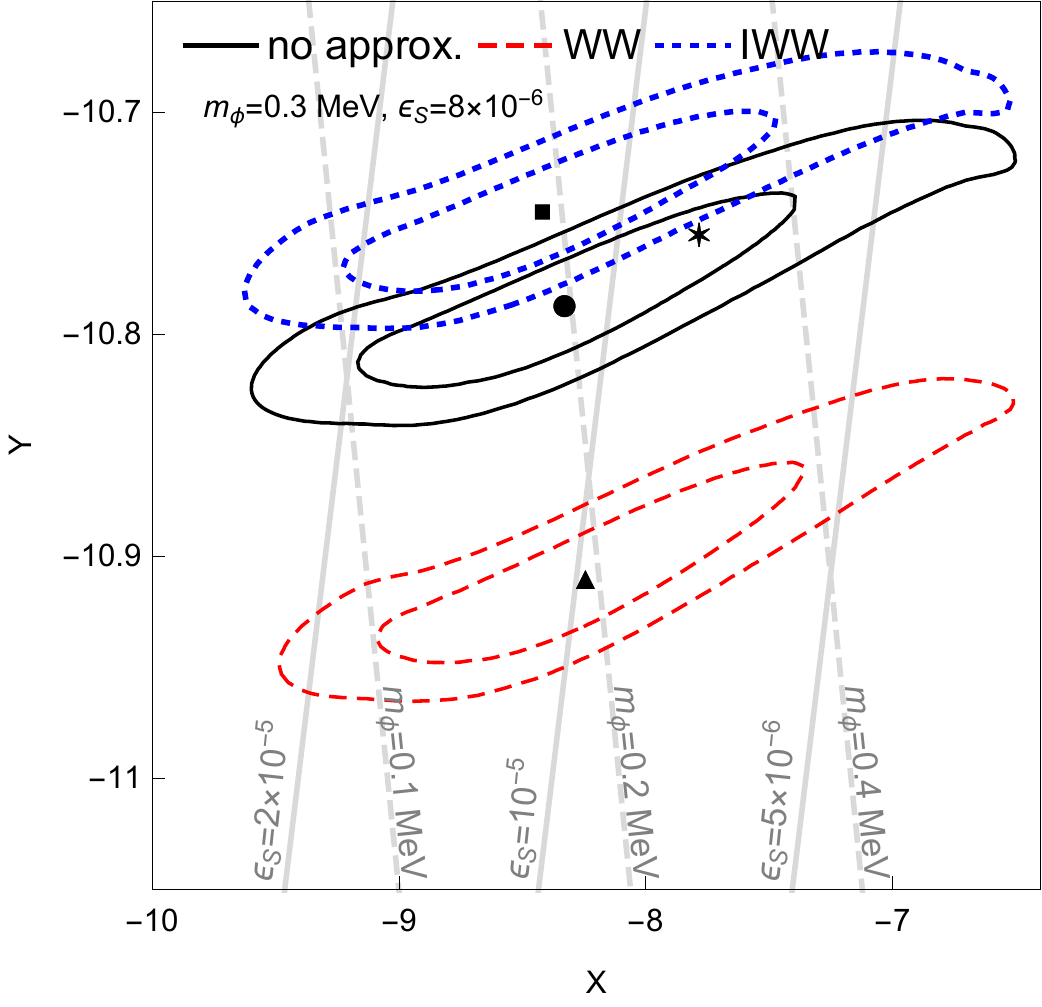}}
\caption{\label{fig:pseudo_exp_3} Assuming the scalar boson exists with $(m_\phi,\epsilon)=(0.3{\rm\;MeV,8\times 10^{-6}})$ and is observed in E137 with 137 times luminosity. (a)--(c) See the caption in figure \ref{fig:pseudo_exp_1} for details. (d) Change of coordinate of $\chi^2$ fit plot: $X=\ln\frac{m_0}{\rm 1\;GeV}+\ln\frac{m_\phi}{m_0}\cos\theta-\ln\frac{\epsilon}{\epsilon_0}\sin\theta$ and $Y=\ln\epsilon_0+\ln\frac{m_\phi}{m_0}\sin\theta+\ln\frac{\epsilon}{\epsilon_0}\cos\theta$, where $\theta=42.4^{\circ}$, $m_0=0.1$ MeV, and $\epsilon_0=2\times10^{-5}$. This means to rotate the coordinate $42.4^{\circ}$ with respect to $(m_\phi,\epsilon)=(0.1{\rm\;MeV,2\times 10^{-5}})$.}
\end{figure}

\begin{figure}
\centering
\includegraphics[scale=1.5]{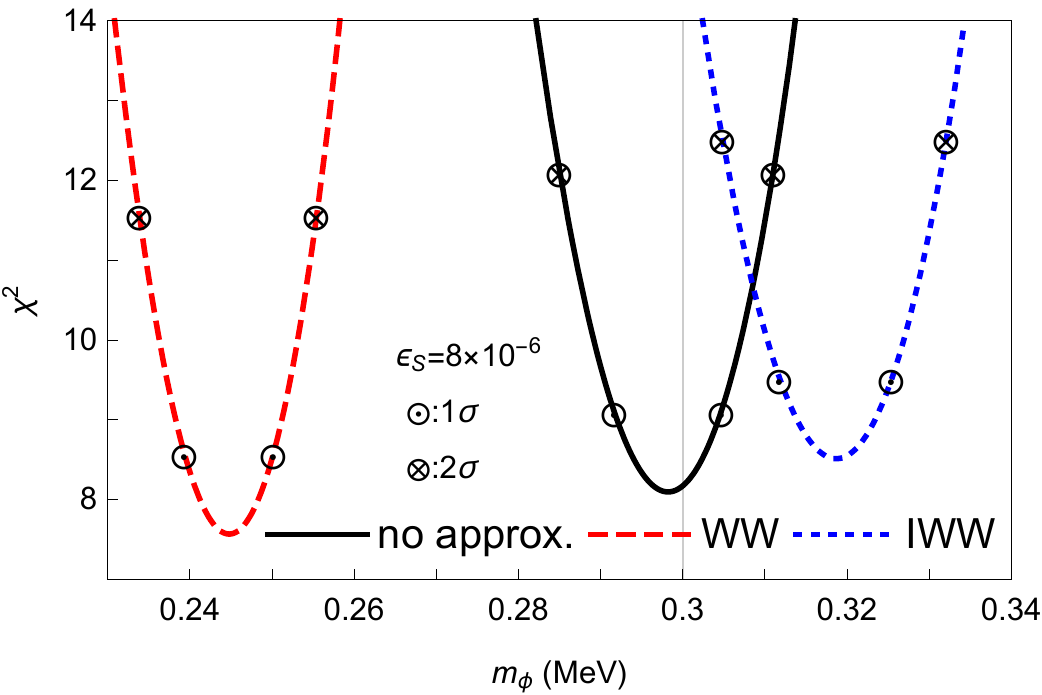}
\caption{\label{fig:pseudo_exp_3_2} Assuming the scalar boson exists with $(m_\phi,\epsilon)=(0.3{\rm\;MeV,8\times 10^{-6}})$ and is observed in E137 with 137 times luminosity. The number of events distribution is the same in figure \ref{fig:pseudo_exp_3}. The value of $\chi^2$ with respect to $m_\phi$ (assuming $\epsilon$ is precisely measured): the black, dashed red, and dotted blue lines correspond to the $\chi^2$ values calculated with no, WW, and IWW approximation. The minimum of $\chi^2$ corresponds the best fit $m_\phi$; the circle dots $\odot$ correspond to 1$\sigma$ range ($\Delta\chi^2=1$); the circle crosses $\otimes$ correspond to 2$\sigma$ range ($\Delta\chi^2=4$). The gray vertical line indicates the true value of $m_\phi$.}
\end{figure}

We show the results of these pseudoexperiments with $(m_\phi,\epsilon)=(110{\rm\;MeV,10^{-7}})$ in figure \ref{fig:pseudo_exp_1}, $(m_\phi,\epsilon)=(200{\rm\;MeV,1.3\times10^{-7}})$ in figure \ref{fig:pseudo_exp_2}, and $(m_\phi,\epsilon)=(0.3{\rm\;MeV,8\times 10^{-6}})$ in figure \ref{fig:pseudo_exp_3}. We see that the ``true'' parameter values lie within the $1\sigma$ allowed regions when fitting with the complete calculation. On the other hand, although using approximation sometimes gives a fairly good estimate of cross section, the result of data fitting lies outside the $2\sigma$ range. It is worth noting that the shape of the $1\sigma$ or $2\sigma$ range is roughly along the exclusion boundary in figure \ref{fig:E137}, because the exclusion boundary is the isocontour of the number of events.

Next, we consider another scenario of the third pseudoexperiment with $(m_\phi,\epsilon)=(0.3{\rm\;MeV,8\times 10^{-6}})$. In this part of parameter space, the allowed coupling and mass can extend over roughly an order of magnitude. To illustrate the usefulness of the complete calculation, we perform fits to this data assuming that there is another experimental result that can sensitively measure the coupling. This would be the case if recently proposed experiments involving decays of radioactive nuclei underground see a nonzero signal~\cite{Izaguirre:2014cza,Liu:2016qwd} and we can use the beam dump experiment to determine the mass precisely. For simplicity, we assume that the other experiment measures the coupling with negligible error. Since there is one parameter to fit, the $1\sigma$ and $2\sigma$ range correspond to $\Delta\chi^2=1$ and $\Delta\chi^2=4$. We show the results in figure \ref{fig:pseudo_exp_3_2}. Again as expected, we see that the ``true'' parameter values lie within the $1\sigma$ allowed region when fitting with the complete calculation. Using the approximations, the ``true'' mass lies outside the $2\sigma$ ranges. We observe that using the complete calculation could be crucial in measuring the mass of a new particle in this region of parameter space.

It is worth to mention that according to \ref{fig:E137Sb}, as expected, the WW approximation works better for the first two pseudoexperiments (higher mass region), and the IWW is better for the third pseudoexperiment (lower mass region).

\section{Discussion}\label{sec:discussion}
In the region where $m_\phi>2m_e$, while the production amplitude, decay length, and the absorption length can differ in detail for particles with different quantum numbers, they are qualitatively similar. The approximations that we have examined deal with the phase space integral and coupling to electromagnetism of the target nucleus. Therefore, as we expected, the exclusion plots for different bosons are similar. On the other hand, where $m_\phi<2m_e$, the decay channels, which are very different for different bosons, result in very different exclusion regions. New results for vector decaying to 3 photons, Eq. (\ref{eq:V decay to photons}), and axial-vector decaying to 4 photons, Eq. (\ref{eq:A decay to photons}), are presented.

Including a coupling to the muon may change the situation for $m_\phi>2m_\mu$~\cite{Liu:2016qwd} due to the opening of a new channel with typically a substantial partial width. A study of the production of vector particles in electron beam dumps that deals with some of the issues we have addressed can be found in Ref.~\cite{Beranek:2013yqa}.

There are some other beam dump experiments using a Cherenkov detector, such as E141 \cite{Riordan:1987aw} and Orsay \cite{Davier:1989wz}. Therefore, their exclusion plots do not extend to the region where $m_\phi<2m_e$. We show the results of the beam dump experiments E141 and Orsay for the scalar boson in Ref.~\cite{Liu:2016qwd}.

We need to consider the LPM effect \cite{Landau:1953um,Landau:1953gr,Migdal:1956tc,Anthony:1995fs} which suppresses particle production cross section below a certain (produced particle) energy. For E137, this energy is about 12 MeV which is much smaller than the energy cutoff of the detector. Therefore we do not consider the LPM effect in our discussion. However, for other experiments (depending on the apparatus), the LPM effect may need to be taken into account.

In this work, we present a complete analysis of beam dump experiments. We show that a brute-force analytical calculation is possible. Software exists using Monte-Carlo simulations, such as \textsc{MadGraph/MadEvent}~\cite{Alwall:2007st} as used in, e.g.,~\cite{Essig:2010xa}, that can calculate the cross section without using approximations. Our work can be used as a consistency check for Monte-Carlo simulations. We show that using the WW approximation can be trusted to an order of magnitude in cross sections and exclusion plots. Additionally our work  allows us to understand the errors introduced by the various common approximations. In certain regions of parameter space different errors partially cancel against each other, leading to results that are accidentally sometimes more accurate than might be expected. However, as we illustrated with several pseudoexperiments in a range of masses, in the event of a nonzero signal, a complete calculation can give very different results from the approximations. This could be useful given the possibility of future electron beam dump experiments~\cite{Wojtsekhowski:2009vz,Abrahamyan:2011gv,Izaguirre:2013uxa,Raggi:2014zpa,Izaguirre:2014bca}.

\chapter{Electrophobic Scalar Bosons and Muonic Puzzles}\label{ch:boson exclusion}

\section{Introduction}

It is known that new physics, which violates lepton universality, is a candidate to explain both proton radius puzzle and muon anomalous magnetic moment problem simultaneously \cite{Pohl:2013yb,TuckerSmith:2010ra,Izaguirre:2014cza}. The pseudoscalar and axial-vector bosons can not be the candidate because they have contributions to $(g-2)_\mu$ with the wrong sign \cite{Carlson:2012pc}. Furthermore we will show that the vector boson can not be the candidate either because the vector boson contribution to the proton radius puzzle is ruled out by the hyperfine splitting of the muonic hydrogen. Therefore, the only candidate is the scalar boson. We proceed assuming that the existence of a scalar boson, $\phi$, resolves the source of proton radius puzzle and $(g-2)_\mu$ problem.

To investigate this hypothetical $\phi$, we make a general assumption that $\phi$ couples to standard model particles through Yukawa interactions
\begin{align}
\mathcal{L}&\supset\sum_f e \epsilon_f\phi\overline{\Psi}_f\Psi_f{\rm\;\,(for\,\,scalar),}\nonumber\\
\mathcal{L}&\supset\sum_f e \epsilon_f\phi_\mu\overline{\Psi}_f\gamma^\mu\Psi_f{\rm\;\,(for\,\,vector)}
\end{align}
where $\epsilon_f=g_f/e$; $e$ is the electric charge of the proton; the subscript $f$ is particle label, which in our case can be taken as electron $e$, muon $\mu$, proton $p$, and neutron $n$. The one boson exchange potential between flavor 1 and 2 is
\begin{align}
V(r)=(-1)^{s+1}\epsilon_{f_1}\epsilon_{f_2}\alpha\frac{e^{-m_\phi r}}{r}
\end{align}
where $s$ is the spin of the boson. Other articles have pursued the same idea, however, using further assumptions, e.g. in \cite{TuckerSmith:2010ra}, $\epsilon_p=\epsilon_\mu$ is assumed; in \cite{Izaguirre:2014cza}, mass-weighted coupling is assumed; $\epsilon_n$ is neglected in both \cite{TuckerSmith:2010ra,Izaguirre:2014cza}. In this chapter, we make no a priori assumptions regarding signs or magnitudes of the coupling constants. If the new physics is the solution to the proton radius puzzle, $\epsilon_\mu$ and $\epsilon_p$ have the same (opposite) sign for scalar (vector) boson. Without loss of generality, we set both $\epsilon_p$ and $\epsilon_\mu$ to be positive for scalar ($\epsilon_p>0$ and $\epsilon_\mu<0$ for vector), and $\epsilon_e$ and $\epsilon_n$ can have either sign.

The outline of this chapter is as follows. In section \ref{sec:experiments}, we discuss the experiments and observables which are used to constrain the parameters of the new boson. In section \ref{sec:exclusions}, we show the exclusion plots for $\epsilon_e$, $\epsilon_\mu$, $\epsilon_p$ and $\epsilon_n$ with respect to $m_\phi$. A discussion is presented in section \ref{sec:electrophobic bosons discussion}

\section{Experiments ans Observables}\label{sec:experiments}

\subsection{Hyperfine Splitting}
The vector boson has contribution to the hyperfine splitting which could be used as a constraint (the scalar boson does not contribute to the hyperfine structure). The additional vector boson contribution to the ground state is given by \cite{Karshenboim:2014tka,Karshenboim:2001yy}
\begin{align}
\frac{\Delta E_{hfs}}{E_{hfs}}=\epsilon_p\epsilon_\mu\frac{2\alpha m_{p\mu}}{\pi m_p}K\left(\frac{m_\phi^2}{m_p^2}\right)
\end{align}
where $m_{p\mu}$ is the reduced mass of the proton and muon, and
\begin{align}
K(s)=2\left(\frac{s}{4}+2\right)\sqrt{\frac{4}{s}-1}\tan^{-1}\sqrt{\frac{4}{s}-1}+\left(\frac{s}{4}+\frac{3}{2}\right)\ln s-\frac{1}{2}.
\end{align}
On the other hand, the current measurement on the muonic hydrogen ground state hyperfine splitting is agreed with the theoretical prediction \cite{Liu:1999iz}
\begin{align}
\left|\frac{\Delta E_{hfs}}{E_{hfs}}\right|<2.88\times 10^{-7}
\end{align}
where we take 2 standard deviation (S.D.).

\subsection{Anomalous Magnetic Moment}
\begin{figure}
\centering
\includegraphics[scale=0.5]{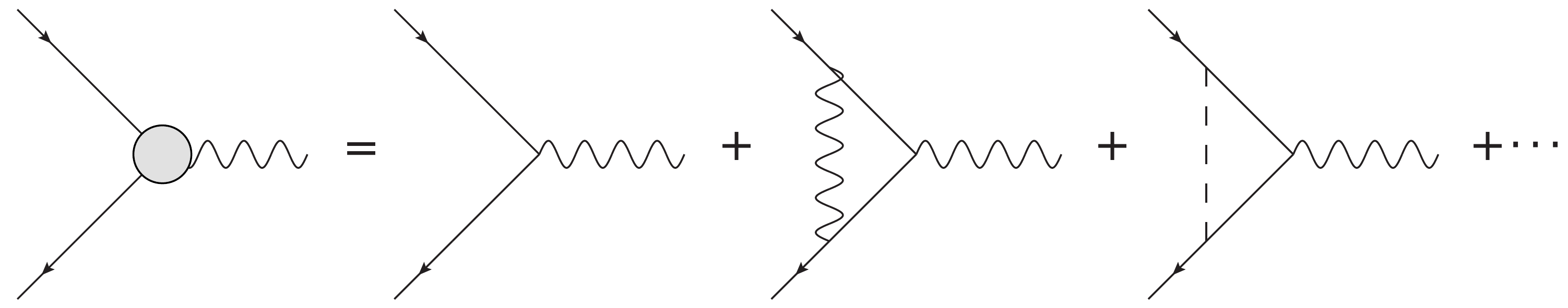}
\caption{\label{fig:g-2} The diagrams to calculate anomalous magnetic moment: The leading contribution of the new boson is the third diagram on the right side.}
\end{figure}

Anomalous magnetic moment of a lepton is shifted by the new boson contribution, see figure \ref{fig:g-2}, by \cite{Jackiw:1972jz,TuckerSmith:2010ra}
\begin{align}\label{eq:g-2}
\Delta a_l=\frac{\alpha}{2\pi}\epsilon_l^2\xi\left(\frac{m_\phi}{m_l}\right)
\end{align}
where $\epsilon_l=g_l/e$ and 
\begin{align}
\xi(x)&=\int_0^1\frac{(1-z)^{2}(1+z)}{(1-z)^{2}+x^{2}z}dz{\rm\;\,(for\,\,scalar),}\nonumber\\
\xi(x)&=\int_0^1\frac{2z(1-z)^{2}}{(1-z)^{2}+x^{2}z}dz{\rm\;\,(for\,\,vector)}.
\end{align}
Ref. \cite{Pospelov:2008zw} emphasizes that the measurement of $(g-2)_e$ is used to extract the fine structure constant. The shift of $(g-2)_e$ caused by $\phi$ results in the shift of measured $\alpha$ by $\Delta\alpha=2\pi\Delta a_e$. Therefore, a measurement of $\alpha$ which is not directly from $(g-2)_e$ and not sensitive to the new boson is needed. 

Currently, one of such experiments is extracting the fine structure constant through the measurement involving $^{87}$Rb atom,
\begin{align}
\alpha^2=\frac{2R_\infty}{c}\frac{m_{\rm Rb}}{m_e}\frac{h}{m_{\rm Rb}}
\end{align}
where the measurements of the Rydberg constant $R_\infty$ \cite{Mohr:2015ccw}, the mass ratio $m_{\rm Rb}/m_e$ \cite{PhysRevA.82.042513}, and $h/m_{\rm Rb}$ \cite{Bouchendira:2010es} are not sensitive to the existence of the new boson. The fine structure constant $\alpha$ deduced from the relation gives 0.66 ppb uncertainty and $\Delta\alpha$ must not exceed it, $\Delta\alpha/\alpha\leq 1.3\times10^{-9}$ (taking 2 S.D.). For $(g-2)_\mu$, the new boson can directly explain the discrepancy, $\Delta a_\mu$.

\subsection{Bhabha Scattering}
Bhabha scattering ($e^+e^-\rightarrow e^+e^-$) is used to search for the new boson by looking for a resonance. Motivated by earlier results from heavy-ion collisions near the Coulomb barrier, a group~\cite{Tsertos:1989gv} used a clean time-stable monoenergetic positron beam incident on a metallic Be foil. No resonances were observed at the 97\% C.L. within the experimental sensitivity of 0.5 b eV/sr (c.m.) for the energy-integrated differential cross section. Given the small value of $\epsilon_e$ the only relevant process is the $s$-channel exchange of a $\phi$ boson (the interference effect gives zero contribution at the resonance energy). Using a narrow width approximation, the energy-integrated differential cross section in c.m. frame is given by
\begin{align}\label{eq:Bhabha}
\int d\sqrt{s}\frac{d\sigma}{d\Omega}&=\epsilon_e^2\frac{\alpha\pi}{4m_\phi}\sqrt{1-\frac{4m_e^2}{m_\phi^2}}{\rm\;\,(for\,\,scalar),}\nonumber\\
\int d\sqrt{s}\frac{d\sigma}{d\Omega}&=\epsilon_e^2\frac{3\alpha\pi}{8m_\phi}\frac{1+\frac{8m_e^2}{m_\phi^2}+\cos^2\theta\left(1-\frac{4m_e^2}{m_\phi^2}\right)^2}{\left(1+\frac{2m_e^2}{m_\phi^2}\right)\left(1-\frac{4m_e^2}{m_\phi^2}\right)^{1/2}} {\rm\;\,(for\,\,vector)}
\end{align}
where $\theta$ is the scattering angle between $80^\circ$ and $100^\circ$ in the CM frame.

\subsection{Beam Dump Experiments}
Beam dump experiments have long been used to search for light, weakly coupled particles that decay to leptons or photons~\cite{Essig:2013lka,Bjorken:1988as,Bjorken:2009mm}. If coupled to electrons, $\phi$ bosons could be produced in such experiments and decay to $e^+e^-$ or $\gamma\gamma$ pairs depending on its mass. Previous work~\cite{Bjorken:2009mm} simplified the evaluation of this cross section by using the Weizs\"{a}cker-Williams (WW) approximation, by making further approximations to the phase space integral, assuming  that the mass of the new particle is much greater than electron mass,  and can't be used if $m_\phi<2m_e$. Our numerical evaluations (discussed in chapter \ref{ch:beam dump}) do not use these assumptions and thereby allow us to cover the entire mass range. Our analysis uses data from the electron beam dump experiments E137~\cite{Bjorken:1988as}, E141~\cite{Riordan:1987aw}, and Orsay~\cite{Davier:1989wz}.

\subsection{Neutron-Nucleus Scattering}
The neutron-nucleus scattering experiment, which uses $n-^{208}$Pb scattering below 10 keV and assume the new boson couples equally to nucleons, gives a 2 S.D. exclusion constraint on $g_N$ in \cite{Leeb:1992qf}. However, the previous constraint assumed the new particle couples equally to proton and neutron, i.e. $g_N=g_p=g_n$. In our case, we have no such assumption and the new boson can couple to proton and neutron arbitrarily. Therefore we make the following replacement
\begin{align}\label{eq:g_neucleon replacement}
\frac{g_N^2}{4\pi}\rightarrow\alpha\frac{A-Z}{A}\epsilon_n^2+\alpha\frac{Z}{A}\epsilon_p\epsilon_n
\end{align}
in the constraint obtained in \cite{Leeb:1992qf}.

\subsection{NN Scattering Length Difference}
The nucleon-nucleon scattering length difference is defined as $\Delta a=\bar{a}-a_{np}$ where $\bar{a}=(a_{pp}+a_{nn})/2$. $\Delta a^{\rm exp}=5.64(60)$ fm \cite{Machleidt:2001rw} is explained very well by the theory calculation $\Delta a^{\rm th}=5.6(5)$ fm \cite{Ericson:1983vw}. The existence of the new boson gives an extra contribution 
\begin{align}
\Delta a_{\phi}=\bar{a}a_{np}M\int_0^\infty\Delta V\bar{u}u_{np}dr
\end{align}
where M is the average of the nucleon mass;
\begin{align}
\Delta V=(-1)^{s+1}\frac{1}{2}\alpha(\epsilon_p-\epsilon_n)^2\frac{e^{-m_\phi r}}{r};
\end{align}
$u(r)$ is zero energy $^1S_0$ wave function normalized so that $u(r)\rightarrow(1-r/a)$ as $r\rightarrow\infty$. We calculate $u(r)$ using the gaussian model \cite{Brown:1975us}. $\Delta a_{\phi}$ can not be greater than 1.6 fm (taking 2 S.D.) to spoil the existing agreement.

\subsection{Nucleon Binding Energy in (N=Z) Nuclear Matter}
\begin{figure}
\centering
\includegraphics[scale=1]{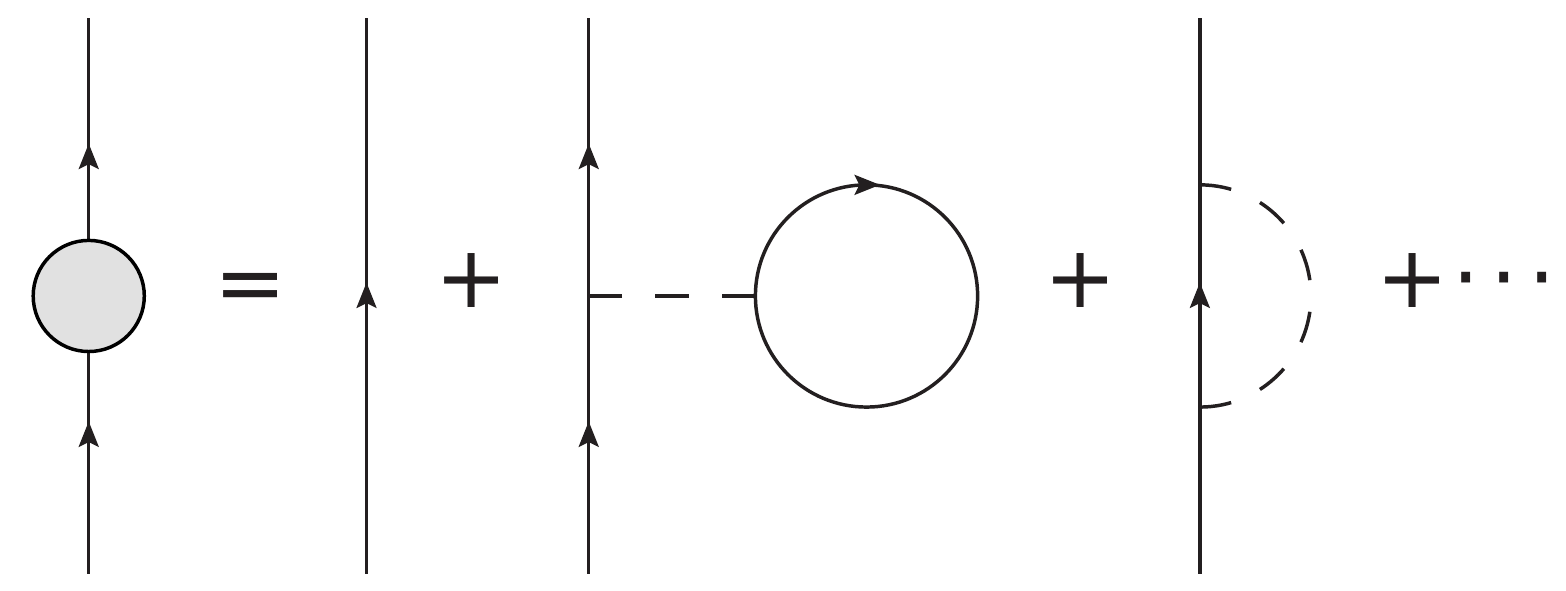}
\caption{\label{fig:self-energy} The diagrams to calculate the nucleon self-energy in matter: On the right side, the second and third diagrams are the Hartree and Fock diagrams, respectively.}
\end{figure}

Nucleon binding energy in (N=Z) nuclear matter is the volume term per nucleon in the semi-empirical mass formula. It gains an additional contribution due to the new boson. Using the Hartree approximation (the contribution from the Fock approximation is negligible if $m_\phi<100$ MeV), see figure \ref{fig:self-energy} and Refs. \cite{Mattuck:1976xt,SanjayReddy}, we have
\begin{align}
\delta B_N=\frac{g_N(g_p+g_n)\rho}{2m_\phi^2}
\end{align}
where $g_N$ is $g_p$ or $g_n$, and $\rho\approx0.08{\rm\;fm^{-3}}$ is the proton (neutron) density. The averaged nucleon binding energy in nuclear matter is given by
\begin{align}
\delta B=\frac{\delta B_p+\delta B_n}{2}=\frac{(g_p+g_n)^2\rho}{4m_\phi^2}
\end{align}
which must not exceed 1 MeV to avoid problems with existing understandings of nuclear physics.

\subsection{$^3$He$-^3$H Binding Energy Difference}
The $^3$He$-^3$H binding energy difference is 763.76 keV, which is nicely explained by the effect of Coulomb interaction (693 keV) and charge asymmetry (about 68 keV) \cite{Friar:1969zz,Friar:1978mr,Coon:1987kt,Miller:1990iz}. Other contributions must not exceed about 30 keV to spoil the agreement. The contribution to the binding energies difference from the new boson can be estimated by using Yukawa potential and the wave function extracted from elastic electron-nuclei scattering
\begin{align}
\frac{2\alpha}{\sqrt{3}\pi}\int_0^\infty q^2dq\frac{\epsilon_p^2-\epsilon_n^2}{q^2+m_\phi^2}F(q^2)<30{\rm\;keV}
\end{align}
where $F=F_S+F_V$; $F_S$ and $F_V$ are determined through $2F^{^3{\rm He}}=(2G_E^p+G_E^n)F_S+(G_E^p-G_E^n)F_V$ and $F^{^3{\rm H}}=(G_E^p+2G_E^n)F_S-(G_E^p-G_E^n)F_V$; $G_E^p$ ($G_E^n$) is the proton (neutron) electric form factor; $F^{^3{\rm He}}$ ($F^{^3{\rm H}}$) is $^3$He ($^3$H) charge form factor.

\subsection{Lamb Shift}
Lamb shift in the 2S-2P transition of lepton-nucleus bound state due to the new boson using first order perturbation theory is given by \cite{TuckerSmith:2010ra}
\begin{align}\label{eq:Lamb shift}
\delta E_L^{l\rm N}=(-1)^{s+1}\frac{\alpha}{2a_{lN}}\epsilon_l[Z\epsilon_p+(A-Z)\epsilon_n] f(a_{l\rm N}m_\phi)
\end{align}
where $f(x)=x^2/(1+x)^4$; the Bohr radius $a_{l\rm N}=(\alpha\,m_{l\rm N})^{-1}$; $m_{l\rm N}$ is the reduced mass of lepton-nucleus system. The current extremely high level of agreement between calculated and experimentally measured energy levels of hydrogen \cite{Eides:2000xc}. An empirical rule was developed in~\cite{Miller:2011yw,Miller:2012ht,Miller:2012ne}. To avoid spoiling this agreement, the new boson contribution must be no more than the experimental uncertainty, 7 kHz. Therefore $\delta E_L^{\rm H}<14$ kHz (taking 2 S.D.). For muonic hydrogen, if the new boson is the missing piece of the proton radius puzzle, $\delta E_L^{\mu\rm H}=-0.307(56)$ meV \cite{Pohl:2010zza,Antognini:1900ns,Mohr:2015ccw,Antognini:2015moa}. 

For muonic deuterium ($\mu$D), the contribution to the muonic Lamb shift caused by the new boson, $\delta E_L^{\mu\rm D}=-0.409(69)$ meV \cite{Antognini:2015moa,Krauth:2015nja}, can be extracted from the difference of the charge radius of deuteron from the muonic Lamb shift measurement $r^\mu_{\rm D}=2.12562(78)$ fm \cite{Pohl1:2016xoo} and CODATA (2014) $r_{\rm D}=2.1213(25)$ \cite{Mohr:2015ccw} from electronic experiments. 

The preliminary unpublished results on the Lamb shift of muonic helium-4 ion ($\mu^4$He$^+$) have important implications for the new boson. For $\mu^4$He${^+}$, $\delta E_L^{\mu^4{\rm He}^+}=-1.4(1.5)$ meV \cite{Antognini:2015moa}, which is extracted from the muonic Lamb shift measurement $r^\mu_{^4\rm He}=1.677(1)$ fm \cite{AldoAntognini} and elastic electron scattering $r_{^4\rm He}=1.681(4)$ fm \cite{Sick:2014yha}.

\subsection{A1 at MAMI and BABAR 2014}
The A1 Collaboration at Mainz Microtron \cite{Merkel:2014avp} used the electron beam hitting $^{181}$Ta foils to look for the dark photon production with the similar process at figure \ref{fig:2 to 3}. In 2014, BABAR \cite{Lees:2014xha} searched for dark photon in the reaction $e^+e^-\to \gamma A'$, $A'\to l^+l^-$ ($l=e,\;\mu$). The data was collected at $\Upsilon(4S)$, $\Upsilon(3S)$, and $\Upsilon(2S)$ resonance peaks.

Both experiments reported null results, and give constraints above tens of MeV.

\section{Exclusions}\label{sec:exclusions}

\subsection{ruling out the vector boson}
\begin{figure}
\centering
\includegraphics[scale=1.4]{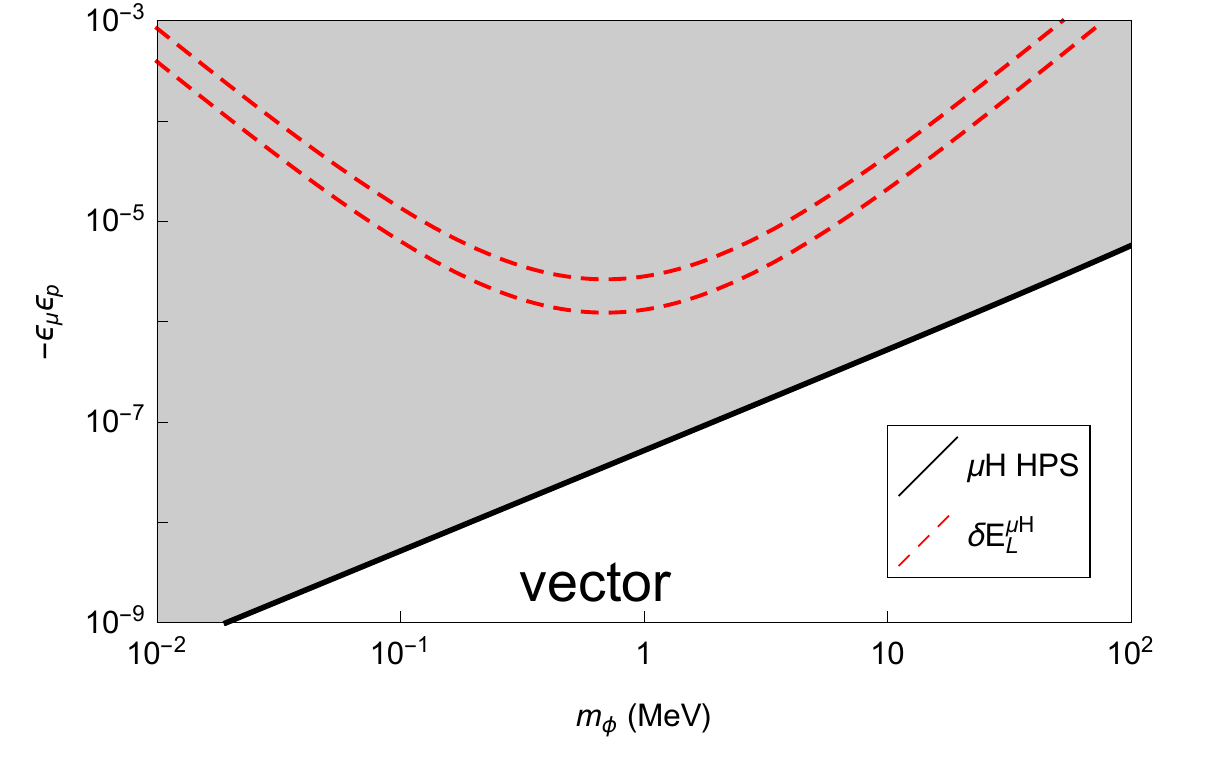}
\caption{\label{fig:vector_exclusion} Exclusion of the vector boson (shaded regions are excluded): the region above the black line is excluded by the muonic hydrogen ground state hyperfine splitting, and between the two dashed red lines is the allowed region for vector boson to solve the proton radius puzzles. The vector boson is ruled out.}
\end{figure}

In figure \ref{fig:vector_exclusion}, using the constraint of the muonic hydrogen ground state hyperfine splitting, the allowed parameter space of the vector boson to solve the proton radius puzzle is completely inside the exclusion region. Therefore the vector boson is ruled out.

\subsection{constraints on $\epsilon_p$ and $\epsilon_n$}
\begin{figure}
\centering
\includegraphics[scale=1.4]{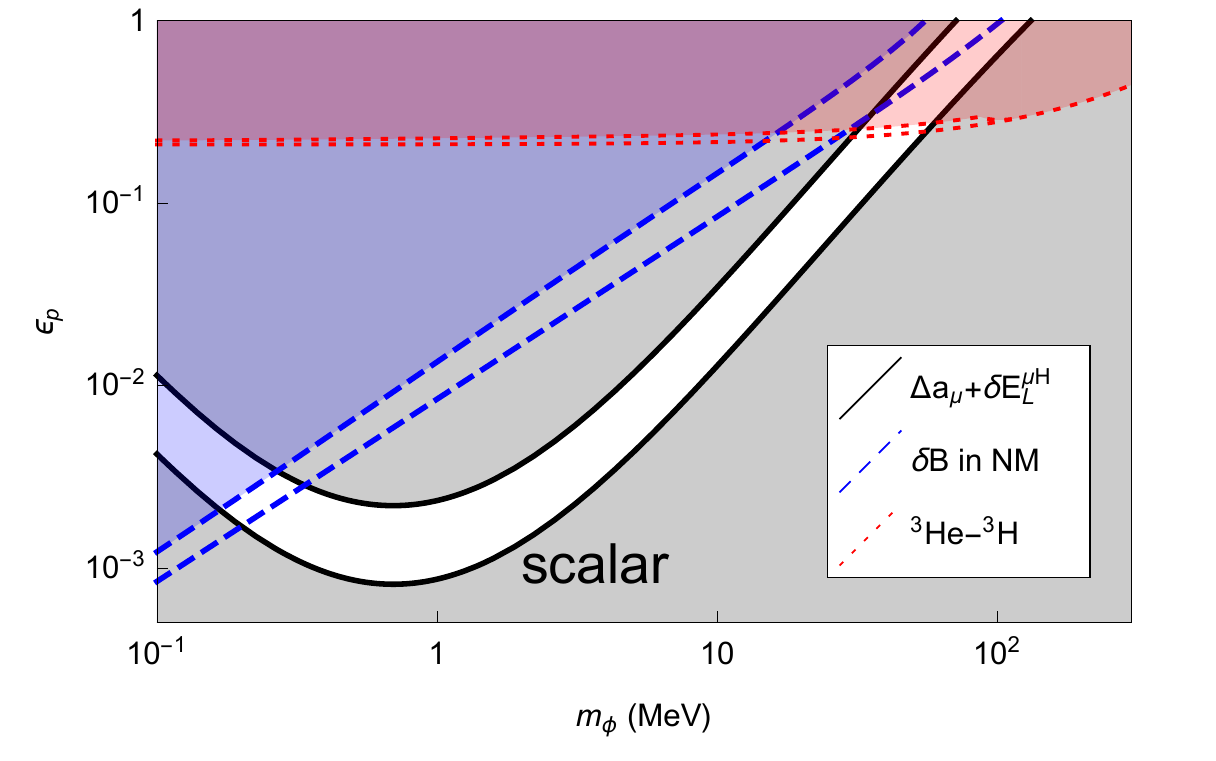}
\caption{\label{fig:gp_exclusion} Exclusion plot for $\epsilon_p$ (shaded regions are excluded): between the black lines are allowed $\epsilon_p$ as a function of $m_\phi$ (\ref{eq:epsilon p}) by solving $(g-2)_\mu$ and proton radius puzzles with scalar boson as the explanation; the dashed blue and dotted red lines correspond to the constraints from the nucleon binding energy in nuclear matter and the $^3$He$-^3$H binding energy difference; the isolated lines are taking $\epsilon_n=0$ and the shaded regions are excluded with the constraint on $\epsilon_n/\epsilon_p$ in figure \ref{fig:Rnp_exclusion}.}
\end{figure}

\begin{figure}
\centering
\includegraphics[scale=1.4]{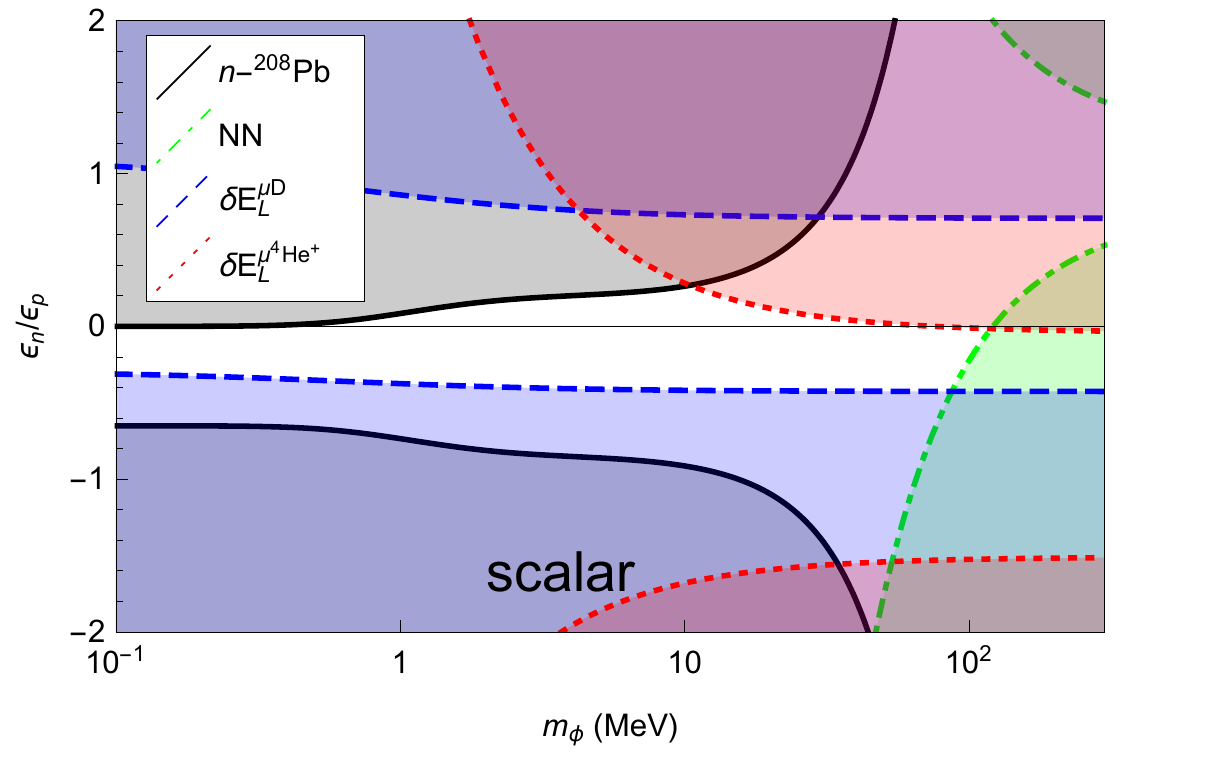}
\caption{\label{fig:Rnp_exclusion} Exclusion plot for $\epsilon_n/\epsilon_p$ (shaded regions are excluded): the black, dashed blue, dotted red, and dotted dashed green lines correspond to the constraints from $n-^{208}$Pb scattering, $\mu$D Lamb shift, $\mu^4$He${^+}$ Lamb shift, and NN scattering length difference. Since the measurement of $\mu^4$He Lamb shift is preliminary, we use 3 S.D..}
\end{figure}

Using the above experiments, we begin to constrain the parameter space of the scalar boson. The $(g-2)_\mu$ experiment shows $\epsilon_\mu$ as a function of $m_\phi$ (\ref{eq:g-2}); the Lamb shift of muonic hydrogen shows $\epsilon_p$ as a function of $\epsilon_\mu$ and $m_\phi$ (\ref{eq:Lamb shift}). Solving these two equations, we obtain allowed $\epsilon_p$ value as a function of $m_\phi$ (taking 2 S.D.)
\begin{align}\label{eq:epsilon p}
\epsilon_p=\frac{\delta E^{\mu \rm H}_L a_{\mu\rm H}}{f(a_{\mu\rm H}m_\phi)}\sqrt{\frac{2\xi(m_\phi/m_\mu)}{\alpha\pi\Delta a_\mu}}.
\end{align}
which is shown as solid black lines in figure \ref{fig:gp_exclusion}.

Bofore we further constrain $\epsilon_p$, we need to constrain the relation between $\epsilon_n$ and $\epsilon_p$ to separate their contributions in various of experiments involving nucleons. Using the neutron experiments, NN scattering length difference, and (\ref{eq:epsilon p}); using the $\mu$D and $\mu^4$He Lamb shift, and comparing them to $\mu$H to obtain the ratio $\epsilon_n/\epsilon_p$ derived from (\ref{eq:Lamb shift})
\begin{align}
(A-Z)\frac{\epsilon_n}{\epsilon_p}=\frac{\delta E_L^{\mu\rm N}}{\delta E_L^{\mu\rm H}}\frac{a_{\mu{\rm N}}}{a_{\mu\rm H}}\frac{f(a_{\mu\rm H} m_\phi)}{f(a_{\mu\rm N} m_\phi)}-Z,
\end{align}
we constrain the ratio $\epsilon_n/\epsilon_p$ in figure \ref{fig:Rnp_exclusion}. Since the measurement of $\mu^4$He Lamb shift is preliminary, we use 3 S.D. as a constraint.

Next, using the nucleon binding energy in nuclear matter, the $^3$He$-^3$H binding energy difference, and the constraints on $\epsilon_n/\epsilon_p$, we constrain the parameter space in the $\epsilon_p-m_\phi$ plane in a finite range. The allowed mass range is $0.167<m_\phi\;(\rm MeV)<59.8$ for the scalar boson, see figure \ref{fig:gp_exclusion}.

\subsection{constraints on $\epsilon_\mu$}
\begin{figure}
\centering
\includegraphics[scale=1.4]{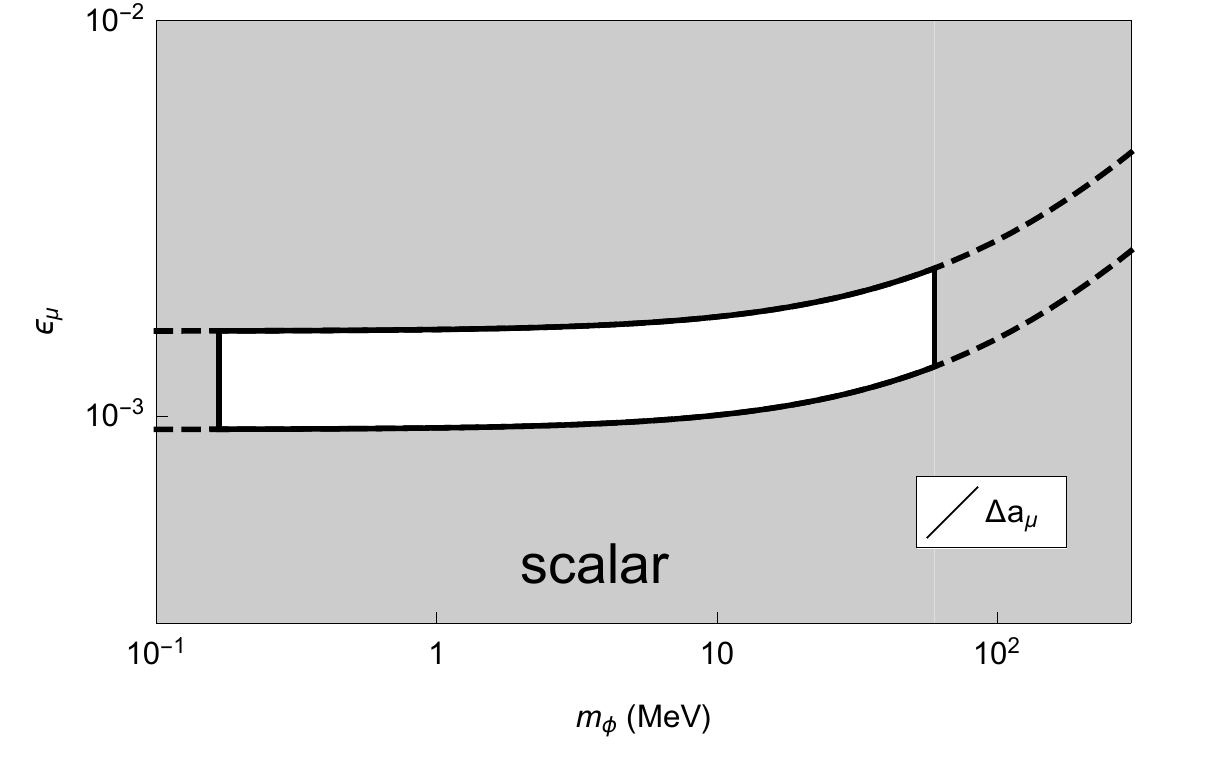}
\caption{\label{fig:gmu_exclusion} Exclusion plot for $\epsilon_\mu$ (shaded region is excluded): between the solid and dashed lines are from $(g-2)_\mu$, Eq. (\ref{eq:epsilon mu}) taking 2 S.D.. The region between the dashed lines is excluded by allowed scalar boson mass in figure \ref{fig:gp_exclusion}.}
\end{figure}

The allowed $\epsilon_\mu$ value as a function of $m_\phi$ from $(g-2)_\mu$ experiment (\ref{eq:g-2}) is
\begin{align}\label{eq:epsilon mu}
\epsilon_\mu=(-1)^{s}\sqrt{\frac{2\pi\Delta a_\mu}{\alpha\xi(m_\phi/m_\mu)}}.
\end{align}
Combining with the allowed the new boson masses range from figure \ref{fig:gp_exclusion}, we obtain the allowed parameter space in the $\epsilon_\mu-m_\phi$ plane in figure \ref{fig:gmu_exclusion}.

\subsection{constraints on $\epsilon_e$}
\begin{figure}
\centering
\includegraphics[scale=1.4]{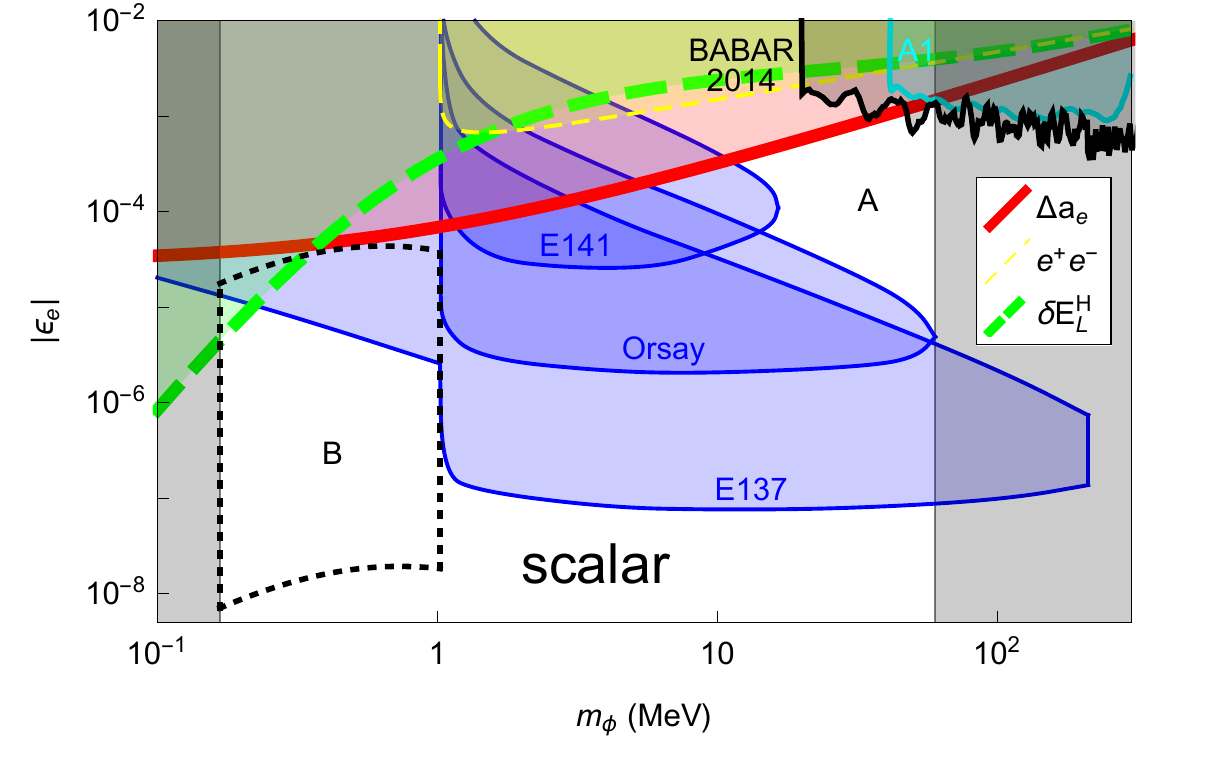}
\caption{\label{fig:ge_exclusion} Exclusion plot for $\epsilon_e$ (shaded regions are excluded): the thick red, thin blue, and dashed yellow, and thick dashed green lines correspond to the constraints from electron anomalous magnetic moment $(g-2)_e$, beam dump experiments, resonance of Bhabha scattering, and Lamb shift of hydrogen with Eq. (\ref{eq:epsilon p}). A1 at MAMI and BABAR 2014 constraints are at the upper right corner. Between the two vertical gray lines are allowed scalar boson mass range from figure \ref{fig:gp_exclusion}. The regions A and dotted region B are covered by proposed experiments (see section \ref{sec:electrophobic bosons discussion} for more detail).}
\end{figure}

\begin{figure}
\centering
\includegraphics[scale=1.4]{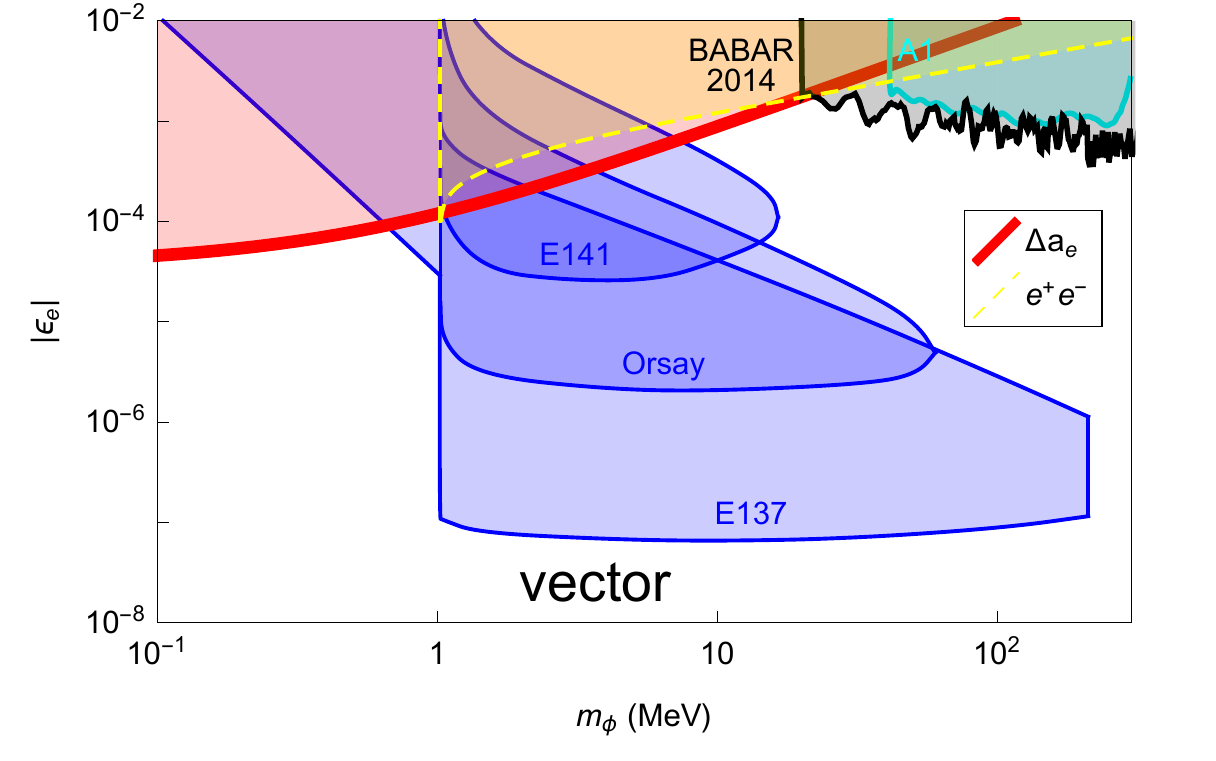}
\caption{\label{fig:ge_v_exclusion} Exclusion plot for $\epsilon_e$ (shaded regions are excluded): the thick red, thin blue, thin dashed yellow lines correspond to the constraints from electron anomalous magnetic moment $(g-2)_e$, beam dump experiments, and resonance of Bhabha scattering. A1 at MAMI and BABAR 2014 constraints are at the upper right corner.}
\end{figure}

In figure \ref{fig:ge_exclusion}, we present the constraints on the scalar boson coupling to electrons, $\epsilon_e$, as a function of $m_\phi$ from electron anomalous magnetic moment, resonance of Bhabha scattering, beam dump experiments (E137, E141, and Orsay), A1 at MAMI, BABAR 2014, and hydrogen Lamb shift with (\ref{eq:epsilon p}). In addition we indicate (via two vertical lines) the allowed mass ranges for $\phi$, taken from figure \ref{fig:gp_exclusion}.

In figure \ref{fig:ge_v_exclusion}, we show the constraints on the vector boson couplings to electron for general purpose.

\section{Discussion}\label{sec:electrophobic bosons discussion}

In figure \ref{fig:ge_exclusion}, we label two allowed regions in the $(m_\phi,\epsilon_e)$ plane in figure \ref{fig:ge_exclusion}: A, where $10~{\rm MeV}\lesssim m_\phi\lesssim 70~{\rm MeV}$, $10^{-6}\lesssim\epsilon_e\lesssim10^{-3}$, and B, where $100~{\rm keV}\lesssim m_\phi\lesssim 1~{\rm MeV}$, $10^{-8}\lesssim\epsilon_e\lesssim10^{-5}$. There are a number of planned electron scattering experiments that will be sensitive to the scalar bosons with parameters in Region A, such as, e.g., APEX~\cite{Essig:2010xa,Abrahamyan:2011gv}, HPS~\cite{Battaglieri:2014hga}, DarkLight~\cite{Freytsis:2009bh,Cowan:2013mma}, VEPP-3~\cite{Wojtsekhowski:2012zq}, and MAMI or MESA~\cite{Beranek:2013yqa}. As studied in Ref.~\cite{Izaguirre:2014cza}, region B can be probed by looking for scalars produced in the nuclear de-excitation of an excited state of $^{16}$O. We have translated this region of couplings $10^{-11}\leq\epsilon_p\epsilon_e\leq10^{-7}$ from Ref.~\cite{Izaguirre:2014cza} to show on our plot by taking $\epsilon_p\rightarrow\epsilon_p+\epsilon_n$ (since $^{16}$O has an equal number of neutrons and protons), using $\epsilon_n/\epsilon_p$ from figure \ref{fig:Rnp_exclusion} and fixing $\epsilon_p$ according to Eq. (\ref{eq:epsilon p}).

We do not show limits derived from stellar cooling that are sensitive to $m_\phi\lesssim 200~\rm keV$~\cite{An:2013yfc} since the lower bound on the mass is similar to the one we have derived. Additionally, we note that constraints from cooling of supernovae do not appear in figure \ref{fig:ge_exclusion}. This is because the required value of $g_p$ is always large enough to keep any new bosons produced trapped in supernova, rendering cooling considerations moot \cite{Rrapaj:2015wgs}. The effects of this scalar boson may have cosmological consequences, beyond the scope of this discussion.

We summarize some conclusions of the parameter space discussed above:
\begin{enumerate}
\item The vector boson is ruled out by including the constraint of the muonic hydrogen ground state hyperfine splitting.

\item The range of allowed $m_\phi$ is widened from a narrow region around 1~MeV in~\cite{TuckerSmith:2010ra} to the region from 167 keV to 59.8 MeV for scalar boson allowing $\epsilon_p\ne\epsilon_\mu$ .

\item We carefully deal with $\epsilon_n$ instead of neglecting it. In particular, as seen in figure \ref{fig:gp_exclusion}, allowing $\epsilon_n$ to be of the opposite sign of $\epsilon_p$ opens up the parameter space.

\item The constraint on $\epsilon_e$ at $m_\phi=1$ MeV is improved by two order of magnitude for scalar and one order of magnitude for vector compared with \cite{TuckerSmith:2010ra} by using the electron beam dump experiments.

\item Near the maximum allowed $m_\phi$, the allowed couplings are relatively large providing ample opportunity to test this solution.

\item All of the allowed values of $\epsilon_e$ are smaller than the required value of $\epsilon_\mu$, thus the name electrophobic boson is applicable.

\end{enumerate}


Our discussion thus far has been purely phenomenological, with no particular UV completion in mind to relate the couplings of fermions with the same quantum numbers (here the electron and muon). From the model-building point of view, there are motivations that the couplings of scalars to fermions in the same family are mass-weighted--in particular, for the leptons, $|\epsilon_\mu/\epsilon_e|=(m_\mu/m_e)^n$ with $n\ge 1$. This is because, generally, coupling fermions to new scalars below the electroweak scale leads to large flavor-changing neutral currents (FCNCs) that are very strongly constrained, e.g. in the lepton sector by null searches for $\mu\to e$ conversion, $\mu\to 3e$, or $\mu\to e\gamma$. A phenomenological ansatz for the structure of the scalars' couplings to fermions that avoids this problem is that its Yukawa matrix is proportional to that of the Higgs. This scenario has been termed minimal flavor violation (MFV), see e.g.~\cite{Cirigliano:2005ck}. In that case, both the Higgs and scalar couplings are simultaneously diagonalized and new FCNCs are absent. The main phenomenological consequence of this is that scalars' coupling to a lepton is proportional to a power of that lepton's mass, $\epsilon_\ell\propto m_\ell^n$ with $n\ge 1$. In the context of a given model, i.e. for fixed $n$, we can relate figures \ref{fig:gmu_exclusion} and \ref{fig:ge_exclusion}. Region A largely corresponds to $0<n\lesssim 1$, which, is less well-motivated from a model building perspective. $1\lesssim n\lesssim 2$ is well-motivated and fits into Region B. To obtain $\epsilon_e\lesssim10^{-7}$, $n\gtrsim 2$ is required.

Building  a complete model, valid at high energy scales, leading  to interactions at low energies is not our purpose. However, we outline one simple possibility. In the lepton sector, couplings to scalar could arise through mixing obtained via a lepton-specific two Higgs doublet model, which would automatically yield MFV~\cite{Batell:2016ove}. In the quark sector, coupling to a light boson via mixing with a Higgs is very tightly constrained by null results in $K\to\pi$ and $B$ meson decays (see, e.g.,~Ref.~\cite{Batell:2009jf}) decays. However, as in Ref.~\cite{Fox:2011qd}, heavy vector-like quarks that couple to scalar and mix primarily with right-handed quarks of the first generation due to a family symmetry are a possibility. The coupling strength of $\phi$ to $u$ and $d$ quarks could differ leading to different couplings to neutrons and protons. If, e.g., $g_d/g_u\sim-0.7$ then $g_n/g_p\sim-0.3$, which, as we see in figure \ref{fig:Rnp_exclusion}, is comparatively less constrained.

The existence of a scalar boson that couples to muons and protons does account for the proton radius puzzle. The coupling to electrons is found here to be so small that for most of the parameter space it is hard to detect by a variety of experiments involving electrons, except in region A and B in figures \ref{fig:ge_exclusion} and \ref{fig:ge_v_exclusion}. Although the coupling to muons and protons is sufficiently large to resolve the proton radius puzzle, these couplings are also sufficiently small to prevent detection with currently feasible experiments. Using nucleon experiments are difficult, because the new physics signal is usually buried under strong interaction and it also needs to separate the contributions of proton and neutron carefully. However, near the maximum allowed $m_\phi$, the allowed $\epsilon_p$ is near 0.3 which may be probable by using low energy proton experiments, such as threshold $\phi$ production in $pp$ interaction. Proton or muon beam dump experiments could also be used~\cite{Gardner:2015wea}. For experiments involving muons, one might think of using muon beam dump experiments, such as the COMPASS experiment as proposed in~\cite{Essig:2010gu}. The MUSE experiment~\cite{Gilman:2013eiv} plans to measure  $\mu^\pm$ and $e^\pm$-p elastic scattering at low energies. Our hypothesis regarding the $\phi$ leads to a prediction for the MUSE experiment even though its direct effect on the scattering will be very small~\cite{Liu:2015sba}: the MUSE experiment will observe  the same `large' value of the proton radius for all of the probes.   Another possibility is to study the spectroscopy of muonium (the bound state of $e^-$ and $\mu^+$) or true muonium (the bound state of $\mu^-$ and $\mu^+$). Perhaps the best way to test the existence of this particle would be an improved measurement of the muon anomalous magnetic moment \cite{Hertzog:2015jru}. The existence of a particle with such a limited role may seem improbable, considering the present state of knowledge. However, such an existence is not ruled out.

\chapter{Conclusion}\label{ch:conclusion}
Motivated by proton radius puzzle and muon anomalous magnetic moment discrepancy, we propose new physics to solve these two puzzles. In chapter \ref{ch:LN scattering}, we begin to search for a new scalar boson by using polarized lepton-nucleon elastic scattering. However, we showed that it is very difficult and probably not achievable experimentally in the next few decades. We expect that it is similar with the vector boson. Therefore, in chapter \ref{ch:beam dump}, we move on to the beam dump experiments. We calculated the cross section without approximations in the phase space integral and we include the production of new scalar, pseudoscalar, vector, and axial-vector bosons for general purpose. 
In chapter \ref{ch:boson exclusion}, we pursue the idea that the new physics can solve two muonic problems simultaneously, and we showed that the scalar boson is the only candidate. We have done a general model independent analysis, showed the constraints of $\epsilon_e$, $\epsilon_\mu$, $\epsilon_p$, $\epsilon_n$, and $m_\phi$ for both, and pointed out that where to look for such new physics in the future experiments. 

\bibliographystyle{unsrt}
\bibliography{uwthesis}

\begin{thebibliography}{100}

\bibitem{Pohl:2010zza}
Randolf Pohl et~al.
\newblock {The size of the proton}.
\newblock {\em Nature}, 466:213--216, 2010.

\bibitem{Antognini:1900ns}
Aldo Antognini et~al.
\newblock {Proton Structure from the Measurement of $2S-2P$ Transition
  Frequencies of Muonic Hydrogen}.
\newblock {\em Science}, 339:417--420, 2013.

\bibitem{Mohr:2015ccw}
Peter~J. Mohr, David~B. Newell, and Barry~N. Taylor.
\newblock {CODATA Recommended Values of the Fundamental Physical Constants:
  2014}.
\newblock 2015.

\bibitem{Hill:2011wy}
Richard~J. Hill and Gil Paz.
\newblock {Model independent analysis of proton structure for hydrogenic bound
  states}.
\newblock {\em Phys. Rev. Lett.}, 107:160402, 2011.

\bibitem{Pohl:2013yb}
Randolf Pohl, Ronald Gilman, Gerald~A. Miller, and Krzysztof Pachucki.
\newblock {Muonic hydrogen and the proton radius puzzle}.
\newblock {\em Ann. Rev. Nucl. Part. Sci.}, 63:175--204, 2013.

\bibitem{Hill:2010yb}
Richard~J. Hill and Gil Paz.
\newblock {Model independent extraction of the proton charge radius from
  electron scattering}.
\newblock {\em Phys. Rev.}, D82:113005, 2010.

\bibitem{TuckerSmith:2010ra}
David Tucker-Smith and Itay Yavin.
\newblock {Muonic hydrogen and MeV forces}.
\newblock {\em Phys. Rev.}, D83:101702, 2011.

\bibitem{Blum:2013xva}
Thomas Blum, Achim Denig, Ivan Logashenko, Eduardo de~Rafael, B.~Lee~Roberts,
  Thomas Teubner, and Graziano Venanzoni.
\newblock {The Muon (g-2) Theory Value: Present and Future}.
\newblock 2013.

\bibitem{Davier:2010nc}
Michel Davier, Andreas Hoecker, Bogdan Malaescu, and Zhiqing Zhang.
\newblock {Reevaluation of the Hadronic Contributions to the Muon g-2 and to
  alpha(MZ)}.
\newblock {\em Eur. Phys. J.}, C71:1515, 2011.
\newblock [Erratum: Eur. Phys. J.C72,1874(2012)].

\bibitem{Hagiwara:2011af}
Kaoru Hagiwara, Ruofan Liao, Alan~D. Martin, Daisuke Nomura, and Thomas
  Teubner.
\newblock {(g-2)\underline{\textcolor{white}{a}}mu and
  alpha(M\underline{\textcolor{white}{a}}Z$^\wedge$2) re-evaluated using new
  precise data}.
\newblock {\em J. Phys.}, G38:085003, 2011.

\bibitem{Izaguirre:2014cza}
Eder Izaguirre, Gordan Krnjaic, and Maxim Pospelov.
\newblock {Probing New Physics with Underground Accelerators and Radioactive
  Sources}.
\newblock {\em Phys. Lett.}, B740:61--65, 2015.

\bibitem{Blunden:2005ew}
P.~G. Blunden, W.~Melnitchouk, and J.~A. Tjon.
\newblock {Two-photon exchange in elastic electron-nucleon scattering}.
\newblock {\em Phys. Rev.}, C72:034612, 2005.

\bibitem{Afanasev:2005mp}
Andrei~V. Afanasev, Stanley~J. Brodsky, Carl~E. Carlson, Yu-Chun Chen, and Marc
  Vanderhaeghen.
\newblock {The Two-photon exchange contribution to elastic electron-nucleon
  scattering at large momentum transfer}.
\newblock {\em Phys. Rev.}, D72:013008, 2005.

\bibitem{Arrington:2007ux}
J.~Arrington, W.~Melnitchouk, and J.~A. Tjon.
\newblock {Global analysis of proton elastic form factor data with two-photon
  exchange corrections}.
\newblock {\em Phys. Rev.}, C76:035205, 2007.

\bibitem{Carlson:2007sp}
Carl~E. Carlson and Marc Vanderhaeghen.
\newblock {Two-Photon Physics in Hadronic Processes}.
\newblock {\em Ann. Rev. Nucl. Part. Sci.}, 57:171--204, 2007.

\bibitem{SoLID:2013fta}
SoLID Collaboration.
\newblock {SoLID (Solenoidal Large Intensity Device) Preliminary Conceptual
  Design Report,
  \url{http://hallaweb.jlab.org/12GeV/SoLID/files/solid_precdr.pdf}}.
\newblock 2014.

\bibitem{MUSE}
MUSE Collaboration.
\newblock {Technical Design Report for the Paul Scherrer Institute Experiment
  R-12-01.1: Studying the Proton ``Radius'' Puzzle with $\mu p$ Elastic
  Scattering,
  \url{http://www.physics.rutgers.edu/~rgilman/elasticmup/muse_tdr_mediumfont.pdf}}.
\newblock 2014.

\bibitem{Rock:2001zi}
Stephen Rock.
\newblock {Measuring the phases of G(E) and G(M) of the nucleon}.
\newblock {\em eConf}, C010430:W14, 2001.
\newblock [,238(2001)].

\bibitem{Arrington:2011kb}
John Arrington, Kees de~Jager, and Charles~F. Perdrisat.
\newblock {Nucleon Form Factors: A Jefferson Lab Perspective}.
\newblock {\em J. Phys. Conf. Ser.}, 299:012002, 2011.

\bibitem{Punjabi:2014tna}
Vina Punjabi and Charles~F. Perdrisat.
\newblock {The Proton Form Factor Ratio Measurements at Jefferson Lab}.
\newblock {\em EPJ Web Conf.}, 66:06019, 2014.

\bibitem{Rosenbluth:1950yq}
M.~N. Rosenbluth.
\newblock {High Energy Elastic Scattering of Electrons on Protons}.
\newblock {\em Phys. Rev.}, 79:615--619, 1950.

\bibitem{Akhiezer:1974em}
A.~I. Akhiezer and Mikhail.P. Rekalo.
\newblock {Polarization effects in the scattering of leptons by hadrons}.
\newblock {\em Sov. J. Part. Nucl.}, 4:277, 1974.
\newblock [Fiz. Elem. Chast. Atom. Yadra4,662(1973)].

\bibitem{Arnold:1980zj}
R.~G. Arnold, Carl~E. Carlson, and Franz Gross.
\newblock {Polarization Transfer in Elastic electron Scattering from Nucleons
  and Deuterons}.
\newblock {\em Phys. Rev.}, C23:363, 1981.

\bibitem{Essig:2013lka}
Rouven Essig et~al.
\newblock {Working Group Report: New Light Weakly Coupled Particles}.
\newblock 2013.

\bibitem{Bjorken:2009mm}
James~D. Bjorken, Rouven Essig, Philip Schuster, and Natalia Toro.
\newblock {New Fixed-Target Experiments to Search for Dark Gauge Forces}.
\newblock {\em Phys. Rev.}, D80:075018, 2009.

\bibitem{Andreas:2012mt}
Sarah Andreas, Carsten Niebuhr, and Andreas Ringwald.
\newblock {New Limits on Hidden Photons from Past Electron Beam Dumps}.
\newblock {\em Phys. Rev.}, D86:095019, 2012.

\bibitem{vonWeizsacker:1934nji}
C.~F. von Weizsacker.
\newblock {Radiation emitted in collisions of very fast electrons}.
\newblock {\em Z. Phys.}, 88:612--625, 1934.

\bibitem{Williams:1935dka}
E.~J. Williams.
\newblock {Correlation of certain collision problems with radiation theory}.
\newblock {\em Kong. Dan. Vid. Sel. Mat. Fys. Med.}, 13N4(4):1--50, 1935.

\bibitem{Kim:1973he}
Kwang~Je Kim and Yung-Su Tsai.
\newblock {IMPROVED WEIZSACKER-WILLIAMS METHOD AND ITS APPLICATION TO LEPTON
  AND W BOSON PAIR PRODUCTION}.
\newblock {\em Phys. Rev.}, D8:3109, 1973.

\bibitem{Tsai:1973py}
Yung-Su Tsai.
\newblock {Pair Production and Bremsstrahlung of Charged Leptons}.
\newblock {\em Rev. Mod. Phys.}, 46:815, 1974.
\newblock [Erratum: Rev. Mod. Phys.49,521(1977)].

\bibitem{Tsai:1986tx}
Yung-Su Tsai.
\newblock {AXION BREMSSTRAHLUNG BY AN ELECTRON BEAM}.
\newblock {\em Phys. Rev.}, D34:1326, 1986.

\bibitem{Bjorken:1988as}
J.~D. Bjorken, S.~Ecklund, W.~R. Nelson, A.~Abashian, C.~Church, B.~Lu, L.~W.
  Mo, T.~A. Nunamaker, and P.~Rassmann.
\newblock {Search for Neutral Metastable Penetrating Particles Produced in the
  SLAC Beam Dump}.
\newblock {\em Phys. Rev.}, D38:3375, 1988.

\bibitem{Landau:1948kw}
L.~D. Landau.
\newblock {On the angular momentum of a system of two photons}.
\newblock {\em Dokl. Akad. Nauk Ser. Fiz.}, 60(2):207--209, 1948.

\bibitem{Yang:1950rg}
Chen-Ning Yang.
\newblock {Selection Rules for the Dematerialization of a Particle Into Two
  Photons}.
\newblock {\em Phys. Rev.}, 77:242--245, 1950.

\bibitem{Zhemchugov:2014dza}
E.~V. Zhemchugov.
\newblock {On $Z \to \gamma \gamma$ decay and cancellation of axial anomaly in
  $Z \to \gamma \gamma$ transition amplitude for massive fermions}.
\newblock {\em Phys. Atom. Nucl.}, 77:11, 2014.

\bibitem{Pospelov:2008jk}
Maxim Pospelov, Adam Ritz, and Mikhail~B. Voloshin.
\newblock {Bosonic super-WIMPs as keV-scale dark matter}.
\newblock {\em Phys. Rev.}, D78:115012, 2008.

\bibitem{Anastasiou:2003gr}
Charalampos Anastasiou, Kirill Melnikov, and Frank Petriello.
\newblock {A new method for real radiation at NNLO}.
\newblock {\em Phys. Rev.}, D69:076010, 2004.

\bibitem{Asatrian:2012tp}
H.~M. Asatrian, A.~Hovhannisyan, and A.~Yeghiazaryan.
\newblock {The phase space analysis for three and four massive particles in
  final states}.
\newblock {\em Phys. Rev.}, D86:114023, 2012.

\bibitem{DeJager:1987qc}
H.~De Vries, C.~W.~De Jager, and C.~De Vries.
\newblock {Nuclear charge and magnetization density distribution parameters
  from elastic electron scattering}.
\newblock {\em Atom. Data Nucl. Data Tabl.}, 36:495, 1987.

\bibitem{atomic_form_factor}
P.~J. Brown, A.~G. Fox, E.~N. Maslen, M.~A. O'Keefe, and B.~T.~M. Willis.
\newblock {\em International Tables for Crystallography}.
\newblock 2006.
\newblock ch. 6.1, pp. 554-595.

\bibitem{Tsai:1966js}
Yung-Su Tsai and Van Whitis.
\newblock {THICK TARGET BREMSSTRAHLUNG AND TARGET CONSIDERATION FOR SECONDARY
  PARTICLE PRODUCTION BY ELECTRONS}.
\newblock {\em Phys. Rev.}, 149:1248--1257, 1966.

\bibitem{Pospelov:2008zw}
Maxim Pospelov.
\newblock {Secluded U(1) below the weak scale}.
\newblock {\em Phys. Rev.}, D80:095002, 2009.

\bibitem{Bouchendira:2010es}
Rym Bouchendira, Pierre Clade, Saida Guellati-Khelifa, Francois Nez, and
  Francois Biraben.
\newblock {New determination of the fine structure constant and test of the
  quantum electrodynamics}.
\newblock {\em Phys. Rev. Lett.}, 106:080801, 2011.

\bibitem{Eides:2000xc}
Michael~I. Eides, Howard Grotch, and Valery~A. Shelyuto.
\newblock {Theory of light hydrogen - like atoms}.
\newblock {\em Phys. Rept.}, 342:63--261, 2001.

\bibitem{Liu:2016qwd}
Yu-Sheng Liu, David McKeen, and Gerald~A. Miller.
\newblock {Electrophobic Scalar Boson and Muonic Puzzles}.
\newblock {\em Phys. Rev. Lett.}, 117(10):101801, 2016.

\bibitem{Beranek:2013yqa}
T.~Beranek, H.~Merkel, and M.~Vanderhaeghen.
\newblock {Theoretical framework to analyze searches for hidden light gauge
  bosons in electron scattering fixed target experiments}.
\newblock {\em Phys. Rev.}, D88:015032, 2013.

\bibitem{Riordan:1987aw}
E.~M. Riordan et~al.
\newblock {A Search for Short Lived Axions in an Electron Beam Dump
  Experiment}.
\newblock {\em Phys. Rev. Lett.}, 59:755, 1987.

\bibitem{Davier:1989wz}
M.~Davier and H.~Nguyen~Ngoc.
\newblock {An Unambiguous Search for a Light Higgs Boson}.
\newblock {\em Phys. Lett.}, B229:150--155, 1989.

\bibitem{Landau:1953um}
L.~D. Landau and I.~Pomeranchuk.
\newblock {Limits of applicability of the theory of bremsstrahlung electrons
  and pair production at high-energies}.
\newblock {\em Dokl. Akad. Nauk Ser. Fiz.}, 92:535--536, 1953.

\bibitem{Landau:1953gr}
L.~D. Landau and I.~Pomeranchuk.
\newblock {Electron cascade process at very high-energies}.
\newblock {\em Dokl. Akad. Nauk Ser. Fiz.}, 92:735--738, 1953.

\bibitem{Migdal:1956tc}
Arkady~B. Migdal.
\newblock {Bremsstrahlung and pair production in condensed media at
  high-energies}.
\newblock {\em Phys. Rev.}, 103:1811--1820, 1956.

\bibitem{Anthony:1995fs}
P.~L. Anthony et~al.
\newblock {An Accurate measurement of the Landau-Pomeranchuk-Migdal effect}.
\newblock {\em Phys. Rev. Lett.}, 75:1949--1952, 1995.

\bibitem{Alwall:2007st}
Johan Alwall, Pavel Demin, Simon de~Visscher, Rikkert Frederix, Michel Herquet,
  Fabio Maltoni, Tilman Plehn, David~L. Rainwater, and Tim Stelzer.
\newblock {MadGraph/MadEvent v4: The New Web Generation}.
\newblock {\em JHEP}, 09:028, 2007.

\bibitem{Essig:2010xa}
Rouven Essig, Philip Schuster, Natalia Toro, and Bogdan Wojtsekhowski.
\newblock {An Electron Fixed Target Experiment to Search for a New Vector Boson
  A' Decaying to e+e-}.
\newblock {\em JHEP}, 02:009, 2011.

\bibitem{Wojtsekhowski:2009vz}
B.~Wojtsekhowski.
\newblock {Searching for a U-boson with a positron beam}.
\newblock {\em AIP Conf. Proc.}, 1160:149--154, 2009.

\bibitem{Abrahamyan:2011gv}
S.~Abrahamyan et~al.
\newblock {Search for a New Gauge Boson in Electron-Nucleus Fixed-Target
  Scattering by the APEX Experiment}.
\newblock {\em Phys. Rev. Lett.}, 107:191804, 2011.

\bibitem{Izaguirre:2013uxa}
Eder Izaguirre, Gordan Krnjaic, Philip Schuster, and Natalia Toro.
\newblock {New Electron Beam-Dump Experiments to Search for MeV to few-GeV Dark
  Matter}.
\newblock {\em Phys. Rev.}, D88:114015, 2013.

\bibitem{Raggi:2014zpa}
Mauro Raggi and Venelin Kozhuharov.
\newblock {Proposal to Search for a Dark Photon in Positron on Target
  Collisions at DA$\Phi$NE Linac}.
\newblock {\em Adv. High Energy Phys.}, 2014:959802, 2014.

\bibitem{Izaguirre:2014bca}
Eder Izaguirre, Gordan Krnjaic, Philip Schuster, and Natalia Toro.
\newblock {Testing GeV-Scale Dark Matter with Fixed-Target Missing Momentum
  Experiments}.
\newblock {\em Phys. Rev.}, D91(9):094026, 2015.

\bibitem{Carlson:2012pc}
Carl~E. Carlson and Benjamin~C. Rislow.
\newblock {New Physics and the Proton Radius Problem}.
\newblock {\em Phys. Rev.}, D86:035013, 2012.

\bibitem{Karshenboim:2014tka}
Savely~G. Karshenboim, David McKeen, and Maxim Pospelov.
\newblock {Constraints on muon-specific dark forces}.
\newblock {\em Phys. Rev.}, D90(7):073004, 2014.
\newblock [Addendum: Phys. Rev.D90,no.7,079905(2014)].

\bibitem{Karshenboim:2001yy}
Savely~G. Karshenboim and Valery~A. Shelyuto.
\newblock {Hadronic vacuum polarization contribution to the muonium hyperfine
  splitting}.
\newblock {\em Phys. Lett.}, B517:32--36, 2001.

\bibitem{Liu:1999iz}
Weiwen Liu et~al.
\newblock {High precision measurements of the ground state hyperfine structure
  interval of muonium and of the muon magnetic moment}.
\newblock {\em Phys. Rev. Lett.}, 82:711--714, 1999.

\bibitem{Jackiw:1972jz}
R.~Jackiw and Steven Weinberg.
\newblock {Weak interaction corrections to the muon magnetic moment and to
  muonic atom energy levels}.
\newblock {\em Phys. Rev.}, D5:2396--2398, 1972.

\bibitem{PhysRevA.82.042513}
Brianna~J. Mount, Matthew Redshaw, and Edmund~G. Myers.
\newblock Atomic masses of
  $^{6}\mathrm{Li},^{23}\mathrm{Na},^{39,41}\mathrm{K},^{85,87}\mathrm{Rb}$,
  and $^{133}\mathrm{Cs}$.
\newblock {\em Phys. Rev. A}, 82:042513, Oct 2010.

\bibitem{Tsertos:1989gv}
Haralabos Tsertos, C.~Kozhuharov, P.~Armbruster, P.~Kienle, B.~Krusche, and
  K.~Schreckenbach.
\newblock {High Sensitivity Measurements of the Excitation Function for Bhabha
  Scattering at {MeV} Energies}.
\newblock {\em Phys. Rev.}, D40:1397, 1989.

\bibitem{Leeb:1992qf}
H.~Leeb and J.~Schmiedmayer.
\newblock {Constraint on hypothetical light interacting bosons from low-energy
  neutron experiments}.
\newblock {\em Phys. Rev. Lett.}, 68:1472--1475, 1992.

\bibitem{Machleidt:2001rw}
R.~Machleidt and I.~Slaus.
\newblock {The Nucleon-nucleon interaction: Topical review}.
\newblock {\em J. Phys.}, G27:R69--R108, 2001.

\bibitem{Ericson:1983vw}
Torleif Erik~Oskar Ericson and G.~A. Miller.
\newblock {Charge Dependence of Nuclear Forces}.
\newblock {\em Phys. Lett.}, B132:32, 1983.
\newblock [Conf. Proc.C830821V2,107(1983)].

\bibitem{Brown:1975us}
G.~E. Brown and A.~D. Jackson.
\newblock {The Nucleon-Nucleon Interaction. 2.}
\newblock 1975.

\bibitem{Mattuck:1976xt}
R.~D. Mattuck.
\newblock {\em {A Guide to Feynman Diagrams in the Many Body Problem (Second
  Edition)}}.
\newblock 1976.

\bibitem{SanjayReddy}
{\em Discussion with Sanjay Reddy and Martin J. Savage}.

\bibitem{Friar:1969zz}
J.~L. Friar.
\newblock {THE He-3 - H-3 CHARGE FORM-FACTOR AND THE He-3 COULOMB ENERGY}.
\newblock {\em Nucl. Phys.}, A156:43--52, 1970.

\bibitem{Friar:1978mr}
James~Lewis Friar and B.~F. Gibson.
\newblock {Coulomb Energies in S Shell Nuclei and Hypernuclei}.
\newblock {\em Phys. Rev.}, C18:908, 1978.

\bibitem{Coon:1987kt}
S.~A. Coon and R.~C. Barrett.
\newblock {$\rho - \Omega$ Mixing in Nuclear Charge Asymmetry}.
\newblock {\em Phys. Rev.}, C36:2189--2194, 1987.

\bibitem{Miller:1990iz}
G.~A. Miller, B.~M.~K. Nefkens, and I.~Slaus.
\newblock {Charge symmetry, quarks and mesons}.
\newblock {\em Phys. Rept.}, 194:1--116, 1990.

\bibitem{Miller:2011yw}
G.~A. Miller, A.~W. Thomas, J.~D. Carroll, and J.~Rafelski.
\newblock {Natural Resolution of the Proton Size Puzzle}.
\newblock {\em Phys. Rev.}, A84:020101, 2011.

\bibitem{Miller:2012ht}
Gerald~A. Miller, Anthony~W. Thomas, and Jonathan~D. Carroll.
\newblock {Nuclear Quasi-Elastic Electron Scattering Limits Nucleon Off-Mass
  Shell Properties}.
\newblock {\em Phys. Rev.}, C86:065201, 2012.

\bibitem{Miller:2012ne}
Gerald~A. Miller.
\newblock {Proton Polarizability Contribution: Muonic Hydrogen Lamb Shift and
  Elastic Scattering}.
\newblock {\em Phys. Lett.}, B718:1078--1082, 2013.

\bibitem{Antognini:2015moa}
A.~Antognini et~al.
\newblock {Experiments towards resolving the proton charge radius puzzle}.
\newblock {\em EPJ Web Conf.}, 113:01006, 2016.

\bibitem{Krauth:2015nja}
Julian~J. Krauth, Marc Diepold, Beatrice Franke, Aldo Antognini, Franz
  Kottmann, and Randolf Pohl.
\newblock {Theory of the n=2 levels in muonic deuterium}.
\newblock {\em Annals Phys.}, 366:168--196, 2016.

\bibitem{Pohl1:2016xoo}
Randolf Pohl et~al.
\newblock {Laser spectroscopy of muonic deuterium}.
\newblock {\em Science}, 353(6300):669--673, 2016.

\bibitem{AldoAntognini}
{\em Reported at NuPECC 2015 by Aldo Antognini for the CREMA collaboration}.

\bibitem{Sick:2014yha}
Ingo Sick.
\newblock {Zemach moments of $^3$He and $^4$He}.
\newblock {\em Phys. Rev.}, C90(6):064002, 2014.

\bibitem{Merkel:2014avp}
H.~Merkel et~al.
\newblock {Search at the Mainz Microtron for Light Massive Gauge Bosons
  Relevant for the Muon g-2 Anomaly}.
\newblock {\em Phys. Rev. Lett.}, 112(22):221802, 2014.

\bibitem{Lees:2014xha}
J.~P. Lees et~al.
\newblock {Search for a Dark Photon in $e^+e^-$ Collisions at BaBar}.
\newblock {\em Phys. Rev. Lett.}, 113(20):201801, 2014.

\bibitem{Battaglieri:2014hga}
M.~Battaglieri et~al.
\newblock {The Heavy Photon Search Test Detector}.
\newblock {\em Nucl. Instrum. Meth.}, A777:91--101, 2015.

\bibitem{Freytsis:2009bh}
Marat Freytsis, Grigory Ovanesyan, and Jesse Thaler.
\newblock {Dark Force Detection in Low Energy e-p Collisions}.
\newblock {\em JHEP}, 01:111, 2010.

\bibitem{Cowan:2013mma}
Ray~F. Cowan.
\newblock {Experimental concept and design of DarkLight, a search for a heavy
  photon}.
\newblock {\em AIP Conf. Proc.}, 1563:126--130, 2013.

\bibitem{Wojtsekhowski:2012zq}
B.~Wojtsekhowski, D.~Nikolenko, and I.~Rachek.
\newblock {Searching for a new force at VEPP-3}.
\newblock 2012.

\bibitem{An:2013yfc}
Haipeng An, Maxim Pospelov, and Josef Pradler.
\newblock {New stellar constraints on dark photons}.
\newblock {\em Phys. Lett.}, B725:190--195, 2013.

\bibitem{Rrapaj:2015wgs}
Ermal Rrapaj and Sanjay Reddy.
\newblock {Nucleon-nucleon bremsstrahlung of dark gauge bosons and revised
  supernova constraints}.
\newblock 2015.

\bibitem{Cirigliano:2005ck}
Vincenzo Cirigliano, Benjamin Grinstein, Gino Isidori, and Mark~B. Wise.
\newblock {Minimal flavor violation in the lepton sector}.
\newblock {\em Nucl. Phys.}, B728:121--134, 2005.

\bibitem{Batell:2016ove}
Brian Batell, Nicholas Lange, David McKeen, Maxim Pospelov, and Adam Ritz.
\newblock {The Leptonic Higgs Portal}.
\newblock 2016.

\bibitem{Batell:2009jf}
Brian Batell, Maxim Pospelov, and Adam Ritz.
\newblock {Multi-lepton Signatures of a Hidden Sector in Rare B Decays}.
\newblock {\em Phys. Rev.}, D83:054005, 2011.

\bibitem{Fox:2011qd}
Patrick~J. Fox, Jia Liu, David Tucker-Smith, and Neal Weiner.
\newblock {An Effective Z'}.
\newblock {\em Phys. Rev.}, D84:115006, 2011.

\bibitem{Gardner:2015wea}
S.~Gardner, R.~J. Holt, and A.~S. Tadepalli.
\newblock {New Prospects in Fixed Target Searches for Dark Forces with the
  SeaQuest Experiment at Fermilab}.
\newblock 2015.

\bibitem{Essig:2010gu}
Rouven Essig, Roni Harnik, Jared Kaplan, and Natalia Toro.
\newblock {Discovering New Light States at Neutrino Experiments}.
\newblock {\em Phys. Rev.}, D82:113008, 2010.

\bibitem{Gilman:2013eiv}
R.~Gilman et~al.
\newblock {Studying the Proton "Radius" Puzzle with p Elastic Scattering}.
\newblock 2013.

\bibitem{Liu:2015sba}
Yu-Sheng Liu and Gerald~A. Miller.
\newblock {Polarized Lepton-Nucleon Elastic Scattering and a Search for a Light
  Scalar Boson}.
\newblock {\em Phys. Rev.}, C92(3):035209, 2015.

\bibitem{Hertzog:2015jru}
David~W. Hertzog.
\newblock {Next Generation Muon g-2 Experiments}.
\newblock {\em EPJ Web Conf.}, 118:01015, 2016.

\end{thebibliography}



\end{document}